\documentclass[aps,prd,longbibliography,reprint,twocolumn,amsmath,amssymb,amsfonts,showpacs,superscriptaddress]{revtex4-2}

\usepackage{ifpdf}
\usepackage{dcolumn}
\usepackage{enumerate}
\usepackage{enumitem}
\usepackage{bm}
\usepackage{dsfont}
\usepackage{here}
\usepackage{bm}
\usepackage{float}
\floatstyle{plaintop}
\restylefloat{table}
\ifpdf

      \usepackage[pdftex]{graphicx}  
      \usepackage[pdftex]{epsfig}

\usepackage[usenames]{color}
\definecolor{navyblue}{rgb}{0.0, 0.0, 0.5}
\definecolor{ferrarired}{rgb}{1.0, 0.11, 0.0}
\definecolor{persianblue}{rgb}{0.11, 0.22, 0.73}

\usepackage[pdftex]{hyperref}
\hypersetup{
		colorlinks=true,
		citecolor=blue,%
		linkcolor=ferrarired,%
		urlcolor=persianblue,%
        filecolor=blue,
        linktoc=page
	}

\else

      \usepackage[dvips]{graphicx}  
      \usepackage[dvips]{epsfig}

      \usepackage[dvipdfm]{hyperref}
\hypersetup{
		colorlinks=true,
		citecolor=blue,%
		linkcolor=ferrarired,%
		urlcolor=persianblue,%
        filecolor=blue,
        linktoc=page
	}

\fi

\usepackage{xfrac}

\usepackage{amsmath, amsfonts}
\usepackage{nccmath}

\DeclareMathAlphabet{\mathpzc}{OT1}{pzc}{m}{it}

\usepackage{tablefootnote}

\makeatletter
\def\l@subsubsection#1#2{}
\makeatother

\begin{document}

\title{Gravitational waves in massive gravity: Waveforms generated \\ by a particle plunging into a black hole and the excitation \\ of quasinormal modes and quasibound states}

\author{Mohamed \surname{Ould~El~Hadj}}\email{med.ouldelhadj@gmail.com}

\affiliation{France}

\date{\today}

\begin{abstract}
With the aim of testing massive gravity in the context of black hole physics, we investigate the gravitational radiation emitted by a massive particle plunging into a Schwarzschild black hole from slightly below the innermost stable circular orbit. To do so, we first construct the quasinormal and quasibound resonance spectra of the spin-2 massive field for odd and even parity. Then, we compute the waveforms produced by the plunging particle and study their spectral content. This allows us to highlight and interpret important phenomena in the plunge regime, including (i) the excitation of quasibound states, with particular emphasis on the amplification and slow decay of the post-ringdown phase of the even-parity dipolar mode due to \emph{harmonic resonance};  (ii) during the adiabatic phase, the waveform emitted by the plunging particle is very well described by the waveform emitted by the particle living on the innermost stable circular orbit, and (iii) the regularized waveforms and their unregularized counterparts constructed from the quasinormal mode spectrum are in excellent agreement. Finally, we construct, for arbitrary directions of observation and, in particular, outside the orbital plane of the plunging particle, the regularized multipolar waveforms, i.e., the waveforms constructed by summing over partial waveforms. 
\end{abstract}

\maketitle

\tableofcontents

\section{Introduction}
\label{sec_1}

Massive gravity, an extension of general relativity in which the graviton---a hypothetical quantum particle mediating gravitational interactions---acquires mass, offers a solid theoretical framework for addressing fundamental questions in cosmology and astrophysics. By modifying gravitational interactions on large scales, it naturally explains the accelerated expansion of the Universe without invoking dark energy or a cosmological constant~\cite{deRham:2014zqa,Hinterbichler:2011tt}.  

Significant progress has been made since the original Fierz-Pauli theory~\cite{Fierz:1939ix,Fierz:1939zz}, which was plagued by inconsistencies such as a discontinuity with general relativity in the limit where the graviton mass is taken to zero, and the presence of a ghost problem. The  \emph{ghost-free} formulation of massive gravity~\cite{deRham:2010ik,Hassan:2011hr} now provides a consistent framework with broad applications. While its ability to reproduce the accelerated expansion of the Universe has been extensively studied (see for e.g., Refs \cite{Akrami:2018ylq,Heisenberg:2018vsk} and references therein), its implications for black hole physics remain an open and exciting area of investigation. Nevertheless, notable contributions in this area include Ref.~\cite{Volkov:2013roa} and references therein for articles dealing with BH solutions in massive gravity, and Refs.~\cite{Babichev:2013una,Brito:2013wya,Brito:2013yxa} for important considerations on the problem of BH stability in massive gravity.  

With the advent of next-generation gravitational wave detectors, such as LISA \cite{LISA:17}, the study of extreme mass ratio inspirals (EMRIs) has gained particular importance (see e.g., \cite{Cárdenas-Avendaño2024} and references therein). These systems, in which a compact stellar mass object spirals into a supermassive black hole (BH), provide a unique opportunity to test general relativity and its potential modifications in the strong-field regime~\cite{Liu:2023onj}. 

In this article, we investigate the possibility to test massive gravity. More specifically, we focus on the Fierz-Pauli theory~\cite{Fierz:1939ix,Fierz:1939zz}, a field theory thoroughly studied by Brito \emph{et al.} in Ref.~\cite{Brito:2013wya}. Our study is set in the framework of black hole physics, analyzing the radiation emitted by a ``particle'' plunging into a Schwarzschild BH from just below the innermost stable circular orbit (ISCO). Assuming an extreme mass ratio, where the BH is much more massive than the particle, the emitted radiation can be studied through the framework of BH perturbation theory. This problem is of fundamental importance within the framework of Einstein's general relativity and has been extensively studied in the literature (see, for example, Refs.~\cite{Detweiler:1979xr,Oohara:1984bw,Buonanno:1998gg,Buonanno:2000ef,Ori:2000zn,Baker:2001nu,Blanchet:2005rj,Campanelli:2006gf,Damour:2007xr,Sperhake:2007gu,Mino:2008at,Hadar:2009ip,Hadar:2011vj,Price:2013paa,
Hadar:2014dpa,Hadar:2015xpa}). The \emph{plunge regime} represents the final phase in the evolution of a stellar-mass object orbiting a supermassive BH and is crucial for understanding the late-time dynamics of binary BH systems. The waveform produced during this regime encodes key information about the BH's final properties. Moreover, the Schwarzschild BH, a fundamental solution of Einstein's general relativity,  is also central to the study of massive gravity \cite{Volkov:2013roa,Tasinato:2013rza,Li:2016fbf}. To our knowledge, no work has addressed this fundamental problem in its entirety. However, in a recent study~\cite{Cardoso:2023dwz}, Cardoso \emph{et al.} analyzed the excitation of dipole modes in gravity theories within the framework of the EMRI problem (see also Ref.\cite{Dong:2020odp} for studies on gravitational wave echoes and Ref.\cite{Konoplya:2023fmh} for the asymptotic tails of massive gravitons). In a previous paper~\cite{Decanini:2014kha} (see also Ref.\cite{Decanini:2014bwa}, which includes a more detailed analysis with analytical results and extensions to other bosonic fields, as well as Ref.\cite{Decanini:2016ifm}), we partially addressed this issue by focusing on specific aspects related to the excitation of quasinormal modes (QNMs). Additionally, in~\cite{Decanini:2015yba}, we considered a toy model where the massive spin-$2$ perturbations were replaced by a massive scalar field and a linear coupling between the particle and this field. We computed the quadrupolar waveform produced by the plunging particle and analyzed its spectral content. This allowed us to describe the excitation of both the QNMs and the quasibound states (QBSs) of the BH, and to demonstrate the influence of the field mass on the amplitude of the emitted signal. In particular, we studied the contribution of the part of the signal that is produced when the particle moves along quasicircular orbits near the ISCO. As expected, the phenomena identified with the toy model are confirmed in the more physical scenario investigated in this article. Furthermore, the study presented here has led to new and original results that enrich our understanding of the problem.

Our paper is organized as follows. In Sec. \ref{sec_2},  we give a brief overview of the Schwarzschild metric and then introduce the geodesic equations describing the trajectory of a massive particle plunging into a Schwarzschild BH. In Sec. \ref{sec_3}, we focus on gravitational perturbations in massive spin-2 fields, starting with a recall of the linearized field equations of Fierz-Pauli theory in the Schwarzschild background~\cite{Brito:2013wya}, which can be obtained, e.g., by linearization of the pathology-free bimetric theory of Hassan, Schmidt-May, and von Strauss~\cite{Hassan:2012wr}, an extension, in curved spacetime, of the fundamental work of de Rham, Gabadadze, and Tolley~\cite{deRham:2010ik,deRham:2010kj}. We derive the master equations for both odd- and even-parity sectors, including the source terms associated with the plunging particle. These derivations, as well as the conventions and notations used, are detailed in Appendix \ref{appendix_A}. In Sec. \ref{sec_4}, we numerically construct the quasinormal and quasibound resonance spectra of the spin-$2$ massive field. This is achieved by solving the homogeneous coupled differential equations for each parity sector under appropriate boundary conditions.  Using an extended version of the Hill determinant method, adapted to matrix-valued systems (described in Appendix \ref{appendix_B}), we present the complete QNM spectrum for the even-parity sector for the first time. In addition, for the even-parity monopole mode, we identified two new branches, one associated with the quasinormal frequency spectrum and the other with the quasibound frequency spectrum.

In Sec. \ref{sec_5}, we study the gravitational waves generated by a massive particle plunging into a Schwarzschild BH from slightly below the ISCO. We begin by deriving the theoretical expressions for the emitted waveforms, considering both even- and odd-parity gravitational perturbations for arbitrary ($\ell, m$) modes, governed by the master equations. Using Green's matrix techniques in the frequency domain, we solve the two coupled master equations governing the ($\ell \geq 2$) odd-parity perturbations and the three coupled equations for the ($\ell \geq 2$) even-parity perturbations. For the odd-parity dipole mode ($\ell = 1$), we solve a single master equation, while the even-parity monopole ($\ell = 0$) and dipole ($\ell = 1$) modes are treated as a system of two coupled equations. The source terms for these systems are constructed from the closed-form expression of the particle's plunge trajectory. From the resulting waveforms, we extract the QNM contributions corresponding to the BH's gravitational ringing (or ringdown). Then, by summing the partial waveforms for the even and odd polarization sectors, we construct the multipolar waveforms for different polarizations. Finally, we detail the numerical methods used to compute the waveforms. The exact waveforms, obtained theoretically as integrals over the Schwarzschild radial coordinate, diverge strongly near the ISCO. For odd-parity perturbations, numerical regularization is performed using Levin's algorithm \cite{Levin1996}. For even-parity perturbations, however, a preliminary reduction of the divergence by successive integration by parts is required, extending the method we developed in our previous work for a charged particle plunging from the ISCO into a Schwarzschild BH \cite{Folacci:2018vtf} and in the case of a massive particle \cite{Folacci:2018cic}. Details of this regularization method are given in Appendix \ref{appendix_C}.

In Sec.~\ref{sec_6} we present our numerical results of the waveforms produced by the plunging particle, focusing on their different phases. We first display the regularized waveforms and their spectral content, highlighting the excitation of QBSs. In particular, our results show the \emph{resonant behavior} of the even-parity dipole mode due to \emph{harmonic resonance}, leading to a strong amplification of its QBS mode. We also study the adiabatic phase of waveforms generated by a particle in circular motion near the ISCO. We find that these waveforms are accurately described by those emitted by the particle living on the ISCO. In addition, we compare the regularized waveforms with their unregularized counterparts constructed from the QNM spectrum only. Finally, we display the emitted multipolar waveforms obtained by summing over (\(\ell, m\)) partial modes for arbitrary observation directions, in particular outside the orbital plane of the plunging particle. The main results obtained in this article are summarized in the conclusion (Sec.~\ref{sec_7}).

The appendixes contain additional technical details to supplement the main text. Appendix \ref{appendix_A} gives the full derivation of the perturbation equations for massive spin-2 fields, covering the structure of the gravitational perturbations and the stress-energy tensor. Appendix \ref{appendix_B} details the numerical methods for resolving the resonance spectra, including the matrix-valued Hill determinant approach and its application to both parity sectors. Appendix \ref{appendix_C} discusses the regularization techniques for divergent partial wave amplitudes, and Appendix \ref{appendix_D} derives the source terms and waveforms for a massive particle on a circular orbit, dealing with both parity sectors.

Throughout this article, we adopt units such that $G=c=1$ and we use the geometrical conventions of Ref.~\cite{Misner:1974qy}.

\section{The Schwarzschild BH and the plunging massive particle}
\label{sec_2}

Let us recall that the exterior region of a Schwarzschild BH with mass $M$ is defined by the metric
\begin{equation}\label{Metric_Schwarzschild}
ds^2 = -f(r) \, dt^2 + f(r)^{-1} \, dr^2 + r^2 \, d\sigma_2^2
\end{equation}
where $f(r) = 1 - \frac{2M}{r}$, and $d\sigma_2^2 = d\theta^2 + \sin^2 \theta \, d\varphi^2$ denotes the metric on the unit 2-sphere $S^2$. The Schwarzschild coordinates $(t, r, \theta, \varphi)$ satisfy the following ranges: $t \in ]-\infty, +\infty[$, $r \in ]2M, +\infty[$, $\theta \in [0, \pi]$, and $\varphi \in [0, 2\pi]$. Additionally, we introduce the tortoise coordinate $r_\ast \in ]-\infty, +\infty[$, defined by the relation $dr/dr_\ast = f(r)$, which is explicitly given by $r_\ast(r) = r + 2M \ln[r/(2M) - 1]$. This function $r_\ast = r_\ast(r)$ defines a bijection from the interval $]2M, +\infty[$ to $]-\infty, +\infty[$.

\begin{figure}[htbp]
 \includegraphics[scale=0.60]{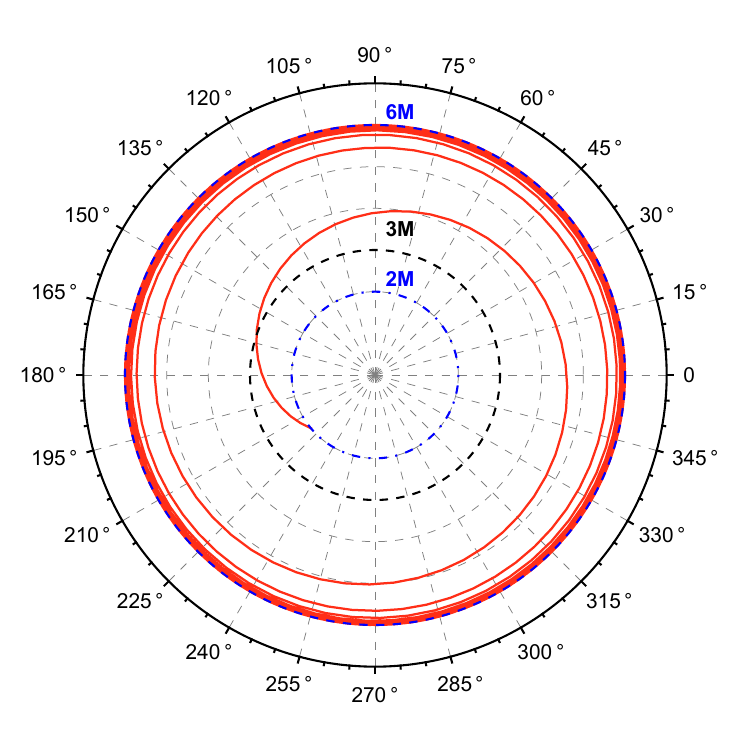}
\caption{\label{Fig:Trajectoire}  The plunge trajectory is obtained from Eq.~\eqref{plunge_phi_trajectory_alternate}, assuming the particle starts at \( r = r_\text{ISCO}(1- \epsilon) \) with \( \epsilon = 10^{-4} \) and \( \varphi_0 = 0 \). The ISCO (\( r = 6M \)), horizon (\( r = 2M \)), and photon sphere (\( r = 3M \)) are marked by blue dashed, blue dot-dashed, and black dashed lines, respectively.}
 \end{figure}
 
We refer to the coordinates of the timelike geodesic $\gamma$ followed by the plunging particle as $t_{p}(\tau)$, $r_{p}(\tau)$, $\theta_{p}(\tau)$, and $\varphi_{p}(\tau)$, where $\tau$ indicates the proper time of the particle, with $m_0$ being its mass. Assuming the particle's trajectory lies within the black hole's equatorial plane, we have $\theta_{p}(\tau) = \pi/2$ without loss of generality. The equations governing the geodesic $\gamma$ are detailed in \cite{Chandrasekhar:1985kt}. The time and azimuthal components of the 4-velocity are determined by the conserved energy and angular momentum per unit mass, respectively,
\begin{subequations}
\label{geodesic_eq}
\begin{eqnarray}
& & f(r_{p})\frac{dt_{p}}{d\tau}=\widetilde{E}, \label{geodesic_eq_part_1} \\
& & r_{p}^{2}\frac{d\varphi_{p}}{d\tau}=\widetilde{L}, \label{geodesic_eq_part_2}
\end{eqnarray}
while the radial component is derived from the normalization condition of the 4-velocity to unity
\begin{equation}
\label{geodesic_eq_part_3}
\left(\frac{dr_{p}}{d\tau}\right)^{2} +\frac{\widetilde{L}^{2}}{r_{p}^{2}}f(r_{p})-\frac{2M}{r_{p}}
=\widetilde{E}^{2}-1.
\end{equation}
\end{subequations}
Here, $\widetilde{E}$ and $\widetilde{L}$ correspond to the particle's energy and angular momentum per unit mass, both of which are conserved quantities determined at the ISCO ($r_\text{\tiny{ISCO}} = 6M$) and given by 
\begin{equation} \label{E_L_isco}
\widetilde{E}=\frac{2\sqrt{2}}{3} \quad \text{and} \quad \widetilde{L}=2\sqrt{3}M.
\end{equation}

To determine the relation between $t_p$ and $r_p$, we integrate equation (\ref{geodesic_eq_part_3}). Using equation (\ref{geodesic_eq_part_1}), we transform the derivative with respect to $\tau$ into a derivative with respect to $t_p$. Considering the condition given by (\ref{E_L_isco}), the integration yields
\begin{eqnarray}
\label{trajectory_plung}
\frac{t_{p}(r)}{2M}&=&\frac{2\sqrt{2}\left(r-24M\right)}{2M\left(6M/r-1\right)^{1/2}}-22\sqrt{2}
\tan^{-1}\left[\left(6M/r-1\right)^{1/2}\right]\nonumber\\
&&+ 2\tanh^{-1}\left[\frac{1}{\sqrt{2}}\left(6M/r-1\right)^{1/2}\right]+ \frac{t_{0}}{2M}.
\end{eqnarray}

Similarly, to find the relation between $\varphi_p$ and $r_p$, we use Eqs.~(\ref{geodesic_eq_part_2})~and~(\ref{geodesic_eq_part_3}). Taking into account the condition from (\ref{E_L_isco}), we obtain
\begin{equation}
\label{trajectory_plung_phi}
\varphi_{p}(r)=-\frac{2\sqrt{3}}{\left(6M/r-1\right)^{1/2}}+\varphi_{0},
\end{equation}
where $t_0$ and $\varphi_0$ are arbitrary integration constants.Using Eq.~(\ref{trajectory_plung_phi}), the spatial trajectory of the plunging
particle can be described as
\begin{equation}
\label{plunge_phi_trajectory_alternate}
r_{p}(\varphi)=\frac{6M}{1+\frac{12}{(\varphi-\varphi_{0})^{2}}}.
\end{equation}
This trajectory is shown in Fig.~\ref{Fig:Trajectoire}

\section{Gravitational perturbations in massive spin-$2$ fields}
\label{sec_3}

The Fierz-Pauli theory in a Schwarzschild spacetime is governed by the following equations of motion~\cite{Cardoso:2023dwz,Cardoso2018} (see also the Supplemental Material of \cite{Cardoso2018}):
\begin{equation}
\label{eq_Fierz-Pauli}
\begin{aligned}
\Box h_{\mu\nu} + & 2R_{\mu\rho\nu\sigma} h^{\rho\sigma} - \mu^2 h_{\mu\nu} = \\
&\quad-16\pi \bigg(\mathcal{T}_{\mu\nu} 
- \frac{1}{3} g_{\mu\nu} \mathcal{T}^{\rho}_{\hphantom{\rho}\rho} 
 + \frac{1}{3\mu^2} \nabla_{\mu}\nabla_{\nu} \mathcal{T}^{\rho}_{\hphantom{\rho}\rho}\bigg),
\end{aligned}
\end{equation}
\begin{equation}\label{eq_Fierz-Pauli_contraintes_1}
\nabla^{\mu}h_{\mu\nu}=-\nabla_{\nu}\left(\frac{16\pi}{3\mu^2}  \mathcal{T}^{\rho}_{\hphantom{\rho}\rho}\right),
\end{equation}
\begin{equation}\label{eq_Fierz-Pauli_contraintes_2}
h=-\frac{16\pi}{3\mu^2} \mathcal{T}.
\end{equation}
Here \( h_{\mu\nu} \) denotes the massive spin-2 field perturbation, and \( \mu \) represents the mass of the graviton. The operator \( \Box = g^{\mu\nu}\nabla_\mu \nabla_\nu \) is the covariant d'Alembertian, while \( R_{\mu\rho\nu\sigma} \) is the Riemann tensor associated with the Schwarzschild background, satisfying \(\nabla_\sigma \nabla_\mu h^\sigma_\nu = -R^\tau_{\nu\sigma\mu} h^\sigma_\tau\). \( \mathcal{T}_{\mu\nu} \) denotes the stress-energy tensor, with its trace defined as \( \mathcal{T} = \mathcal{T}^\rho_{\hphantom{\rho}\rho} \). Similarly, \( h = g^{\mu\nu}h_{\mu\nu} \) denotes the trace of the perturbation field.

The gravitational waves emitted by the Schwarzschild black hole, excited by the plunging particle, are described by the perturbation field \(h_{\mu\nu}\)  and the stress-energy tensor associated with the massive particle \( \mathcal{T}_{\mu\nu} \) is given by
\begin{subequations} \label{SET_moving_tot}
\begin{fleqn}
\begin{eqnarray}
& & \mathcal{T}^{\mu \nu} (x) = m_0 \int_{\gamma}d\tau\, \frac{dx^\mu_p(\tau)}{d\tau} \frac{dx^\nu_p(\tau)}{d\tau} \frac{\delta^{4}(x-x_p(\tau))}{\sqrt{-g(x)}} \\
& & \phantom{T^{\mu \nu} (x)} = m_0 \frac{dx^\mu_p}{d\tau}(r) \frac{dx^\nu_p}{d\tau} (r) \left[\frac{ dr_p}{d\tau} (r)\right]^{-1}\nonumber \\
& & \qquad \qquad \quad \times \frac{\delta[t-t_p(r)] \delta[\theta- \pi/2] \delta[\varphi-\varphi_p(r)]}{r^2 \sin \theta}, \label{SET_moving}
\end{eqnarray}
\end{fleqn}
\end{subequations}
where \(x = (t, r, \theta, \varphi)\) denotes a spacetime location in Schwarzschild coordinates, \( m_0 \) is the mass of the plunging particle, and \( t_p(r) \) and \( \varphi_p(r) \) represent its trajectory functions as defined in Eqs.~(\ref{trajectory_plung}) and (\ref{trajectory_plung_phi}).

To solve problems \eqref{eq_Fierz-Pauli}--\eqref{eq_Fierz-Pauli_contraintes_2} and \eqref{SET_moving}, as well as to explore the broader topic of gravitational perturbations in BHs, extensive research has been carried out since the seminal works of Regge and Wheeler  \cite{Regge:1957td} and Zerilli \cite{Zerilli:1971wd} (e.g., \cite{Martel:2005ir}, \cite{Nagar:2005ea}; see also \cite{Brito:2013wya,Cardoso:2018zhm} for massive spin-2). Due to spherical symmetry, the tensor field $h_{\mu \nu}$ can be decomposed into a complete basis of spherical tensor harmonics, yielding $h^{\text{(e)}}_{\mu\nu}$ and $h^{\text{(o)}}_{\mu\nu}$. Similarly, the stress-energy tensor $\mathcal{T}_{\mu \nu}$ can be decomposed into $ \mathcal{T}^{\text{(e)}}$ and $ \mathcal{T}^{\text{(o)}}$. In this context, the symbols (e) and (o) denote the even (polar) and odd (axial) components respectively, reflecting whether they have even or odd parity under the antipodal transformation on the unit $2$-sphere $S^2$. Details of perturbation equations and conventions used are given in Appendix \ref{appendix_A}.  

\subsection{Odd-parity sector}
\label{sec_2_1}

The system of coupled equations governing the odd-parity partial modes, which fully describes the axial sector, is derived in Appendix \ref{appendix_A} and can be written as follows:
\begin{eqnarray}
   && \left[\frac{d^2}{dr^{2}_*}+\omega^2-V^{(\phi)}_{\ell}\right]\phi_{\omega\ell m} +\alpha^{(\phi)}\psi_{\omega\ell m}=S^{(\phi)}_{\omega\ell m},\label{Système_odd_freq_1}\\
   \nonumber\\[-2ex]
   && \left[\frac{d^2}{dr^{2}_*}+\omega^2-V^{(\psi)}_{\ell}\right]\psi_{\omega\ell m}+\alpha^{(\psi)}\phi_{\omega\ell m}=S^{(\psi)}_{\omega\ell m}.\label{Système_odd_freq_2}
\end{eqnarray}

Here, the potential  $V^{(\phi)}_{\ell}$ is given by 
\begin{equation}\label{potentiel_phi}
V^{(\phi)}_{\ell}(r)=f(r)\left(\mu^2+\frac{\Lambda+6}{r^2}-\frac{16M}{r^3}\right)
\end{equation}
and for the potential $V^{(\psi)}_{\ell}$ we have 
\begin{equation}\label{potentiel_psi}
V^{(\psi)}_{\ell}(r)=f(r)\left(\mu^2+\frac{\Lambda}{r^2}+\frac{2M}{r^3}\right)
\end{equation}
with the coupling terms defined as
\begin{equation}\label{feed_back_odd}
  \begin{aligned}
    \alpha^{(\phi)}(r) &= \frac{\Lambda}{r^2} f(r)\left(1-\frac{3M}{r}\right), \\
    \alpha^{(\psi)}(r) &= \frac{4}{r^2} f(r),
  \end{aligned}
\end{equation}
where  $\Lambda = (\ell-1)(\ell+2) = \ell(\ell+1)-2$.

In Eqs.~\eqref{Système_odd_freq_1}--\eqref{Système_odd_freq_2}, the functions $\phi_{\ell m}(r) = f(r) h^{\ell m}_{r}$ and   $\psi_{\ell m}(t,r) = h^{\ell m}/r$  represent combinations of the axial perturbations (see \ref{appendix_A_3}). The source terms $S^{(\phi)}_{\omega\ell m}(r)$ and $S^{(\psi)}_{\omega\ell m}(r)$ are constructed from the components of the stress-energy tensor, expressed in the basis of tensor spherical harmonics [see \eqref{source_phi} and \eqref{source_psi}]. Using the stress-energy tensor \eqref{SET_moving} and orthonormalization properties of (scalar, vector, and tensor) spherical harmonics \cite{Martel:2005ir}, we have
\begin{equation}\label{soure_phi_tot}
S_{\omega \ell m}^{(\phi)}(r)  = -\frac{16\sqrt{6 \pi} B(\ell, m)}{\Lambda + 2}  \frac{m_0 M}{r^2} f(r) e^{i\left[\omega t_p(r)- m \varphi_p(r)\right]}
\end{equation}
and
\begin{multline}\label{soure_psi_tot}
 S_{\omega \ell m}^{(\psi)}(r) = -i m\frac{576 \sqrt{2\pi}B(\ell, m) }{\Lambda (\Lambda + 2)} \frac{m_0 M^2}{r^3} \\ 
                                 \frac{f(r)}{\left(6M/r-1\right)^{3/2}} e^{i\left[\omega t_p(r) - m\varphi_p(r)\right]}
\end{multline}
where 
\begin{multline}\label{B_l_m}
 B(\ell,m) = \frac{2^{m+1}}{\sqrt{\pi}}\sqrt{\frac{2\ell+1}{4\pi}\frac{(\ell-m)!}{(\ell+m)!}} \\
             \times\frac{\Gamma\left[\frac{(\ell+m)}{2}+1\right]}{\Gamma\left[\frac{(\ell-m-1)}{2}+1\right]} \sin\left[\frac{\pi}{2}(\ell+m)\right]. 
\end{multline}
Here, it is important to remark that $B(\ell,m) $, and thus the source terms \eqref{soure_phi_tot} and \eqref{soure_psi_tot}, vanish when $\ell + m$ is even.\\

\subsubsection{Odd-parity dipole mode}

It should be noted that the monopole mode ($\ell = 0$) does not exist in the case of odd-parity (see Appendix \ref{appendix_A}). Regarding the dipole mode ($\ell = 1$), the angular functions $X^{\ell m}_{\theta\theta}$, $X^{\ell m}_{\varphi\varphi}$, and $X^{\ell m}_{\theta\varphi}$ vanish, and the coupled system \eqref{Système_odd_freq_1}--\eqref{Système_odd_freq_2} reduces to a single differential equation 
\begin{equation}\label{dipole_odd_parity}
\left[\frac{d^2}{dr^{2}_*}+\omega^2-V_1^{(\phi)}(r)\right]\phi_{\omega 10}(r) =S^{(\phi)}_{\omega 10}(r)
\end{equation} 
where the potential  $ V^{(\phi)}_{1}(r)$ is given by 
\begin{equation}
\label{potentiel_phi}
V^{(\phi)}_{1}(r)=f(r)\left(\mu^2+\frac{6}{r^2}-\frac{16M}{r^3}\right)
\end{equation}
and we have for the source term  
\begin{equation}
\label{source_dipole}
S^{(\phi)}_{\omega 1 0}(r) =  -12\sqrt{2} \,\frac{m_0 M}{r^2} f(r) e^{i\omega t_p(r)}.
\end{equation}

\subsection{Even-parity sector}
\label{sec_2_2}

The polar equations are more complicated and are detailed in Appendix \ref{appendix_A}. The polar sector is thoroughly characterized by a system of three coupled ordinary differential equations
\begin{fleqn}
\begin{eqnarray}
  && f(r)^2  \frac{d^2 K}{dr^2}+ \alpha^{(K)}_1\frac{d K}{dr}+ \alpha^{(K)}_2 K +C^{(K)} = S^{(K)}, \label{Système_even_freq_1} \\
   \nonumber\\[-2ex]
  && f(r)^2  \frac{d^2 H_{r}}{dr^2}+ \alpha^{(H)}_1\frac{d H_{r}}{dr}+ \alpha^{(H)}_2 H_{r} +C^{(H)} = S^{(H)},  \label{Système_even_freq_2}\\
   \nonumber\\[-2ex]
  && f(r)^2  \frac{d^2 G}{dr^2}+ \alpha^{(G)}_1\frac{d G}{dr}+ \alpha^{(G)}_2 G +C^{(G)} = S^{(G)},\label{Système_even_freq_3}
\end{eqnarray}
\end{fleqn}
where the coupling terms are defined as
\begin{fleqn}
\begin{eqnarray}
    C^{(K)} &=& \alpha^{(K)}_3 \frac{d H_{r}}{dr}+\alpha^{(K)}_4 H_{r} + \alpha^{(K)}_5 \frac{d G}{dr}+\alpha^{(K)}_6 G, \label{C_k}\\
    C^{(H)} &=&  \alpha^{(H)}_3 \frac{d K}{dr}+\alpha^{(H)}_4 K + \alpha^{(H)}_5 \frac{d G}{dr}+\alpha^{(H)}_6 G,        \label{C_H} \\
    C^{(G)} &=&  \alpha^{(G)}_3 \frac{d K}{dr}+\alpha^{(G)}_4 K + \alpha^{(G)}_5 \frac{d H_{r}}{dr}+\alpha^{(G)}_6 H_{r}. \label{C_G}
\end{eqnarray}
\end{fleqn}
The radial functions $\alpha^{(K)}_i, \alpha^{(H)}_i$ and $\alpha^{(G)}_i$ are given by Eqs.~\eqref{alpha_K_1}--\eqref{alpha_G_6} in Appendix \ref{appendix_A}. It is important to note that the $H$ in the exponents of the coefficients here refers to the $H_r$ component.\\

Similarly, the source terms in Eqs.~\eqref{Système_even_freq_1}--\eqref{Système_even_freq_3} have been constructed from the components of the stress-energy tensor, expressed in the basis of tensor spherical harmonics [see \eqref{S_K_even_tot}, \eqref{S_H_even_tot}, and \eqref{S_G_even_tot}]. Using the orthonormalization properties of (scalar, vector, and tensor) spherical harmonics and the expression for the stress-energy tensor of a massive particle \eqref{SET_moving}, we obtained
\begin{widetext}
\begin{multline}\label{source_K_main}
S^{(K)}_{\omega \ell m}(r) = -\frac{4 m_0 \sqrt{\pi} A(\ell, m) f(r)}{ \mathcal{B}^{(K)}_{\omega\ell m}(r)} 
\Bigg[ \mathcal{C}^{(K)}_{\omega\ell m}(r) + \frac{\mathcal{D}^{(K)}_{\omega\ell m}(r)}{R(r)^3} 
+ \frac{\mathcal{E}^{(K)}_{\omega\ell m}(r)}{R(r)^{5/2}} \\ 
+ \frac{\mathcal{F}^{(K)}_{\omega\ell m}(r)}{R(r)^{3/2}} 
+ \mathcal{I}^{(K)}_{\omega\ell m}(r) R(r)^{1/2} 
+ \mathcal{J}^{(K)}_{\omega\ell m}(r) R(r)^{3/2} \Bigg] e^{i\bigl[\omega t_p(r) - m \varphi_p(r)\bigr]}
\end{multline}

\begin{multline}\label{source_H_main}
S^{(H)}_{\omega \ell m}(r) = \frac{8\, m_0 \sqrt{\pi}\, A(\ell, m) }{ \mathcal{B}^{(H)}_{\omega\ell m}(r)} 
\Bigg[ \mathcal{C}^{(H)}_{\omega\ell m}(r) + \frac{\mathcal{D}^{(H)}_{\omega\ell m}(r)}{R(r)^3} 
+ \frac{\mathcal{E}^{(H)}_{\omega\ell m}(r)}{R(r)^{5/2}} \\ 
+ \frac{\mathcal{F}^{(H)}_{\omega\ell m}(r)}{R(r)^{3/2}} 
+ \mathcal{I}^{(H)}_{\omega\ell m}(r) R(r)^{1/2} 
+ \mathcal{J}^{(H)}_{\omega\ell m}(r) R(r)^{3/2} \Bigg] e^{i\bigl[\omega t_p(r) - m \varphi_p(r)\bigr]}
\end{multline}
and 
\begin{equation}\label{source_G_main}
  S^{(G)}_{\omega \ell m}(r) = - 8 m_0 \sqrt{2 \pi} A(l, m) \frac{f(r)}{r^4 R(r)^{3/2}} \left[\frac{1}{\mu^2} - \frac{36 M^2 (\Lambda + 2 - 2 m^2)}{\Lambda (\Lambda + 2)} \right] e^{i\bigl[\omega t_p(r) - m \varphi_p(r)\bigr]},
\end{equation}
\end{widetext}
where 
\begin{multline}\label{A_l_m}
 A(\ell,m) = \frac{2^{m}}{\sqrt{\pi}}\sqrt{\frac{2\ell+1}{4\pi}\frac{(\ell-m)!}{(\ell+m)!}} \\
             \times\frac{\Gamma\left[\frac{1}{2} (l + m + 1)\right]}{\Gamma\left[\frac{1}{2} (l - m) + 1\right]} \cos\left[\frac{\pi}{2}(\ell+m)\right] 
\end{multline}
and
\begin{equation}\label{R}
  R(r) =\frac{6M}{r} - 1.
\end{equation}
Note that $A(\ell,m)$, and thus the source terms \eqref{source_K_main}, \eqref{source_H_main}, and \eqref{source_G_main}, vanish when $\ell + m$ is odd. In expressions~\eqref{source_K_main} and \eqref{source_H_main}, the functions  \(\mathcal{B}_{\omega \ell m}^{(i)}, \mathcal{C}_{\omega \ell m}^{(i)}, \mathcal{D}_{\omega \ell m}^{(i)}, \mathcal{E}_{\omega \ell m}^{(i)}, \mathcal{F}_{\omega \ell m}^{(i)}, \mathcal{I}_{\omega \ell m}^{(i)}, \mathcal{J}_{\omega \ell m}^{(i)}\), where \(i = K, H\), are provided by Eqs.~\eqref{B_K_S}--\eqref{J_K_S} and \eqref{B_H_S}--\eqref{J_H_S}, respectively.

\subsubsection{Even-parity monopole}
\label{sec_2_2_1}

Unlike the odd-parity case, the monopole mode $\ell = 0$ exists in the even-parity sector and is governed by a pair of coupled differential equations, 
\begin{fleqn}
\begin{eqnarray}
  && f(r)^2  \frac{d^2 K}{dr^2}+ \alpha^{(K)}_1\frac{d K}{dr}+ \alpha^{(K)}_2 K +C^{(K)} = S^{(K)}, \label{l_0_even_freq_1} \\
   \nonumber\\[-2ex]
  && f(r)^2  \frac{d^2 H_{tr}}{dr^2}+ \alpha^{(H)}_1\frac{d H_{tr}}{dr}+ \alpha^{(H)}_2 H_{tr} +C^{(H)} = S^{(H)} \label{l_0_even_freq_2}
\end{eqnarray}
\end{fleqn}
where the coupling terms are given by 
\begin{eqnarray}
  C^{(K)} &=& \alpha^{(K)}_3 \frac{d H_{tr}}{dr}+\alpha^{(K)}_4 H_{tr}, \label{C_k_l_0} \\
  C^{(H)} &=&  \alpha^{(H)}_3 \frac{d K}{dr}+\alpha^{(H)}_4 K. \label{C_H_l_0}
\end{eqnarray}

Here, the radial functions $\alpha^{(K)}_i$ and $\alpha^{(H)}_i$ are given by Eqs. \eqref{alpha_K_1_l_0}--\eqref{alpha_Htr_4_l_0} in Appendix \ref{appendix_A_4_1}. Note  that the $H$ in the exponents of the coefficients here refers to the $H_{tr}$ component.

The source terms in ~\eqref{l_0_even_freq_1} and \eqref{l_0_even_freq_2} are derived from the components of the stress-energy tensor expressed in terms of tensor spherical harmonics [see \eqref{S_K_monopole} and \eqref{S_H_monopole}]. By using the orthonormalization properties of scalar, vector, and tensor spherical harmonics, as well as the expression for the stress-energy tensor of a massive particle \eqref{SET_moving}, we obtained
\begin{widetext}
\begin{multline}\label{source_K_monopole_even}
S_{\omega}^{(K)}(r) =\frac{4 \sqrt{2} m_0 f(r)}{9 r^7 \mu^2} \Biggl[\frac{18 i \sqrt{2} r^4 \omega}{R(r)^3} 
+ \frac{-72 M r^2 + 972 M^3 f(r)}{R(r)^{5/2}} \\
+ \frac{648 M^3 + 54 M^2 r (-4 + 3 r^2 \mu^2) + r^3 (16 + 9 r^2 \mu^2) - 216 M^2 r f(r)}{R(r)^{3/2}}
- 9 M r^2 R(r)^{1/2} - 2 r^3 R(r)^{3/2} \Biggr] e^{i\omega t_p(r)},
\end{multline}
\begin{multline}\label{source_H_monopole_even}
S_{\omega}^{(H)}(r) =-\frac{4 m_0}{9 r^6 \mu^2} \Biggl[12 r^4 \mu^2 - \frac{36 r^4 \omega^2}{R(r)^3} + 
\frac{36 i \sqrt{2} M \omega \Bigl(-2 r^2 + 27 M^2 f(r)\Bigr)}{R(r)^{5/2}} \\
+ \frac{i \sqrt{2} \omega \Bigl(108 M^3 + 54 M^2 r + 9 M r^2 + 16 r^3 - 486 M^2 r f(r)\Bigr)}{R(r)^{3/2}} 
- 9 i \sqrt{2} M r^2 \omega R(r)^{1/2} - 2 i \sqrt{2} r^3 \omega R(r)^{3/2} \Biggr]  e^{i\omega t_p(r)}.
\end{multline}
\end{widetext}

It is important to note that the system of Eqs.~\eqref{l_0_even_freq_1} and \eqref{l_0_even_freq_2} governing the monopole can be reduced to a single differential equation by combining the  $K^{\ell m}$ and $H^{\ell m}_{tr}$  components using the generalized version of the Berndtson-Zerilli transformation (see Ref.~\cite{Brito:2013wya} for more details). However, we chose to work with the complete system of equations, since this combination leads to a ``loss of information,'' specifically the presence of a new branch of quasinormal frequencies and another branch of quasibound frequencies that do not appear in the equation obtained after the combination.

\subsubsection{Even-parity dipole mode}
\label{sec_2_2_2}

For the even-parity dipole mode ($\ell = 1$), the component $G^{\ell m}$ vanishes and the system \eqref{Système_even_freq_1}--\eqref{Système_even_freq_3} reduces to a system of two equations
\begin{fleqn}
\begin{eqnarray}
  && f(r)^2  \frac{d^2 K}{dr^2}+ \alpha^{(K)}_1\frac{d K}{dr}+ \alpha^{(K)}_2 K +C^{(K)} = S^{(K)}, \label{Système_even_freq_1_bis} \\
   \nonumber\\[1ex]
  && f(r)^2  \frac{d^2 H_{r}}{dr^2}+ \alpha^{(H)}_1\frac{d H_{r}}{dr}+ \alpha^{(H)}_2 H_{r} +C^{(H)} = S^{(H)}.  \label{Système_even_freq_2_bis}
\end{eqnarray}
\end{fleqn}
The radial functions  $\alpha^{(K)}_i$ and $\alpha^{(H)}_i$, the coupling coefficients $C^{(K)}$ and $C^{(H)}$, as well as  the source terms $S^{(K)}$ and $S^{(H)} $, are given by \eqref{alpha_K_1}--\eqref{alpha_H_6}, \eqref{C_k}, \eqref{C_H}, \eqref{source_K_main} and \eqref{source_H_main}, respectively, with $ \ell = 1$  (\textit{i.e.}, $\Lambda = 0$) and  $m = 1$.

\section{Resonance spectra : quasinormal modes and quasibound states } 
\label{sec_4}

In this section we construct the resonance spectra of the massive spin-2 field, which involves solving the corresponding homogeneous systems of coupled differential equations. For the odd-parity sector, we considered the homogeneous system given by \eqref{Système_odd_freq_1}--\eqref{Système_odd_freq_2}, while for the even-parity sector we solved the system represented by \eqref{Système_even_freq_1}--\eqref{Système_even_freq_3}. In both cases, appropriate boundary conditions have been imposed. For a detailed discussion of the numerical methods used, the reader is referred to Appendix \ref{appendix_B}.

It can be shown that, in general, the solution exhibits an asymptotic behavior at spatial infinity (i.e., as $r_* \to \infty$) given by
\begin{multline}\label{Asymp_Behavior}
  \mathcal{Q}_j(r) \sim  A^{(-)}(\omega) e^{- i \left[p(\omega) r_* + \frac{M \mu^2}{p(\omega)} \ln\left(\frac{r}{M}\right)\right]} \\
  + A^{(+)}(\omega) e^{+ i \left[p(\omega) r_* + \frac{M \mu^2}{p(\omega)} \ln\left(\frac{r}{M}\right)\right]}
\end{multline}
where the function $p(\omega) = (\omega^2 - \mu^2)^{1/2}$ denotes ``the wave number,'' while the coefficients $ A^{(-)}(\omega)$ and $ A^{(+)}(\omega)$ are the complex amplitudes. Two distinct families of modes arise based on their behavior at spatial infinity. The first family consists of quasinormal modes, characterized by purely outgoing waves at infinity and defined by $ A^{(-)}(\omega) = 0$. The second family involves quasibound states, defined by$ A^{(+)}(\omega) = 0$, which are localized within the vicinity of the black hole and decay exponentially at spatial infinity. In both cases, applying boundary conditions at spatial infinity yields a discrete spectrum of allowed  frequencies, $\omega_{\ell n}$, where each frequency is labeled by its angular momentum $\ell$ and overtone number $n$.

\subsection{Numerical results}
\label{sec_4_1}

\subsubsection{Quasinormal modes}
\label{sec_4_1_1}
\begin{figure}[htbp]
 \includegraphics[scale=0.57]{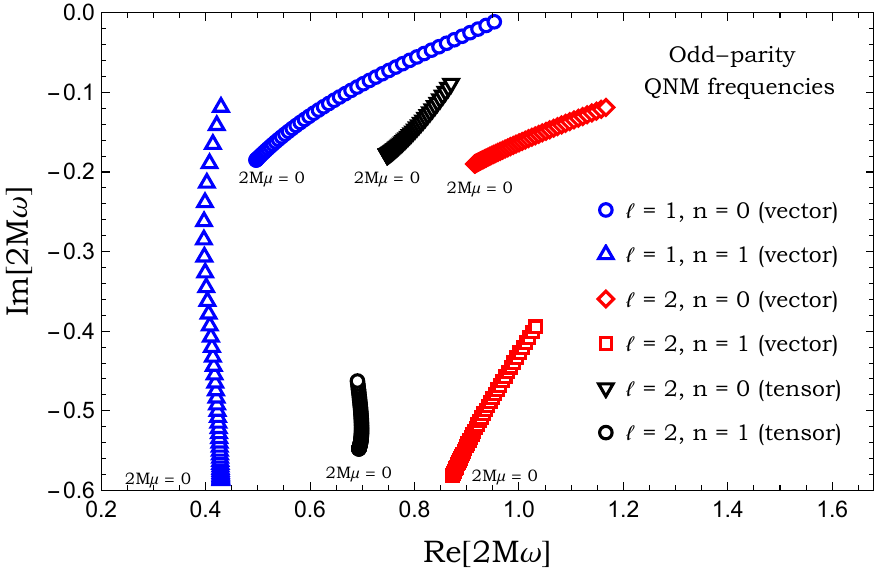}
\caption{\label{Fig:QNMs_Odd-parity} Quasinormal mode frequencies of the odd-parity massive spin-2 field are displayed  for dipole modes ($\ell=1$) and quadrupole modes ($\ell=2$), for a range of field masses, $2M \mu =0,0.02,...,1$. The fundamental ($n=0$) and first overtone ($n=1$) frequencies are shown. In the massless limit, the quasinormal frequencies of the ``vector'' modes correspond to those of the electromagnetic field ($s=1$), while the  ``tensor'' modes match the quasinormal frequencies of the massless gravitational field ($s=2$).}
 \end{figure}
\begin{figure}[htbp]
 \includegraphics[scale=0.57]{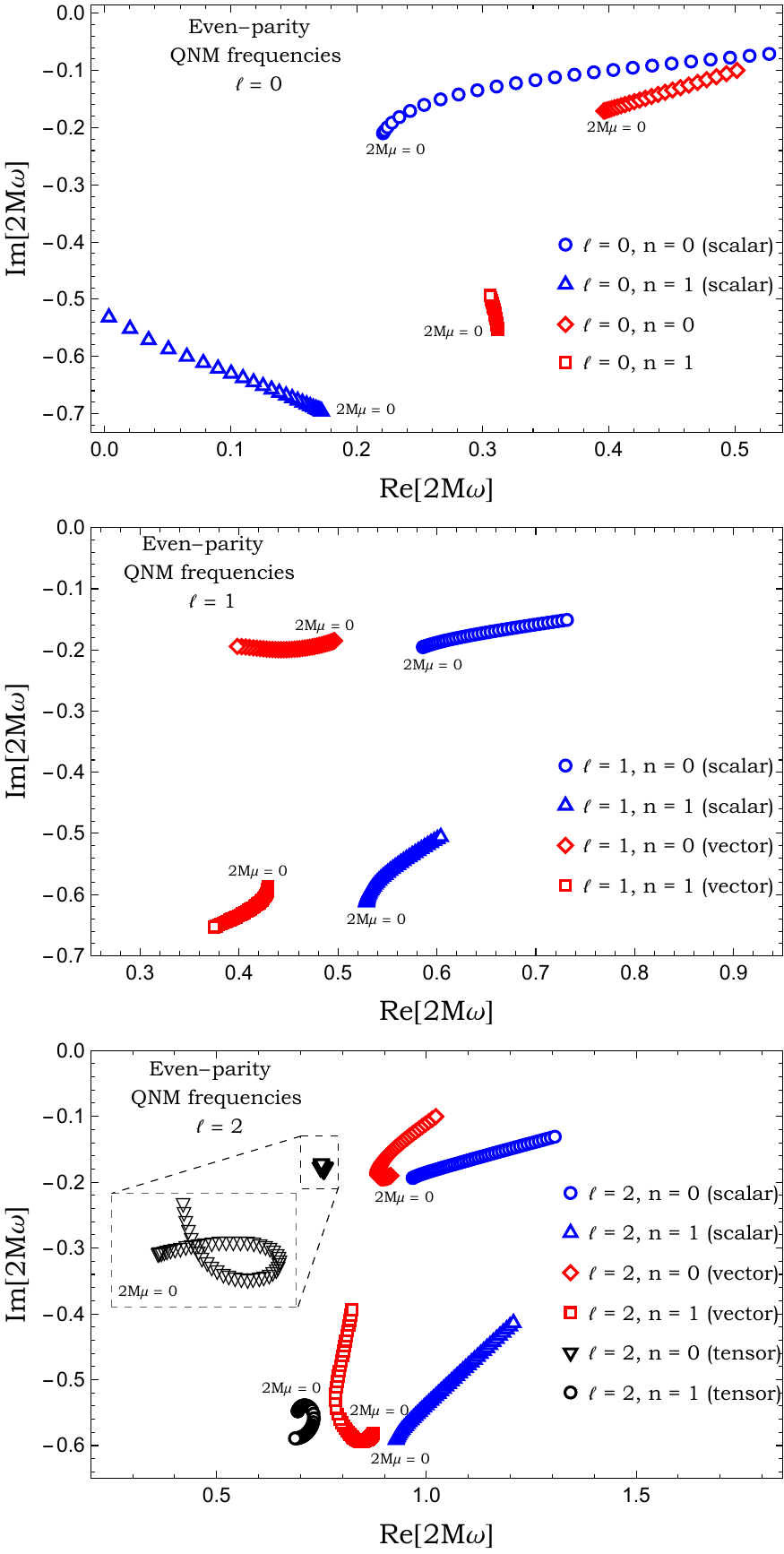}
\caption{\label{Fig:QNMs_Even-parity} Quasinormal mode frequencies of the even-parity massive spin-2 field are displayed. Top: monopole mode ($\ell=0$) for a range of field masses, $2M \mu = 0, 0.02, \ldots, 0.50$. Middle: dipole mode ($\ell=1$) for a range of field masses, $2M \mu = 0, 0.02, \ldots, 0.70$. Bottom: quadrupole mode ($\ell=2$) for a range of field masses, $2M \mu = 0, 0.02, \ldots, 1.16$. In all cases, the fundamental mode ($n=0$) and the first overtone ($n=1$) are plotted. In the massless limit, the quasinormal frequencies of the ``scalar'' modes correspond to those of a scalar field ($s = 0$), the ``vector'' modes correspond to those of the electromagnetic field ($s = 1$), while the ``tensor'' modes correspond to the quasinormal frequencies of the massless gravitational field ($s=2$). It is worth noting that in the case of $\ell = 0$ a new branch appears which does not correspond to the scalar, electromagnetic or gravitational fields in the massless limit.}
 \end{figure}
In Fig.~\ref{Fig:QNMs_Odd-parity}, we display the effect of mass on the QNM frequencies of the odd-parity sector, focusing on the dipole mode ($\ell = 1$) and the quadrupole mode ($\ell = 2$) for both the fundamental ($n = 0$) and the first overtone ($n = 1$). As expected, the modes for any given ($\ell, n$) with $\ell \geq 2$ can be grouped into two distinct branches. These branches are distinguished by their behavior in the massless limit, the ``vector'' modes correspond to the spectrum of the electromagnetic field ($s = 1$), while the ``tensor'' modes, match the QNM spectrum of the massless gravitational field ($s = 2$). For the lower overtones, increasing the mass leads to a decrease in the decay rate, eventually reaching a point where the QNM frequencies vanish. This phenomenon is related to the reduction of the height of the effective potential barrier,  as has already been observed for both the massive scalar field and the massive vector field~\cite{Ohashi:2004wr, Konoplya:2004wg, Konoplya:2005hr, Rosa:2011my}. It should be noted that our results are in agreement with those previously obtained by Brito \textit{et al.}~\cite{Brito:2013wya}.

In Fig.~\ref{Fig:QNMs_Even-parity} we also show the effect of mass on the QNM frequencies in the even-parity sector, focusing on the monopole mode ($\ell = 0$), the dipole mode ($\ell = 1$), and the quadrupole mode ($\ell = 2$), for both the fundamental mode ($n = 0$) and the first overtone ($n = 1$). For the monopole mode, in addition to the branch corresponding to the massless scalar field in the massless limit, a new branch appears that does not correspond to scalar, electromagnetic, or gravitational fields in this limit. For the dipole mode, two branches can be distinguished for any given pair ($\ell, n$). The ``scalar'' modes correspond to the massless scalar field spectrum in the massless limit, while the ``vector'' modes converge to the electromagnetic field spectrum in the same limit. For $\ell \geq 2$ the modes are grouped into three distinct branches. In addition to the vector and ``tensor'' quasinormal modes already observed in the odd-parity sector, which reduce to the electromagnetic and massless gravitational field spectra in the massless limit, we also find the scalar quasinormal mode frequencies, which correspond to the spectrum of the massless scalar field in this limit.

The overall behavior is similar to that of the odd-parity spectra. In particular, as the mass increases, the imaginary part (representing the decay rate) decreases until the quasinormal frequencies disappear. However, this does not hold for the dipole mode vector family, where the real part decreases instead. It is also noteworthy that the tensor family of the quadrupole mode shows minimal variation with mass.

\subsubsection{Quasibound states}
\label{sec_4_1_2}

\begin{figure*}[htbp]
 \includegraphics[scale=0.53]{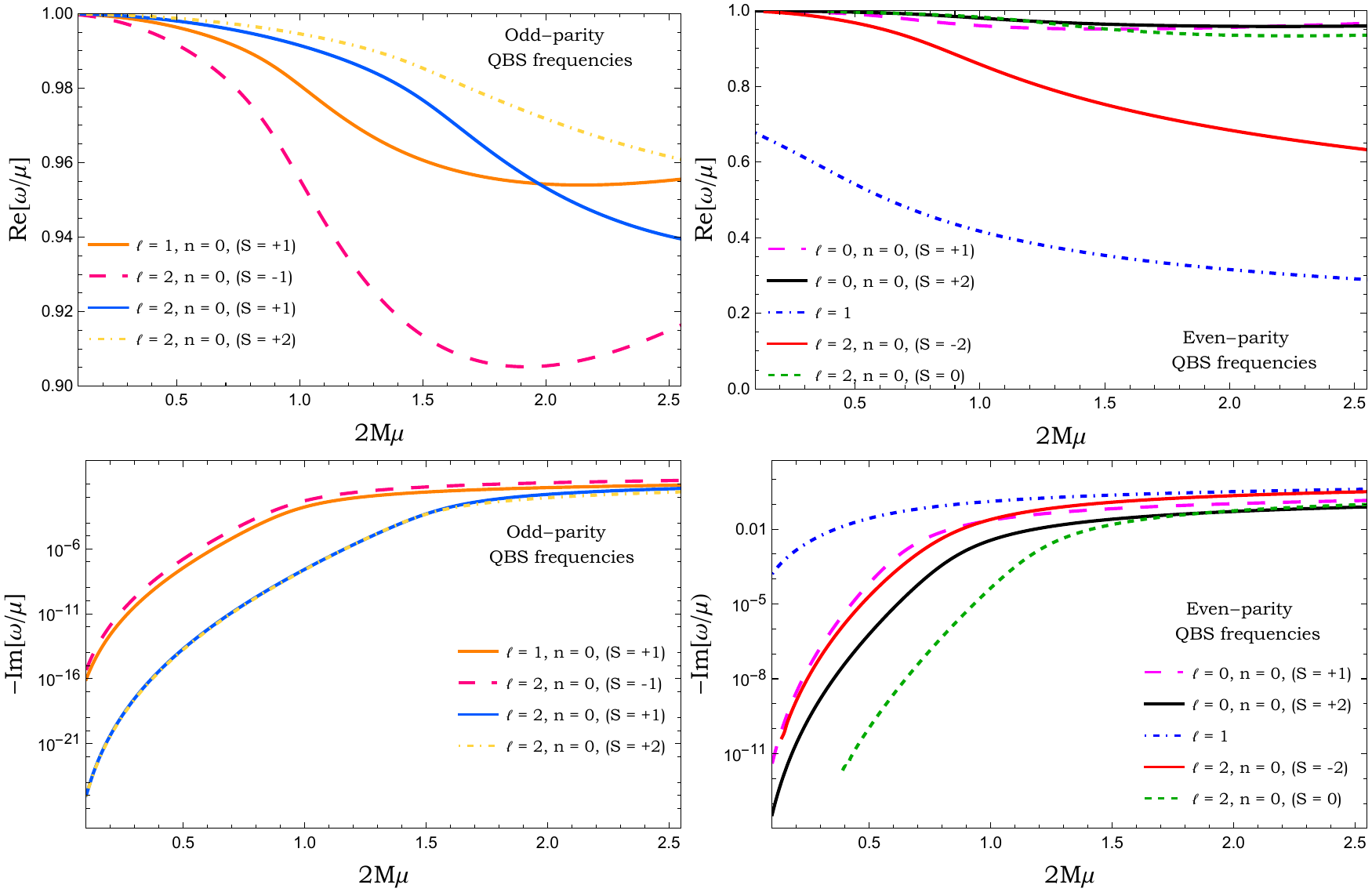}
\caption{\label{Fig:QBSs_Odd_Even-parity} The odd-parity bound state levels of the massive spin-2 field are shown on the left, while the even-parity bound state levels are shown on the right. The top panel shows the real part of the frequency, $\text{Re}(\omega/\mu)$, as a function of the mass coupling $2M\mu$, while the bottom panel shows the negative of the imaginary part, $\text{Im}(\omega/\mu)$, on a logarithmic scale. The modes are labeled according to their angular momentum $\ell$, their number of overtones $n$, and their spin projection $S$, except in the case of the even dipole mode $ \ell = 1$.}
 \end{figure*}

In addition to the QNM spectrum, massive fields can localize near a BH, producing a rich spectrum of QBSs with complex frequencies. These QBSs have been studied for both massive scalar and Proca fields (see Refs.~\cite{Dolan:2007mj,Galtsov1983,Lasenby:2002mc,Galtsov1984,Rosa:2011my}). In this section we present numerical results for the quasibound state spectrum of the massive spin-2 field, obtained using the matrix-valued Hill determinant method. Our results show excellent agreement with those previously investigated in Ref.~\cite{Brito:2013wya} using the direct integration method, and we have extended them by finding other modes. It has also been shown that for massive fields the spectrum is similar to that of the hydrogen atom in the $2M\mu \to 0$ limit
\begin{equation}\label{Spectre_H} 
\text{Re}\left[\omega/\mu\right] \sim 1 - \frac{(M\mu)^2}{2(j + n + 1)^2} 
\end{equation} 
where $j = \ell + S$ is the total angular momentum of the state with spin projections $S = 0, \pm 1, \pm 2$. For a given pair $(\ell, n)$, the total angular momentum $j$ satisfies the quantum mechanical angular momentum addition rule, $|\ell - s| \leq j \leq \ell + s$, where $s$ is the spin of the field.

In Fig.~\ref{Fig:QBSs_Odd_Even-parity} we plot the spectrum of the QBS frequencies as a function of the mass coupling $2M\mu$ for the lowest modes $\ell = 1, 2$ for odd-parity and $\ell = 0, 1, 2$ for even parity, focusing on the fundamental harmonic $n = 0$. As mentioned in Ref.~\cite{Brito:2013wya}, the frequency spectrum of the QBS generally follows that of the hydrogen atom described by Eq.~\eqref{Spectre_H}, with a few exceptions that we will discuss below. For the monopole $\ell = 0$, two branches appear, one compatible with $S = +2$ and consistent with the angular momentum addition rule, giving $j = 2$, while a new branch appears compatible with $S = +1$ but violating the angular momentum addition rule, giving $j = 1$, which does not satisfy the inequality $|\ell - 2| \leq j \leq \ell + 2$. For the dipole mode $\ell = 1$, we show an odd-parity mode that is fully consistent with  $S = +1$ and $j = 2$, in agreement with the angular momentum addition rule. However, for the even-parity dipole mode, as discussed in Ref.\cite{Brito:2013wya}, it is an isolated mode and does not exhibit the small-mass behavior predicted by Eq.\eqref{Spectre_H}. Moreover, using our matrix-valued Hill determinant method, we found only a single fundamental mode for this state, with no overtones. We will discuss this mode later. Finally, we identify five modes characterized by their spin projection $S$ for the $\ell \geq 2$ modes with a given $n$. There are three modes for odd parity, with $S = -1, +1, +2$, and two modes for even parity, with $S = -2, 0$.

\begin{figure}[htbp]
\includegraphics[scale=0.53]{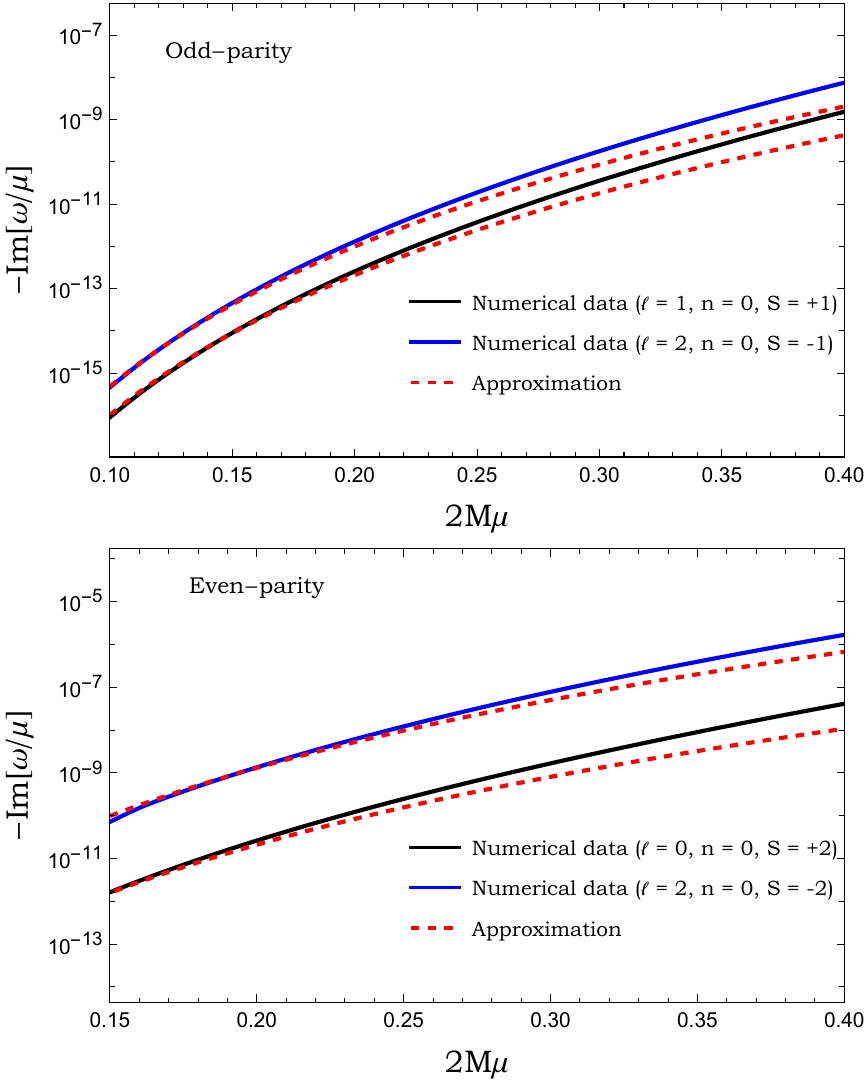}
\caption{\label{Fig:Data_vs_Approx_Im} Comparison between the numerical data and the analytical results for the odd modes ($\ell = 1, n = 0, S = +1$) and ($\ell = 2, n = 0, S = -1$) (top), and the even modes ($\ell = 0, n = 0, S = +2$) and ($\ell = 2, n = 0, S = -2$) (bottom) as a function of the mass coupling $2M\mu$. The solid lines (black and blue) represent the numerical data, while the dashed red line shows the analytical formula from \eqref{Im_Analytic}.}
\end{figure} 

Regarding the imaginary part, in the regime $2M \mu \to 0$, there is a power-law dependence similar to that already found for the massive vector field~\cite{Rosa:2011my}. Specifically, we have 
\begin{equation}\label{Im_Analytic}
  \text{Im}[\omega/\mu] \propto -\left(M\mu\right)^{\eta(\ell,S)}
\end{equation}
with
\begin{equation}\label{eta}
  \eta(\ell,S) = 4\ell + 2S +5
\end{equation}
where the proportionality constant depends on the overtone number and, more generally, on $\ell$ and $S$. As shown in Fig.~\ref{Fig:Data_vs_Approx_Im}, the analytical approximation \eqref{Im_Analytic} accurately describes the odd-parity modes $\ell=1,n=0,S=+1$ and $\ell=2,n=0,S=-1$, as well as the even-parity modes $\ell=0,n=0,S=+2$ and $\ell=2,n=0,S=-2$, with the corresponding proportionality coefficients $0. 021, 0.1, 0.021$, and $1.31$, respectively, and the exponents $\eta(1,+1)=\eta(2,-1)=11$ and $\eta(0,+2)=\eta(2,-2)=9$. It is important to note that for the new branch of quasibound frequencies of the $\ell=0,n=0,S=+1$ mode, its imaginary part also follows, in the small mass limit, Eq.~\eqref{Im_Analytic}, but with an exponent $\eta(0,+2)=9$ and a proportionality constant of $2.36$.

As mentioned above, the even-parity dipole mode is peculiar. In fact, its behavior is completely different from the other modes, which follow the predictions of Eqs. \eqref{Spectre_H} and \eqref{Im_Analytic} in the small mass limit. By fitting the real part for $0.1 \leq2 M \mu \leq 0.50$ we obtain
\begin{equation}\label{l_1_Re_approx}
  \text{Re}[\omega/\mu] \sim 0.72 (1 - M \mu),
\end{equation}
while for the imaginary part as $2M\mu \to 0$ we find 
\begin{equation}\label{l_1_Re_approx}
  \text{Im}[\omega/\mu] \sim -\frac{4}{3} (M \mu)^3.
\end{equation}

\subsubsection{Instability of Schwarzschild black hole}
\label{sec_4_1_3}
\begin{figure}[htbp]
\includegraphics[scale=0.53]{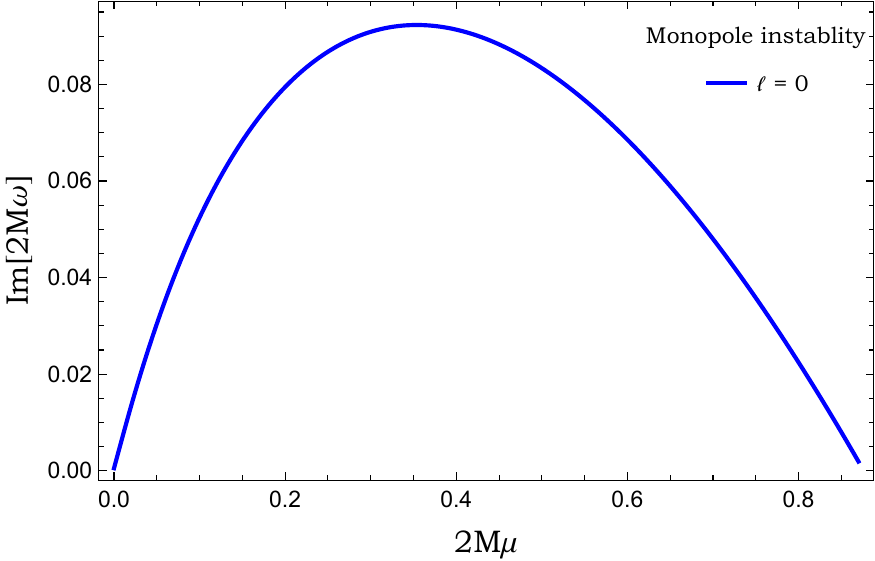}
\caption{\label{Fig:Monopole_Instability} The instability of Schwarzschild black holes under spherically symmetric polar mode of a massive spin-2 field. The plot shows the inverse of the instability timescale, $\text{Im}[\omega] = 1/\tau$, as a function of the mass coupling $2M\mu$.}
\end{figure} 

In this subsection, we do not go into the details or study of instability. Instead, we refer the reader to Sec. IV of Brito \emph{et al.} \cite{Brito:2013wya}, which provides a complete analysis of the subject. Nevertheless, using our algorithm based on the matrix-valued Hill determinant method, we confirm the existence of a Gregory-Laflamme type instability \cite{Babichev:2013una, Brito:2013wya}. This unstable mode, which affects only the spherically symmetric sector \( \ell = 0 \) of the Schwarzschild black hole, is illustrated in Fig.~\ref{Fig:Monopole_Instability} as a function of the mass coupling \( 2M\mu \). It is characterized by a purely imaginary and positive frequency component. Our numerical results show that this instability is significant for small values of \( 2M\mu \) and disappears for \( 2M\mu > 0.87 \). Furthermore, the instability timescale exhibits a strong dependence on the mass coupling \( 2M\mu \). For small values of \( 2M\mu \), our result shows \( \text{Im}[\omega] \sim 0.7\mu \), in agreement with the numerical result of Ref.~\cite{Brito:2013wya} and consistent with the analytical calculation in Ref.~\cite{Camps:2010br}. As already mentioned in Ref. ~\cite{Brito:2013wya}, this linear regime instability cannot describe its nonlinear evolution or its potential final state. However, as suggested by the mode profile shown in Fig.~\ref{Fig:Monopole_Instability}, a Schwarzschild black hole surrounded by a graviton cloud could represent a viable solution to the field equations. The possible effects of this instability on the waveform properties will be discussed in Sec.~\ref{sec_6}.

\section{Gravitational waves generated by the plunging massive particle}
\label{sec_5}

\subsection{Construction of the partial amplitudes : Odd-parity sector}
\label{sec_5_1}

\subsubsection{Odd-parity dipole mode}

To solve Eq.~\eqref{dipole_odd_parity}, which governs the dipole mode ($\ell = 1$), we will use the Green's function machinery (see Ref.~\cite{MorseFeshbach1953} for generalities, and, \text{e.g.}, Ref.~\cite{Breuer:1974uc} for its application in black hole physics). Let us consider the Green's function $G_{\omega} 1(r_*,r'_*)$ defined by 
\begin{equation}\label{dipole_odd_parity_Green_f}
\left[\frac{d^2}{dr^{2}_*}+\omega^2-V^{(\phi)}_{1}(r)\right]G_{\omega 1}(r_*,r'_*)= - \delta(r_* -r'_*)
\end{equation}
which can be written as
\begin{equation}
\label{Green_Function_2}
\hspace{-5pt}G_{\omega 1}(r_*,r'_*)=-\frac{1}{W_{1}(\omega)}
\left\{\,
\begin{aligned}
&\!\!\phi_{\omega 1}^{\mathrm {in}}(r_{*})\,\phi_{\omega 1}^{\mathrm{up}}({r}_{*}'),\!\!&r_{*}<{r}_{\ast}',\\
&\!\!\phi_{\omega 1}^{\mathrm{up}}(r_{*})\,\phi_{\omega 1}^{\mathrm {in}}({r}_{*}'),\!\!&r_{*}>{r}_{\ast}'.
\end{aligned}
\right.
\end{equation}
where $W_{1}(\omega)$ denotes the Wronskian of $\phi_{\omega 1}^{\mathrm {in}}$ and $\phi_{\omega 1}^{\mathrm {up}}$, two linearly independent solutions of the homogeneous equation~\eqref{dipole_odd_parity}.
The function $\phi_{\omega 1}^{\mathrm {in}}$  is characterized by its purely ingoing behavior at the event horizon $r=2M$ (i.e., for $r_\ast \to -\infty$)
\begin{subequations}
\label{bc_in}
\begin{equation}\label{bc_1_in}
\phi^{\mathrm {in}}_{\omega 1} (r)\scriptstyle{\underset{r_\ast \to -\infty}{\sim}} \displaystyle{e^{-i\omega r_\ast}}
\end{equation}
while it exhibits an asymptotic behavior at spatial infinity $r \to +\infty$ (i.e., for $r_\ast \to +\infty$) of the form
\begin{multline}
\phi^{\mathrm{in}}_{\omega 1} (r)\scriptstyle{\underset{r_\ast \to +\infty}{\sim}}
 \displaystyle{\sqrt{\frac{\omega}{p(\omega)}} \left\{A_{1}^{(-)}(\omega) e^{-i\left[p(\omega) r_* + \frac{M \mu^2}{p(\omega)}\ln\left(\frac{r}{M}\right)\right]}\right.} \\
 \displaystyle{\left.+ A_{1}^{(+)}(\omega) e^{+i\left[p(\omega) r_* + \frac{M \mu^2}{p(\omega)}\ln\left(\frac{r}{M}\right)\right]} \right\}}
\end{multline}
\end{subequations}
Similarly, the function $\phi^{\mathrm{up}}_{\omega 1}$ is characterized by its purely outgoing behavior at spatial infinity
\begin{subequations}
\label{bc_up}
\begin{multline}
\phi^{\mathrm{up}}_{\omega 1} (r)\scriptstyle{\underset{r_\ast \to +\infty}{\sim}}
 \displaystyle{\sqrt{\frac{\omega}{p(\omega)}} \, e^{+i\left[p(\omega) r_* + \frac{M \mu^2}{p(\omega)}\ln\left(\frac{r}{M}\right)\right]}} 
\end{multline}
and, at the horizon, it has an asymptotic behavior of the form
\begin{equation}\label{bc_2_up}
\phi^{\mathrm{up}}_{\omega 1}(r) \scriptstyle{\underset{r_\ast \to -\infty}{\sim}}\displaystyle{B_{1}^{(-)}(\omega) e^{-i\omega r_\ast}  + B_{1}^{(+)} (\omega) e^{+i\omega r_\ast}}.
\end{equation}
\end{subequations}
In the previous expressions, the coefficients $A_{1}^{(\pm)}(\omega)$ and $B_{1}^{(\pm)}(\omega)$ are complex amplitudes, while the Wronskian $W_1(\omega)$ is given by 
\begin{equation}
\label{Well}
W_{1}(\omega) = 2i\omega A_{1}^{(-)} (\omega) = 2i\omega B_{1}^{(+)} (\omega).
\end{equation}

Using the Green's function (\ref{Green_Function_2}) and taking into account Eq. \eqref{Well}, we can show that the solution of the equation with the source term \eqref{dipole_odd_parity} is given by
\begin{subequations}\label{G_Sol_ZMetRW_Eq_en_ast}
\begin{eqnarray}
 \phi_{\omega 10}(r) &=&- \int_{-\infty}^{+\infty}d{r}_{*}'\,G_{\omega 1}(r_{*},{r}_{*}')
S^{(\phi)}_{\omega 10}({r}_{*}') \label{G_Sol_Dipole_Eq_Odd_a}\\
  &=& - \int_{2M}^{6M} \frac{dr'}{f(r')} G_{\omega 1}(r,r')
S^{(\phi)}_{\omega 10}(r') \label{G_Sol_Dipole_Eq_Odd_b}
\end{eqnarray}
\end{subequations}
For an observer at a finite distance from the BH and located beyond the source, we obtain
\begin{equation}
\label{Partial_Response_1}
\phi_{\omega 10}(r)= \frac{\phi_{\omega 1}^{\mathrm {up}}(r)}{2i\omega A_{1}^{(-)} (\omega)} \int_{2M}^{6M}\frac{dr'}{f(r')} \,\phi_{\omega 1}^{\mathrm {in}}(r')\,S^{(\phi)}_{\omega 10}(r')
\end{equation}
In the time domain, the dipole mode waveform is given by
\begin{multline}
\label{Partial_Response_2}
\phi_{10}(t,r)=\frac{1}{\sqrt{2}} \int_{-\infty}^{+\infty} d\omega \left(\frac{e^{-i\omega t}}{2i\omega A_{1}^{(-)} (\omega)}\right) \\
 \times \phi_{\omega 1}^{\mathrm {up}}(r) \int_{2M}^{6M}\frac{dr'}{f(r')} \,\phi_{\omega 1}^{\mathrm {in}}(r')\,S^{(\phi)}_{\omega 10}(r').
\end{multline}

\subsubsection{Odd-parity dipole modes ($\ell  \geq 2$)}

In order to solve the system of coupled differential equations \eqref{Système_odd_freq_1} and \eqref{Système_odd_freq_2} governing the partial modes $\ell \geq 2$ of odd parity, we will use the Green's matrix method. It can be shown that the solution for the partial amplitudes can be written in the form \cite{Sisman:2009mk,doi:10.1137/0504011,WilliamReid}
\begin{subequations}
\begin{eqnarray}
  \bm{\Phi}_{\omega \ell m}(r) &=& \int_{-\infty}^{+\infty} dr'_* \bm{\mathcal{G}}_{\omega\ell}(r_*,r'_*) \bm{\mathcal{S}}_{\omega\ell m}(r'_*) \label{Sol_Odd_Coupled_1}\\
                               &=& \int_{2M}^{6M} \frac{dr'}{f(r')} \bm{\mathcal{G}}_{\omega\ell}(r,r') \bm{\mathcal{S}}_{\omega\ell m}(r') \label{Sol_Odd_Coupled_2}  
\end{eqnarray} 
\end{subequations}
where the amplitude vector is 
\begin{equation}
 \bm{\Phi}_{\omega \ell m}(r) = \begin{pmatrix} \phi_{\omega\ell m} \\ \psi_{\omega\ell m}  \end{pmatrix}
\end{equation}
and the source vector  is 
\begin{equation}
 \bm{\mathcal{S}}_{\omega\ell m}(r) = \begin{pmatrix} S^{(\phi)}_{\omega\ell m} \\ S^{(\psi)}_{\omega\ell m}  \end{pmatrix}
\end{equation}
where the Green's matrix is given by 
\begin{equation}
\label{Green_Matrix_odd}
\hspace{-10pt}\bm{\mathcal{G}}_{\omega\ell}(r_*,r'_*)=
\left\{\,
\begin{aligned}
&\hspace{-7pt}      -\bm{U} \bm{W^{(\textbf{in})}}(r_*) \bm{W}^{-1}(r'_*) \bm{L}   ,\!\!&r_{*}<{r}_{\ast}',\\
&\hspace{-7pt}\quad  \bm{U} \bm{W^{(\textbf{up})}}(r_*) \bm{W}^{-1}(r'_*) \bm{L}   ,\!\!&r_{*}>{r}_{\ast}'.
\end{aligned}
\right.
\end{equation}

In expression \eqref{Green_Matrix_odd}, $\bm{W}(r)$ denotes the Wronskian matrix associated with the pair of coupled differential equations \eqref{Système_odd_freq_1} and \eqref{Système_odd_freq_2}. It is constructed from the four independent homogeneous solutions of these equations and is given by
\begin{equation}
\hspace{-10pt}\bm{W} = \begin{pmatrix}
\phi^{(\text{in},0)} & \phi^{(\text{in},1)} & \phi^{(\text{up},0)} & \phi^{(\text{up},1)} \\
\psi^{(\text{in},0)} & \psi^{(\text{in},1)} & \psi^{(\text{up},0)} & \psi^{(\text{up},1)} \\
\partial_{r_*} \phi^{(\text{in},0)} &\partial_{r_*} \phi^{(\text{in},1)} &\partial_{r_*} \phi^{(\text{up},0)}  & \partial_{r_*} \phi^{(\text{up},1)}  \\
\partial_{r_*} \psi^{(\text{in},0)} &\partial_{r_*} \psi^{(\text{in},1)} &\partial_{r_*} \psi^{(\text{up},0)}  & \partial_{r_*} \psi^{(\text{up},1)} 
\end{pmatrix}
\end{equation}
where the solutions ($\phi^{(\text{in},i)}, \psi^{(\text{in},i)}$) are characterized by their purely
ingoing behavior at the event horizon
\begin{align}\label{Cond_Coupled_Phi_Psi_rh_condensed}
&\begin{pmatrix} \phi^{(\text{in}, i)} \\ \psi^{(\text{in}, i)} \end{pmatrix} \underset{r_\ast \to -\infty}{\sim} e^{-i \omega r_*} \begin{pmatrix} \delta_{i0} \\ \delta_{i1} \end{pmatrix},
\end{align}
while, the solutions ($\phi^{(\text{up},i)}, \psi^{(\text{up},i)}$)   exhibit a purely outgoing behavior at spatial infinity
\begin{align}\label{Cond_Coupled_Phi_Psi_inf_condensed}
&\begin{pmatrix} \phi^{(\text{up}, i)} \\ \psi^{(\text{up}, i)} \end{pmatrix} \underset{r_\ast \to +\infty}{\sim} e^{+i \left[p(\omega) r_* + \frac{M\mu^2}{p(\omega)}\ln\left(\frac{r}{M}\right)\right]} \begin{pmatrix} \delta_{i0} \\ \delta_{i1} \end{pmatrix}
\end{align}
Here, $\delta_{ij}$ is the Kronecker delta function, defined as $\delta_{ij} = 1$  when $i = j$ and  $\delta_{ij} = 0$ when $i \neq j$.

The matrices  $\bm{W^{(\textbf{in})}}$  and $\bm{W^{(\textbf{up})}}$  are both constructed from the Wronskian matrix $\bm{W}$. Specifically, they correspond to the ingoing and outgoing solutions, respectively, and are given by
\begin{equation}\label{W_in}
  \bm{W^{(\textbf{in})}} = \bm{W} \begin{pmatrix}
\bm{I} & \bm{0} \\
\bm{0} & \bm{0}
\end{pmatrix}
\end{equation}
and 
\begin{equation}\label{W_up}
  \bm{W^{(\textbf{up})}} = \bm{W} \begin{pmatrix}
\bm{0} & \bm{0} \\
\bm{0} & \bm{I}
\end{pmatrix}
\end{equation}

Finally, the matrix $\bm{U}_{2 \times 4}$  and the matrix $\bm{L}_{4 \times 2}$ act as ``selection matrices'' and are given by
\begin{equation}\label{U_L}
\bm{U} = \begin{pmatrix}
\bm{I} & \bm{0}
\end{pmatrix} \quad \text{and} \quad
\bm{L} = \begin{pmatrix}
\bm{0} \\
\bm{I}
\end{pmatrix}
\end{equation}
where $\bm{I}$ denotes the  $2 \times 2$ identity matrix in the previous expressions.

For an observer at a finite distance from the BH and located beyond the source, we obtain [cf., Eqs.~\eqref{Sol_Odd_Coupled_2}~and~\eqref{Green_Matrix_odd}]
\begin{equation}\label{Sol_Odd_Coupled_inf}
  \bm{\Phi}_{\omega \ell m}(r) = \int_{2M}^{6M} \frac{dr'}{f(r')}  \bm{U} \bm{W^{(\textbf{up})}}(r) \bm{W}^{-1}(r') \bm{L}  \bm{\mathcal{S}}_{\omega\ell m}(r')
\end{equation}  
and, in the time domain, the $\ell \geq 2$ modes waveform are given by
\begin{multline}\label{Sol_Odd_Coupled_inf_time}
  \bm{\Phi}_{\ell m}(t,r) = \frac{1}{\sqrt{2\pi}}\int_{-\infty}^{+\infty} d\omega \, e^{-i \omega t} \\
   \times \int_{2M}^{6M} \frac{dr'}{f(r')}  \, \bm{U} \bm{W^{(\textbf{up})}}(r) \bm{W}^{-1}(r') \bm{L}  \bm{\mathcal{S}}_{\omega\ell m}(r')
\end{multline}  
with the vector amplitude  $\bm{\Phi}_{\ell m}(t,r) = \begin{pmatrix}\phi_{\ell m} & \psi_{\ell m} \end{pmatrix}^\top $.

\subsection{Construction of the partial amplitudes : Even-parity sector}
\label{sec_5_2}

\subsubsection{Even-paity monopole mode}
To construct the partial amplitudes of the even-parity monopole, we have to solve the system of differential equations  \eqref{l_0_even_freq_1} and \eqref{l_0_even_freq_2}, we have applied, \textit{mutatis mutandis}, the Green's matrix method described in the previous section. As a result, the monopole mode  partial amplitudes are obtained for an observer at a finite distance from the black hole, located beyond the source
\begin{equation}\label{Sol_Even_l_0_Coupled_inf_frequency}
  \bm{\Psi}_{\omega 00}(r) = \int_{2M}^{6M} \frac{dr'}{f(r')}  \, \bm{U} \bm{W^{(\textbf{up})}}(r) \bm{W}^{-1}(r') \bm{L}  \bm{\mathcal{S}}_{\omega 00 }(r')
\end{equation}
and,  in the time domain, the waveforms are  given by
\begin{multline}\label{Sol_Even_l_0_Coupled_inf_time}
  \bm{\Psi}_{00}(t,r) = \frac{1}{\sqrt{2}}\int_{-\infty}^{+\infty} d\omega \, e^{-i \omega t} \\
   \times \int_{2M}^{6M} \frac{dr'}{f(r')}  \, \bm{U} \bm{W^{(\textbf{up})}}(r) \bm{W}^{-1}(r') \bm{L}  \bm{\mathcal{S}}_{\omega 00}(r')
\end{multline}
In expression \eqref{Sol_Even_l_0_Coupled_inf_time}, the vector amplitude $\bm{\Psi}_{00}(t,r) = \begin{pmatrix}K  &  H_{tr}\end{pmatrix}^\top$, the source vector is composed of the source terms 
$\bm{\mathcal{S}}_{\omega 00} = \begin{pmatrix}S_\omega^{(K)} & S_{\omega}^{(H)}\end{pmatrix}^\top$, and the  $4 \times 4$ Wronskian matrix $\bm{W}$ is constructed from the independent homogeneous solutions of the system \eqref{l_0_even_freq_1} and \eqref{l_0_even_freq_2}. Two of these homogeneous solutions,  ($K^{(\text{in},i)},H_{tr}^{(\text{in},i)}$), are characterized by their ingoing behavior at the event horizon
\begin{align}\label{Cond_Coupled_K_H_rh_l_0_condensed}
&\begin{pmatrix} K^{(\text{in}, i)} \\ H_{tr}^{(\text{in}, i)} \end{pmatrix} \underset{r_\ast \to -\infty}{\sim} \begin{pmatrix} e^{-i \omega r_*} \\ e^{-i \omega r_* - \frac{r_*}{2M}} \end{pmatrix}  \begin{pmatrix} \delta_{i0} \\ \delta_{i1} \end{pmatrix},
\end{align}
while the other two solutions, ($K^{(\text{up},i)},H_{tr}^{(\text{up},i)}$), exhibit outgoing behavior at spatial infinity
\begin{align}\label{Cond_Coupled_K_H_inf_l_0_condensed}
&\hspace{-15pt}\begin{pmatrix} K^{(\text{up}, i)} \\ H_{tr}^{(\text{up}, i)} \end{pmatrix} \underset{r_\ast \to +\infty}{\sim} e^{+i \left[p(\omega) r_* + \frac{M \mu^2}{p(\omega)} \ln \left( \frac{r}{M} \right)\right] - \ln \left( \frac{r}{2M} \right)} \begin{pmatrix} \delta_{i0} \\ \delta_{i1} \end{pmatrix},
\end{align}
where $i = 0,1$.

\subsubsection{Even-parity dipole mode}

For the dipole mode $\ell = 1$ , which is governed by the system of Eqs. \eqref{Système_even_freq_1_bis} and \eqref{Système_even_freq_2_bis}, the solution is also obtained by using Green's matrix machinery. The dipole mode waveforms, for an observer located beyond the source at a finite distance from the BH, are then given by
\begin{equation}\label{Sol_Even_l_1_Coupled_inf_frequency}
  \bm{\Psi}_{\omega 11}(r) = \int_{2M}^{6M} \frac{dr'}{f(r')}  \, \bm{U} \bm{W^{(\textbf{up})}}(r) \bm{W}^{-1}(r') \bm{L}  \bm{\mathcal{S}}_{\omega 11 }(r')
\end{equation}
and we have in the time domain
\begin{multline}\label{Sol_Even_l_1_Coupled_inf_time}
  \bm{\Psi}_{11}(t,r) = \frac{1}{\sqrt{2\pi}}\int_{-\infty}^{+\infty} d\omega \, e^{-i \omega t} \\
   \times \int_{2M}^{6M} \frac{dr'}{f(r')}  \, \bm{U} \bm{W^{(\textbf{up})}}(r) \bm{W}^{-1}(r') \bm{L}  \bm{\mathcal{S}}_{\omega 11 }(r')
\end{multline}
with the vector amplitude  $\bm{\Psi}_{11}(t,r) = \begin{pmatrix}K & H_r \end{pmatrix}^\top $ and the source vector  $ \bm{\mathcal{S}}_{\omega 11} = \begin{pmatrix}S_\omega^{(K)} & S_\omega^{(H)}\end{pmatrix}^\top$. The  $4 \times 4$ Wronskian matrix $\bm{W}$ is constructed from the independent homogeneous solutions of the system \eqref{Système_even_freq_1_bis} and \eqref{Système_even_freq_2_bis}, where two of these solutions, ($K^{(\text{in},i)},H_{r}^{(\text{in},i)}$), exhibit ingoing behavior at the event horizon
\begin{align}\label{Cond_Coupled_K_H_rh_l_1_condensed}
&\begin{pmatrix} K^{(\text{in}, i)} \\ H_{r}^{(\text{in}, i)} \end{pmatrix} \underset{r_\ast \to -\infty}{\sim} \begin{pmatrix} e^{-i \omega r_*} \\ e^{-i \omega r_* - \frac{r_*}{2M}} \end{pmatrix} \cdot \begin{pmatrix} \delta_{i0} \\ \delta_{i1} \end{pmatrix},
\end{align}
and for the other two solutions, ($K^{(\text{up},i)},H_{r}^{(\text{up},i)}$), the outgoing behavior at spatial infinity
\begin{align}\label{Cond_Coupled_K_H_rh_l_1_condensed}
&\begin{pmatrix} K^{(\text{up}, i)} \\ H_{r}^{(\text{up}, i)} \end{pmatrix} \underset{r_\ast \to +\infty}{\sim} \begin{pmatrix} e^{+i \left[p(\omega) r_* + \frac{M \mu^2}{p(\omega)} \ln \left(\frac{r}{M}\right)\right] - \ln \left(\frac{r}{2M}\right)} \\ e^{+i \left[p(\omega) r_* + \frac{M \mu^2}{p(\omega)} \ln \left(\frac{r}{M}\right)\right]} \end{pmatrix} \begin{pmatrix} \delta_{i0} \\ \delta_{i1} \end{pmatrix}.
\end{align}
where $i = 0,1$.

\subsubsection{Even-parity dipole modes ($\ell \geq 2$)}

In the case of dipole modes ($\ell \geq 2$) governed by the system of three coupled differential equations, Eqs.~\eqref{Système_even_freq_1}--\eqref{Système_even_freq_3}, the Green's matrix method can be extended to obtain the solution, which is given by
\begin{equation}\label{Sol_Even_l_2_Coupled_inf_frequency} 
\bm{\Psi}_{\omega\ell m}(r) = \int_{2M}^{6M} \frac{dr'}{f(r')}  \, \bm{U} \bm{W^{(\textbf{up})}}(r) \bm{W}^{-1}(r') \bm{L}  \bm{\mathcal{S}}_{\omega \ell m }(r')
\end{equation}
that can be written in the time domain 
\begin{multline}\label{Sol_Even_l_2_Coupled_inf_time}
  \bm{\Psi}_{\ell m}(t,r) = \frac{1}{\sqrt{2\pi}}\int_{-\infty}^{+\infty} d\omega \, e^{-i \omega t} \\
   \times \int_{2M}^{6M} \frac{dr'}{f(r')}  \, \bm{U} \bm{W^{(\textbf{up})}}(r) \bm{W}^{-1}(r') \bm{L}  \bm{\mathcal{S}}_{\omega \ell m }(r')
\end{multline}
Here, the vector amplitude is denoted as $ \bm{\Psi}_{\ell m}(t,r) = \begin{pmatrix}K^{\ell m} & H_r^{\ell} & G^{\ell m}\end{pmatrix}^\top$, and the source vector as $  \bm{\mathcal{S}}_{\omega \ell m } = \begin{pmatrix}S^{K}_{\omega\ell m} & S^{H}_{\omega\ell m} & S^{G}_{\omega\ell m}\end{pmatrix}^\top$. The  $6 \times 6$  Wronskian matrix, constructed from the six independent homogeneous solutions of the coupled system \eqref{Système_even_freq_1}–\eqref{Système_even_freq_3}, is expressed as
\begin{widetext}
\begin{equation}
\bm{W} = \begin{pmatrix}
K^{(\text{in},0)} & K^{(\text{in},1)} & K^{(\text{in},2)} & K^{(\text{up},0)} & K^{(\text{up},1)} & K^{(\text{up},2)} \\
H_r^{(\text{in},0)} & H_r^{(\text{in},1)} & H_r^{(\text{in},2)} & H_r^{(\text{up},0)} & H_r^{(\text{up},1)} & H_r^{(\text{up},2)} \\
G^{(\text{in},0)} & G^{(\text{in},1)} & G^{(\text{in},2)} & G^{(\text{up},0)} & G^{(\text{up},1)} & G^{(\text{up},2)} \\
\partial_{r_*} K^{(\text{in},0)} & \partial_{r_*} K^{(\text{in},1)} & \partial_{r_*} K^{(\text{in},2)} & \partial_{r_*} K^{(\text{up},0)} & \partial_{r_*} K^{(\text{up},1)} & \partial_{r_*} K^{(\text{up},2)} \\
\partial_{r_*} H_r^{(\text{in},0)} & \partial_{r_*} H_r^{(\text{in},1)} & \partial_{r_*} H_r^{(\text{in},2)} & \partial_{r_*} H_r^{(\text{up},0)} & \partial_{r_*} H_r^{(\text{up},1)} & \partial_{r_*} H_r^{(\text{up},2)} \\
\partial_{r_*} G^{(\text{in},0)} & \partial_{r_*} G^{(\text{in},1)} & \partial_{r_*} G^{(\text{in},2)} & \partial_{r_*} G^{(\text{up},0)} & \partial_{r_*} G^{(\text{up},1)} & \partial_{r_*} G^{(\text{up},2)}
\end{pmatrix}
\end{equation}
\end{widetext}
while the matrices $\bm{W^{(\textbf{up})}}$, $\bm{U}_{3 \times 6} $ and $\bm{L}_{6 \times 3} $ are provided in \eqref{W_up} and \eqref{U_L}, respectively, with  $\bm{I}$ being the $3 \times 3$ identity matrix.

The solutions $K^{(\text{in},i)}, H_r^{(\text{in},i)}$ and $G^{(\text{in},i)}$, with  $i = 0,1$ et $2$, exhibit ingoing behavior at the horizon
\begin{equation}\label{Cond_Coupled_K_H_G_rh_l}
\begin{aligned}
&\begin{pmatrix} K^{(\text{in},i)} \\ H_r^{(\text{in},i)} \\ G_r^{(\text{in},i)} \end{pmatrix} \scriptstyle{\underset{r_\ast \to -\infty}{\sim}} \begin{pmatrix} e^{-i \omega r_*} \\ e^{-i \omega r_* - \frac{r_*}{2M}} \\ e^{-i \omega r_*} \end{pmatrix}  \begin{pmatrix} \delta_{i0} \\ \delta_{i1} \\ \delta_{i2} \end{pmatrix},
\end{aligned}
\end{equation}
and the solutions $K^{(\text{up},i)}, H_r^{(\text{up},i)}$, and $G^{(\text{up},i)}$ an outgoing behavior at spatial infinity 
\begin{align}\label{Cond_Coupled_K_H_G_inf_l}
&\begin{pmatrix} K^{(\text{up}, i)} \\ H_{r}^{(\text{up}, i)} \\ G_{r}^{(\text{up}, i)}  \end{pmatrix} \underset{r_\ast \to +\infty}{\sim} \begin{pmatrix} e^{+i \left[p(\omega) r_* + \frac{M \mu^2}{p(\omega)} \ln \left(\frac{r}{M}\right)\right] - \ln \left(\frac{r}{2M}\right)} \\ e^{+i \left[p(\omega) r_* + \frac{M \mu^2}{p(\omega)} \ln \left(\frac{r}{M}\right)\right]} \\ e^{+i \left[p(\omega) r_* + \frac{M \mu^2}{p(\omega)} \ln \left(\frac{r}{M}\right)\right] - \ln \left(\frac{r}{2M}\right)} \end{pmatrix} \begin{pmatrix} \delta_{i0} \\ \delta_{i1} \\  \delta_{i2}\end{pmatrix}.
\end{align}

\subsection{Quasinormal ringings due to the plunging massive particle}
\label{sec_5_3}

In this section, we construct the quasinormal ringings associated with the partial wave amplitudes \eqref{Sol_Odd_Coupled_inf_time} and \eqref{Sol_Even_l_2_Coupled_inf_time}, corresponding to odd-parity modes ($\ell \geq 2$) and even-parity modes ($\ell \geq 1$), respectively, which are obtained by summing the contributions of all QNMs obtained from the different branches of the quasinormal frequencies. To extract the specific quasinormal ringings from these partial amplitudes, the contour of the integration over \(\omega\) is deformed according to a standard procedure (see, e.g., Ref.~\cite{Leaver86}). This deformation allows us to capture the zeros of the determinant of the Wronskian matrix, \(\bm{W}(r)\), lying in the lower half of the complex \(\omega\)-plane. These zeros, associated with each parity and angular mode \(\ell\), correspond to the complex frequencies \(\omega_{s\ell n}\) of the \((s, \ell, n)\) QNMs, constructed in Sec.~\ref{sec_4}.

For a given angular mode \(\ell\), the index \(n = 0\) identifies the fundamental QNM (i.e., the least damped mode), while \(n = 1, 2, \dots\) correspond to the overtones. The parameter \(s\) specifies the type of branch: scalar (\(s = 0\)), electromagnetic (\(s = 1\)), or gravitational (\(s = 2\)) in the massless limit. Furthermore, the spectrum of quasinormal frequencies is symmetric with respect to the imaginary axis. Specifically, if \(\omega_{s\ell n}\) is a quasinormal frequency in the fourth quadrant, then \(-\omega_{s\ell n}^*\) is the symmetric frequency in the third quadrant. We then easily get from \eqref{Sol_Odd_Coupled_inf_time} for the odd-parity

\begin{equation}\label{QNM_odd}
  \bm{\Phi}^{\text{QNM}}_{\ell m}(t,r) = \sum_{n = 0}^{+\infty} \sum_{s}  \bm{\Phi}^{\text{QNM}}_{s \ell m n}(t,r) 
\end{equation}
with 
\begin{multline}\label{QNM_odd_n_s}
  \bm{\Phi}^{\text{QNM}}_{s\ell m n}(t,r)  = - i\sqrt{2\pi} \Bigl( \bm{C}^{\text{(o)}}_{s \ell m n}e^{- i \omega_{s \ell n} t} \\ + \bm{D}^{\text{(o)}}_{s \ell m n}e^{+ i \omega^{*}_{s \ell n} t} \Bigr)
\end{multline}
In the previous expression, $\bm{C}^{\text{(o)}}_{s \ell m n}$ and $\bm{D}^{\text{(o)}}_{s \ell m n}$ denote the extrinsic excitation coefficients (see, e.g., Refs.~\cite{Leaver86,Berti06,Zhang13})
which are here defined by
\begin{multline}\label{C_odd_fact} 
\mathbf{C}^{\text{(o)}}_{s \ell m n} = \frac{1}{\frac{d}{d\omega}\det(\mathbf{W}(r_\infty))} 
 \int_{2M}^{6M} \frac{dr'}{f(r')} \bm{U} \bm{W}^{(\text{up})}(r) \\ \operatorname{Cof}(\mathbf{W}(r'))^\top \bm{L}  \bm{\mathcal{S}}_{\omega\ell m}(r')\Bigg|_{\omega = \omega_{s\ell n}}
\end{multline}
and 
\begin{multline}\label{D_odd_fact}
  \mathbf{D}^{\text{(o)}}_{s \ell m n} = \frac{1}{\frac{d}{d\omega}\det(\mathbf{W}(r_\infty))} 
  \int_{2M}^{6M} \frac{dr'}{f(r')} \bm{U} \bm{W}^{(\text{up})}(r)  \\ \operatorname{Cof}(\mathbf{W}(r'))^\top \bm{L}  \bm{\mathcal{S}}_{\omega\ell m}(r') \Bigg|_{\omega = -\omega^{*}_{s\ell n}}
\end{multline}
In expressions \eqref{C_odd_fact} and \eqref{D_odd_fact}, since the determinant of the Wronskian matrix is constant at both the horizon and spatial infinity in the odd-parity case, we compute its derivative with respect to \(\omega\) at a very large value of \(r\). In practice, we take \(r= 100M\). The term \(\operatorname{Cof}(\mathbf{W}(r))^\top\) refers to the transpose of the cofactor matrix of \(\mathbf{W}(r)\).

For the even-parity case, since the determinant of the Wronskian matrix \(\mathbf{W}(r)\) is no longer constant but depends on the variable \(r\), the evaluation is slightly different. Using \eqref{Sol_Even_l_2_Coupled_inf_time} we get
\begin{equation}\label{QNM_even}
  \bm{\Psi}^{\text{QNM}}_{\ell m}(t,r) = \sum_{n = 0}^{+\infty} \sum_{s}  \bm{\Psi}^{\text{QNM}}_{s \ell m n}(t,r) 
\end{equation}
with 
\begin{multline}\label{QNM_even_n_s}
  \bm{\Psi}^{\text{QNM}}_{s\ell m n}(t,r)  = - i\sqrt{2\pi} \Bigl( \bm{C}^{\text{(e)}}_{s \ell m n}e^{- i \omega_{s \ell n} t} \\ + \bm{D}^{\text{(e)}}_{s \ell m n}e^{+ i \omega^{*}_{s \ell n} t} \Bigr)
\end{multline}
where 
\begin{multline}\label{C_even_fact} 
  \mathbf{C}^{\text{(e)}}_{s \ell m n} = 
  \int_{2M}^{6M} \frac{dr'}{f(r')} \bm{U} \bm{W}^{(\text{up})}(r) \\ \frac{\operatorname{Cof}(\mathbf{W}(r'))^\top}{\frac{d}{d\omega}\det(\mathbf{W}(r'))} \bm{L}  \bm{\mathcal{S}}_{\omega\ell m}(r') \Bigg|_{\omega = \omega_{s\ell n}}
\end{multline}
and 
\begin{multline}\label{D_even_fact}
  \mathbf{D}^{\text{(e)}}_{s \ell m n} = 
  \int_{2M}^{6M} \frac{dr'}{f(r')} \bm{U} \bm{W}^{(\text{up})}(r) \\ \frac{\operatorname{Cof}(\mathbf{W}(r'))^\top}{\frac{d}{d\omega}\det(\mathbf{W}(r'))} \bm{L}  \bm{\mathcal{S}}_{\omega\ell m}(r') \Bigg|_{\omega = -\omega^{*}_{s\ell n}}
\end{multline}

\subsection{Multipolar gravitational waveforms}
\label{sec_5_4}

In theories of massive gravity, gravitational waves exhibit additional polarization modes beyond the two familiar transverse-traceless polarizations (\( h_+ \) and \( h_\times \)) derived in general relativity. These extra polarizations arise from the massive nature of the graviton and lead to different physical signatures \cite{dePaula:2004bc,Nishizawa:2009bf,Bessada:2009qw,Tachinami:2021jnf,Poddar:2021yjd}. (See also Refs.~\cite{Eardley1973,Eardley197bis,Isi:2017fbj,Gong:2018ybk,Hyun:2018pgn} and references therein for metric-based theories of gravity in four-dimensional spacetime, which predict up to six polarization modes for gravitational waves. This corresponds to the maximum number of independent degrees of freedom that a spin-2 field can propagate).

In the Fierz-Pauli theory considered here, the massive graviton propagates five physical degrees of freedom \cite{deRham:2014zqa,Blasi:2017pkk}, corresponding to five distinct polarization modes. In addition to the two tensor polarizations of general relativity (\( h_+ \) and \( h_\times \)), there are three additional polarization modes: two vector modes (\( h_x \) and \( h_y \)) and a scalar mode, commonly referred to as the ``breathing'' mode (\( h_b \)) \cite{dePaula:2004bc,Nishizawa:2009bf,Bessada:2009qw,Tachinami:2021jnf,Poddar:2021yjd}. The amplitudes of the gravitational wave polarizations observed far from the BH can be obtained from the partial amplitudes constructed in Secs \ref{sec_5_1} and \ref{sec_5_2}, and they can be expressed as \cite{Isi:2017fbj,Nishizawa:2009bf,Tachinami:2021jnf}
\begin{equation}\label{Polarizations}
  h_\mathfrak{p} = h_\mathfrak{p}^{(\text{e})} + h_\mathfrak{p}^{(\text{o})}
\end{equation}
where $\mathfrak{p}$ represents the five polarization modes ($+, \times, x, y, b$). For the even-parity contributions, we have
\begin{subequations}\label{Polarizations_Even}
\begin{eqnarray}
&& h_+^{(\text{e})} =  \sum_{\ell = 2}^{+\infty} \sum_{m = -\ell}^{+\ell} G^{\ell m}  \left[\frac{1}{2}\left(Y^{\ell m}_{\theta\theta} -\frac{Y^{\ell m}_{\varphi\varphi}}{\sin^2\theta}\right) \right]\\
&& h_\times^{(\text{e})} = \sum_{\ell = 2}^{+\infty} \sum_{m = -\ell}^{+\ell} G^{\ell m}  \frac{Y^{\ell m}_{\theta\varphi}}{\sin\theta}\\
&& h_x^{(\text{e})} = \frac{1}{r} \sum_{\ell = 1}^{+\infty} \sum_{m = -\ell}^{+\ell} H_r^{\ell m} Y_\theta^{\ell m}\\
&& h_y^{(\text{e})} = \frac{1}{r} \sum_{\ell = 1}^{+\infty} \sum_{m = -\ell}^{+\ell} H_r^{\ell m} \frac{Y_\varphi^{\ell m}}{\sin\theta}\\
&& h_b^{(\text{e})} =\sum_{\ell = 0}^{+\infty} \sum_{m = -\ell}^{+\ell} K^{\ell m} Y^{\ell m}
\end{eqnarray}
\end{subequations}
and for the odd-parity contributions, we have
\begin{subequations}\label{Polarizations_Odd}
\begin{eqnarray}
&& h_+^{(\text{o})} = \frac{1}{r} \sum_{\ell = 2}^{+\infty} \sum_{m = -\ell}^{+\ell} \psi^{\ell m}  \left[\frac{1}{2}\left(X^{\ell m}_{\theta\theta} -\frac{X^{\ell m}_{\varphi\varphi}}{\sin^2\theta}\right) \right]\\
&& h_\times^{(\text{o})} =\frac{1}{r} \sum_{\ell = 2}^{+\infty} \sum_{m = -\ell}^{+\ell} \psi^{\ell m}  \frac{X^{\ell m}_{\theta\varphi}}{\sin\theta}\\
&& h_x^{(\text{o})} = \frac{1}{r} \sum_{\ell = 1}^{+\infty} \sum_{m = -\ell}^{+\ell} \phi^{\ell m} X_\theta^{\ell m}\\
&& h_y^{(\text{o})} = \frac{1}{r} \sum_{\ell = 1}^{+\infty} \sum_{m = -\ell}^{+\ell} \phi^{\ell m} \frac{X_\varphi^{\ell m}}{\sin\theta}\\
&& h_b^{(\text{o})} = 0
\end{eqnarray}
\end{subequations}
Note that the scalar breathing mode (\( h_b \)) has no odd-parity contribution, since it results from the condition \( X_{\theta\theta} + \frac{X_{\varphi\varphi}}{\sin^2\theta} = 0 \). This distinguishes it from the vectorial and tensorial modes, which have both even and odd parity components.

\subsection{Numerical methods}
\label{sec_5_5}

In this section, we numerically construct the partial amplitude waveforms for both odd- and even-parity modes, as well as their associated quasinormal ringings.

For the partial amplitude waveforms, we distinguish between two cases: first, the odd-parity dipole mode (\(\ell = 1\)), which is governed by a single differential equation; and second, the odd-parity modes with \(\ell \geq 2\) and the even-parity modes with \(\ell \geq 0\), which are governed by systems of two or three coupled differential equations.

\begin{enumerate}[label=(\roman*)]
   \item \label{item:1} In the case of the odd-parity dipole mode ($\ell = 1$), we determined the functions  $\phi_{\omega 1}^{\text{in}}$ and $\phi_{\omega 1}^{\text{up}}$ as well as the coefficient $A_1^{(-)}(\omega)$ by numerically integrating the homogeneous equation \eqref{dipole_odd_parity} using the Runge-Kutta method. The initialization was performed with Taylor series expansions that converge near the horizon. We then compared the solutions with asymptotic expansions, representing ingoing and outgoing behavior at spatial infinity, which were decoded using Padé summation. Finally, we applied a Fourier transform to 
       obtain the final result $\phi_{10}(t,r)$ which is given by \eqref{Partial_Response_2}.

   \item \label{item:2} In the case of coupled systems, such as the $\ell \geq 2$ modes for odd parity and the \( \ell \geq 0 \) modes for even parity, we have generalized the method described in \ref{item:1}. Consider a system of \( n \) coupled second-order differential equations
        \begin{equation}\label{sys_generic}
         \bm{Y}''(r) + \bm{A}(r) \bm{Y}'(r) + \bm{B}(r) \bm{Y}(r) = \bm{S}(r)
         \end{equation}
         where $\bm{Y} = \begin{pmatrix}Y_1 & Y_2 & \dots & Y_n\end{pmatrix}^\top$ is the amplitude vector, $\bm{S} = \begin{pmatrix}S_1 & S_2 & \dots & S_n\end{pmatrix}^\top$ is the source vector, and $\bm{A}(r) $ and $ \bm{B}(r) $ are matrices of size $n \times n$. The $2n \times 2n$ Wronskian matrix for this system can be constructed from independent solutions of the associated homogeneous problem satisfying the appropriate boundary conditions. At the event horizon, the solutions can generally be expressed as
         \begin{equation}\label{Taylor_generic}
           \bm{Y}(r) = e^{-i \omega r_*} \sum_{k=0}^{+\infty} \bm{a}_k f(r)^k
        \end{equation}
         where $\bm{a}_k = \begin{pmatrix}a_k^{(1)} & a_k^{(2)} & \dots & a_k^{(n)}\end{pmatrix}^\top$ is the vector of Taylor series coefficients for the $k$th term. The boundary conditions for the $n$ independent solutions are specified by the first coefficients $\bm{a}_0$. For the first solution, $\bm{a}_0 = \begin{pmatrix}1 & 0 & 0 & \dots & 0\end{pmatrix}^\top$; for the second solution,$\bm{a}_0 = \begin{pmatrix}0 & 1 & 0 & \dots & 0\end{pmatrix}^\top$; and so on until the $n$th solution, where $\bm{a}_0 = \begin{pmatrix}0 & 0 & 0 & \dots & 1\end{pmatrix}^\top$. We then numerically integrated the homogeneous equations from the horizon using the Runge-Kutta method.
         At the spatial infinity, the solutions can generally be expressed as  
          \begin{equation}\label{Asympt_generic}
         \hspace{+20pt}  \bm{Y}(r) = e^{+i\left[ p(\omega) r_*+ \frac{M \mu^2}{p(\omega)}\ln\left(\frac{r}{M}\right)\right]} \sum_{k=0}^{+\infty} \bm{b}_k  \left(\frac{2M}{r}\right)^k
         \end{equation}
          where $\bm{b}_k = \begin{pmatrix}b_k^{(1)} & b_k^{(2)} & \dots & b_k^{(n)}\end{pmatrix}^\top$ represents the vector of coefficients for the asymptotic expansion at the $k$th term. The boundary conditions for the $n$ independent solutions are determined by the first set of coefficients $\bm{b}_0$ with $\bm{b}_0 = \begin{pmatrix}1 & 0 & 0 & \dots & 0\end{pmatrix}^\top$ for the first solution, $\bm{b}_0 = \begin{pmatrix}0 & 1 & 0 & \dots & 0\end{pmatrix}^\top$ for the second, and for the $n$th solution we have $\bm{b}_0 = \begin{pmatrix}0 & 0 & 0 & \dots & 1\end{pmatrix}^\top$.  We used Padé summation to decode additional information, and then numerically integrated the homogeneous equations inward down to the horizon using the Runge-Kutta method.
                  
         It is important to note that for each solved system we have used \eqref{Taylor_generic} and \eqref{Asympt_generic} with the appropriate boundary conditions. These are specific to each system and are detailed in Sec.~\ref{sec_5_1}.

   \item   The partial amplitudes \eqref{Sol_Even_l_0_Coupled_inf_frequency}, \eqref{Sol_Even_l_1_Coupled_inf_frequency}, and \eqref{Sol_Even_l_2_Coupled_inf_frequency} have been regularized. Indeed, these amplitudes as integrals over the radial Schwarzschild coordinate are strongly divergent near the ISCO. This is due to the behavior of the sources \eqref{source_K_monopole_even} and \eqref{source_H_monopole_even} for monopole ($\ell = 0$) mode as well as \eqref{source_K_main} and \eqref{source_H_main} for  $\ell \geq 1$ modes in the limit $r \to 6M$. The regularization process is described in Appendix \ref{appendix_C}. It consists in replacing the partial amplitudes \eqref{Sol_Even_l_0_Coupled_inf_frequency}, \eqref{Sol_Even_l_1_Coupled_inf_frequency} and \eqref{Sol_Even_l_2_Coupled_inf_frequency} by their counterparts \eqref{Psi_i_regularized} and \eqref{Psi_i_regularized_l_2} and to evaluate the result by using Levin's algorithm \cite{Levin1996}.

  \item We have Fourier transformed the components of  $\bm{\Phi}_{\omega\ell m}(r) $ and $\bm{\Psi}_{\omega\ell m}(r) $ to get the final result.

\end{enumerate}

In the case of quasinormal ringing, we focus on the coupled systems of both parities. To construct the quasinormal ringings associated with the partial wave amplitudes \eqref{Sol_Odd_Coupled_inf_time} for odd-parity modes and \eqref{Sol_Even_l_2_Coupled_inf_time} for even-parity modes, it is necessary to numerically compute the partial amplitudes \(\bm{\Phi}^{\text{QNM}}_{\ell m}(t,r)\) and \(\bm{\Psi}^{\text{QNM}}_{\ell m}(t,r)\), which are given by \eqref{QNM_odd} and \eqref{QNM_even}, respectively. This involves determining the quasinormal frequencies \(\omega_{s \ell n}\) as well as the excitation coefficients \(\mathbf{C}^{\text{(e/o)}}_{s \ell m n}\) and \(\mathbf{D}^{\text{(e/o)}}_{s \ell m n}\). The quasinormal frequencies \(\omega_{s \ell n}\) are obtained through the numerical implementation of the matrix-valued Hill determinant approach (see Sec. \ref{sec_4} for more details). The excitation coefficients \(\mathbf{C}^{\text{(e/o)}}_{s \ell m n}\) and \(\mathbf{D}^{\text{(e/o)}}_{s \ell m n}\), on the other hand, can be computed by constructing the Wronskian matrix \(\bm{W}\) following the procedure outlined above [see item (ii.)]. No regularization is required when evaluating the integrals in Eqs.~\eqref{C_odd_fact}, \eqref{D_odd_fact}, \eqref{C_even_fact}, and \eqref{D_even_fact}. However, special care must be taken to address numerical instabilities that may arise. It is also worth noting that for a given \(\ell\), it is often sufficient to consider only the fundamental quasinormal mode (\(n=0\)) for each branch, as this mode is the least damped and therefore dominates the late-time behavior.

Finally, from the partial amplitudes \(\bm{\Phi}_{\omega\ell m}(r)\) and \(\bm{\Psi}_{\omega\ell m}(r)\), more precisely from their components, it is possible to construct the gravitational wave components \(h_\mathfrak{p}^{(\text{e/o})}\) using the sums~\eqref{Polarizations_Even} and \eqref{Polarizations_Odd}. The even components are constructed using \((\ell, m)\) modes with \(\ell = 2, 3\) for \(h_+^{(\text{e})}\) and \(h_\times^{(\text{e})}\), \(\ell = 1, 2\) for \(h_x^{(\text{e})}\) and \(h_y^{(\text{e})}\), and \(\ell = 0, 1, 2\) for \(h_b^{(\text{e})}\), where \(m = \pm \ell\), which constitute the main contributions. Similarly, the odd components are constructed from \((\ell, m)\) modes with \(\ell = 2, 3\) for \(h_+^{(\text{o})}\) and \(h_\times^{(\text{o})}\), and \(\ell = 1, 2, 3\) for \(h_x^{(\text{o})}\) and \(h_y^{(\text{o})}\), with \(m = \pm (\ell - 1)\).  

It should be noted that all numerical calculations have been performed using \emph{Mathematica}~\cite{Mathematica13}.

\section{Results: Waveforms produced by the plunging particle}
\label{sec_6}

\subsection{Partial waveforms and their spectral content: Excitation of QBSs}
\label{sec_6_1}

\subsubsection{Odd-parity partial waveforms}
\label{sec_6_1_1}

\begin{figure}[htbp]
 \includegraphics[scale=0.53]{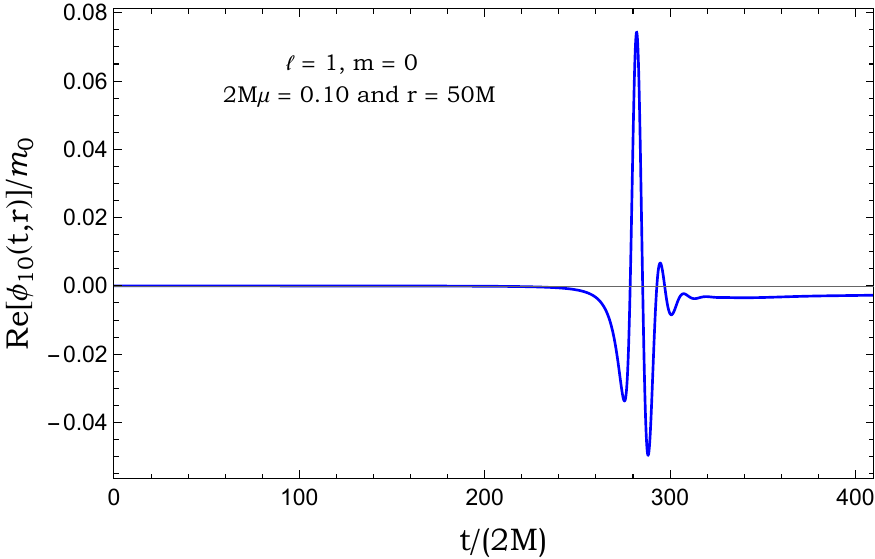}
\caption{\label{Fig:Odd_Waveform_l_1_m_0} Dipolar waveform of the $\phi_{10}$ component produced by the plunging particle. The result is obtained for a massive spin-$2$ field ($2M\mu=0.10$), and the observer is located at $r = 50M$.}
\end{figure}
\begin{figure*}[htbp]
\includegraphics[scale=0.50]{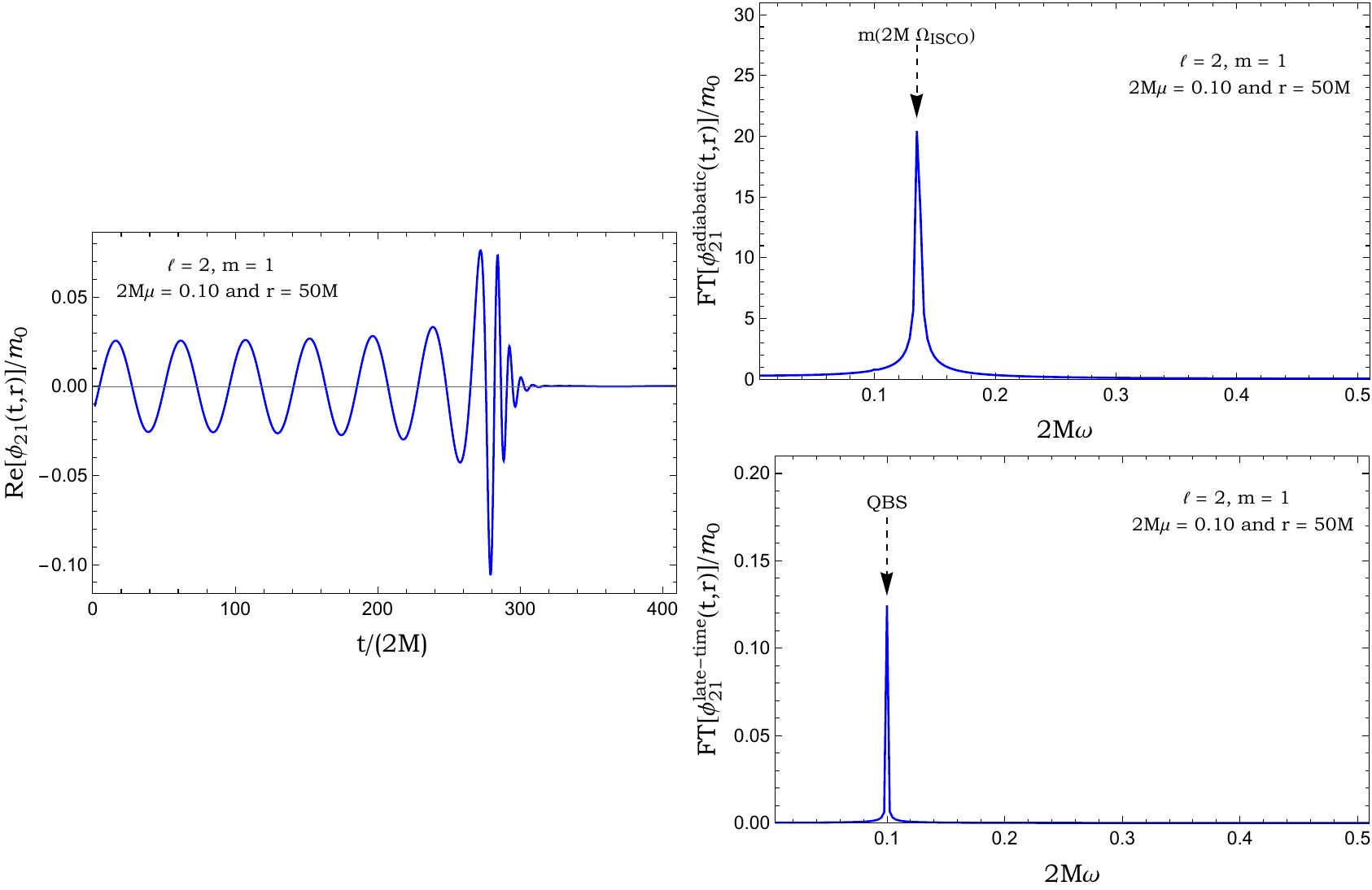}
\caption{\label{Fig:Odd_Phi_Waveform_l_2_m_1}  Quadrupolar waveform of the $\phi_{\ell m}$ component produced by the plunging particle (right panel) and the spectral content of the adiabatic and late-time phases (upper left and lower left panels, respectively). The result is obtained for a massive spin-$2$ field ($2M\mu=0.10$), and the observer is located at $r = 50M$.}
\end{figure*}
\begin{figure*}[htbp]
 \includegraphics[scale=0.50]{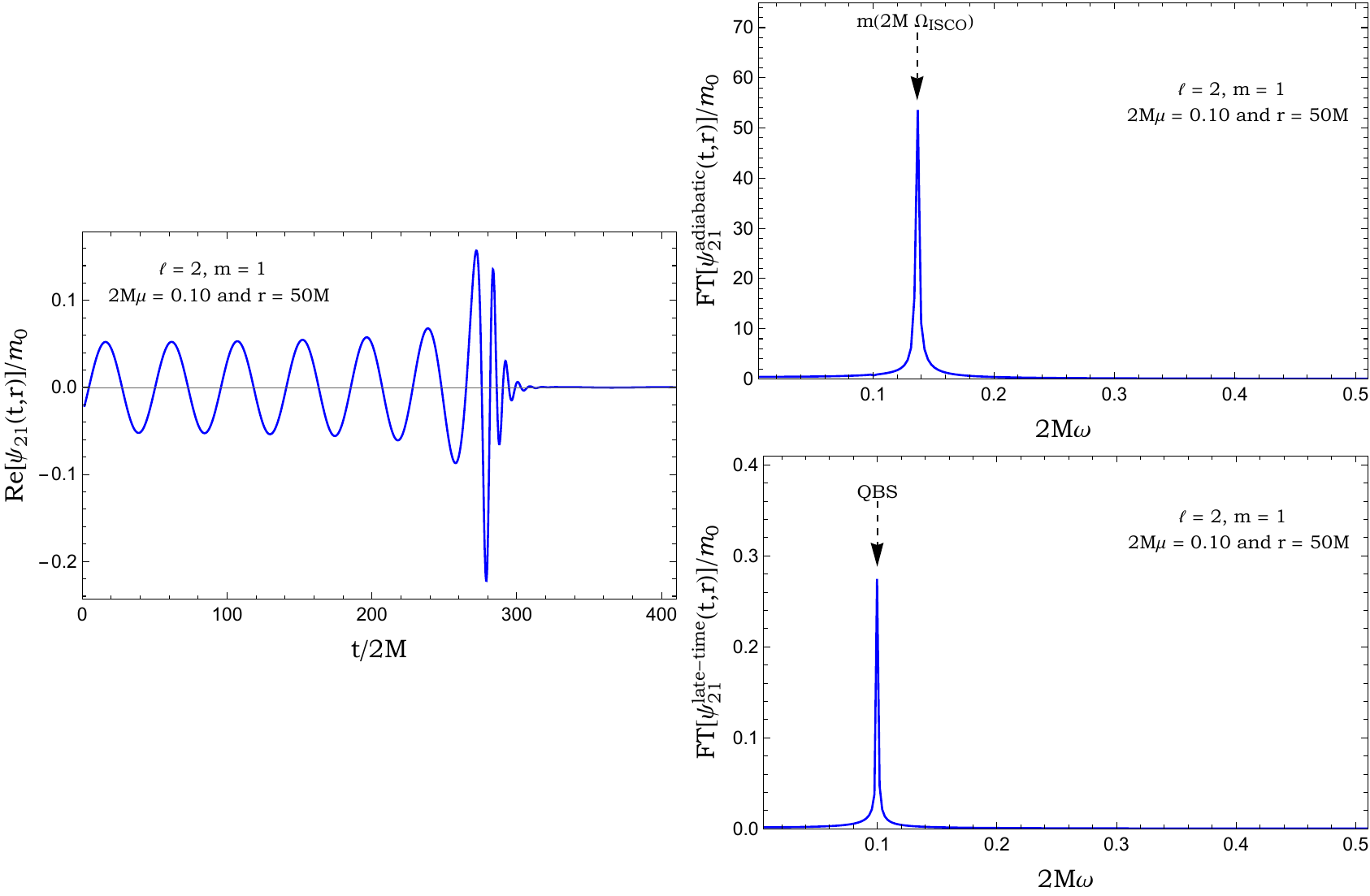}
\caption{\label{Fig:Odd_Psi_Waveform_l_2_m_1} Quadrupolar waveform of the $\psi_{\ell m}$ component produced by the plunging particle (right panel) and the spectral content of the adiabatic and late-time phases (upper left and lower left panels, respectively). The result is obtained for a massive spin-$2$ field ($2M\mu=0.10$), and the observer is located at $r = 50M$.}
\end{figure*}

In Figs.~\ref{Fig:Odd_Waveform_l_1_m_0}, \ref{Fig:Odd_Phi_Waveform_l_2_m_1}, and \ref{Fig:Odd_Psi_Waveform_l_2_m_1} we display the partial waveforms of the odd-parity sector corresponding to the dipole mode $\ell = 1, m = 0$ of the $\phi_{\ell m}$ component and the quadrupole mode $\ell = 2, m = 1$ of the $\phi_{\ell m}$ and $\psi_{\ell m}$ components for a mass coupling parameter $2M\mu = 0.1$. For each of these, we consider that the observer is located at $r = 50M$. The waveforms were obtained by assuming that the particle starts at $r = r_{\text{ISCO}} - \epsilon$, with $\epsilon = 10^{-4}$. In addition, in Eqs. \eqref{trajectory_plung} and \eqref{trajectory_plung_phi} we took $\varphi_0 = 0$ and adjusted $t_0/(2M)$ to shift the interesting part of the signal in the window $t/(2M) \in [0, 410]$.

As with the massive scalar field (see Ref.~\cite{Decanini:2015yba}), the waveform in the massless limit can be decomposed into three phases: (i) the ``adiabatic phase'' corresponding to the quasicircular motion of the particle near the ISCO, (ii) the ringdown phase due to the excitation of QNMs, and (iii) a late-time phase. This decomposition remains generally valid for the massive field (see Figs. \ref{Fig:Odd_Phi_Waveform_l_2_m_1}~and~\ref{Fig:Odd_Psi_Waveform_l_2_m_1}), although the behavior of the signal is now modified by the excitation of QBSs.

We now focus on analyzing the spectral content of the waveform, distinguishing between the adiabatic and late-time phases. The spectral content corresponding to each of these phases can be obtained by applying the Fourier transform, while limiting the time integrations to the phase that is being studied.  Thus, in Figs. \ref{Fig:Odd_Phi_Waveform_l_2_m_1}~and~\ref{Fig:Odd_Psi_Waveform_l_2_m_1}, we show the partial waveform (left panel) and its spectral content in the adiabatic (top right panel) and late-time phases (bottom right panel) for the quadrupole mode ($\ell = 2, m = 1$) of the $\phi_{\ell m}$ and $\psi_{\ell m}$ components, respectively. During the adiabatic phase, a peak is observed at $\omega =  \Omega_{\text{ISCO}}$ for both components, corresponding to the quasicircular motion of the plunging particle near the ISCO, where $\Omega_{\text{ISCO}}$ is given by~\eqref{Angular_Velocity_ISCO}. As for the late-time phase, the spectral analysis reveals a peak at a frequency corresponding to the real part of the complex frequency of the ``first'' long-lived  QBS mode. In fact, in the case of the odd-parity quadrupole mode, three fundamental QBS modes are identified, each corresponding to a given spin projection $S$. The real parts of these three modes are very close, with a difference of $|\Delta\omega| \sim 10^{-5}$. However, the frequency resolution used to construct the waveform, $\delta\omega = 1/1000$, is not sufficient to clearly distinguish the three peaks associated with these modes. Increasing the resolution leads to numerical instabilities, as we are close to the mass of the field.

\subsubsection{Even-parity partial  waveforms}
\label{sec_6_1_2}

\begin{figure*}[htbp]
\includegraphics[scale=0.50]{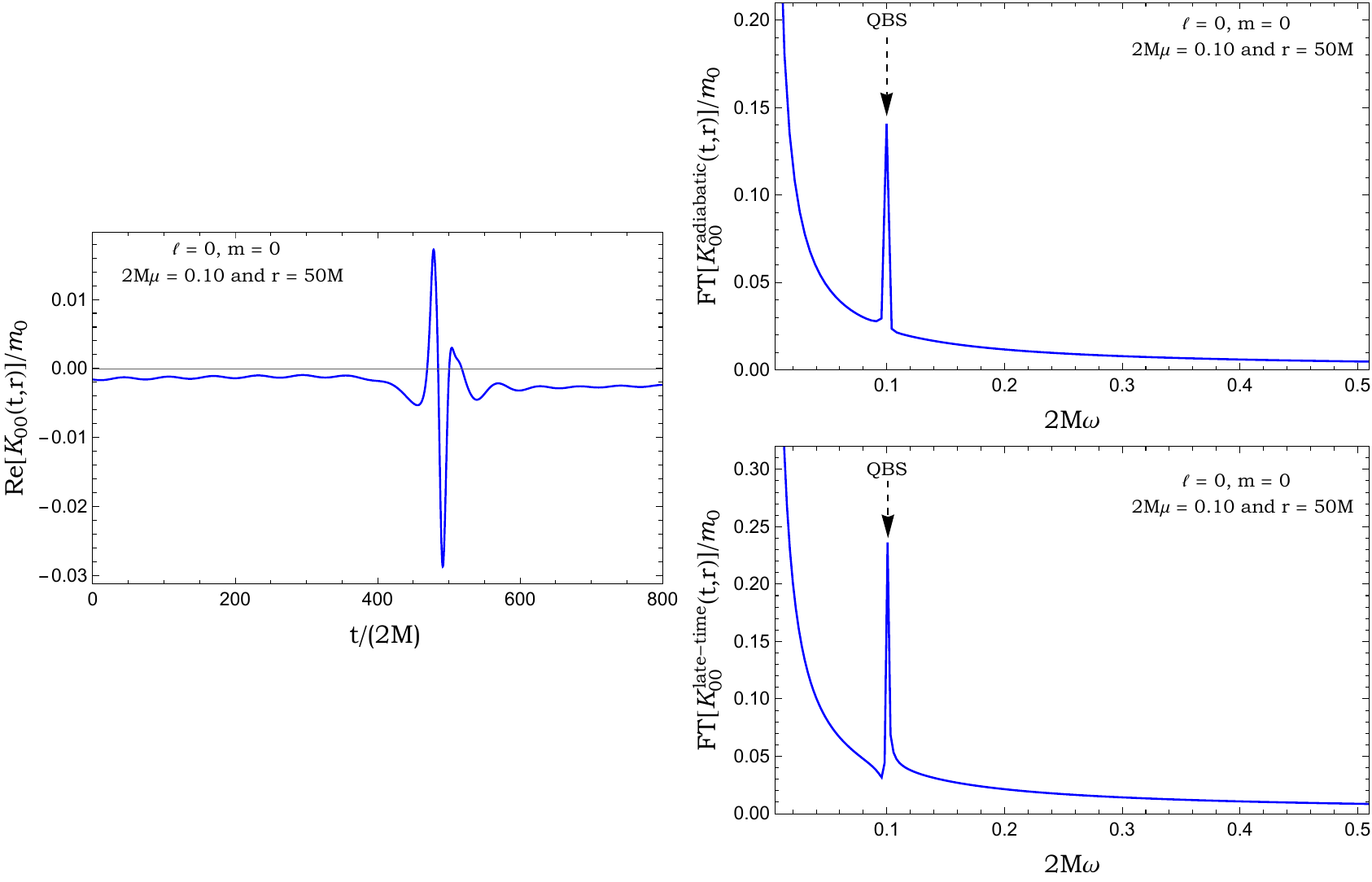}
\caption{\label{Fig:Even_K_Waveform_l_0_m_0}  Monopolar waveform of the $K$ component produced by the plunging particle (right panel) and the spectral content of the adiabatic and late-time phases (upper left and lower left panels, respectively). The result is obtained for a massive spin-$2$ field ($2M\mu=0.10$), and the observer is located at $r = 50M$.}
\end{figure*}
\begin{figure*}[htbp]
 \includegraphics[scale=0.50]{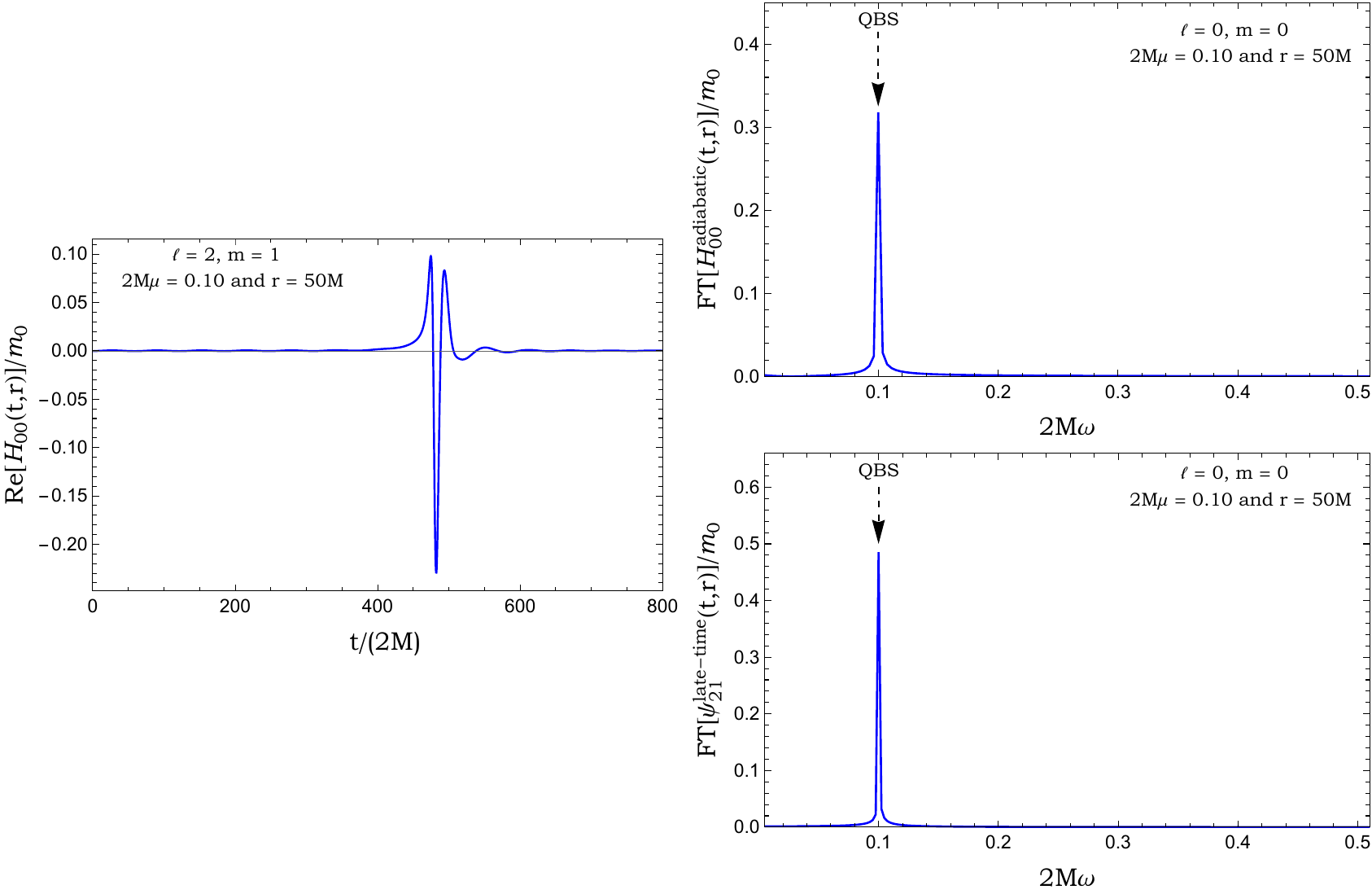}
\caption{\label{Fig:Even_H_Waveform_l_0_m_0} Monopolar waveform of the $H_{tr}$ component produced by the plunging particle (right panel) and the spectral content of the adiabatic and late-time phases (upper left and lower left panels, respectively). The result is obtained for a massive spin-$2$ field ($2M\mu=0.10$), and the observer is located at $r = 50M$.}
\end{figure*}
\begin{figure*}[htbp]
\includegraphics[scale=0.50]{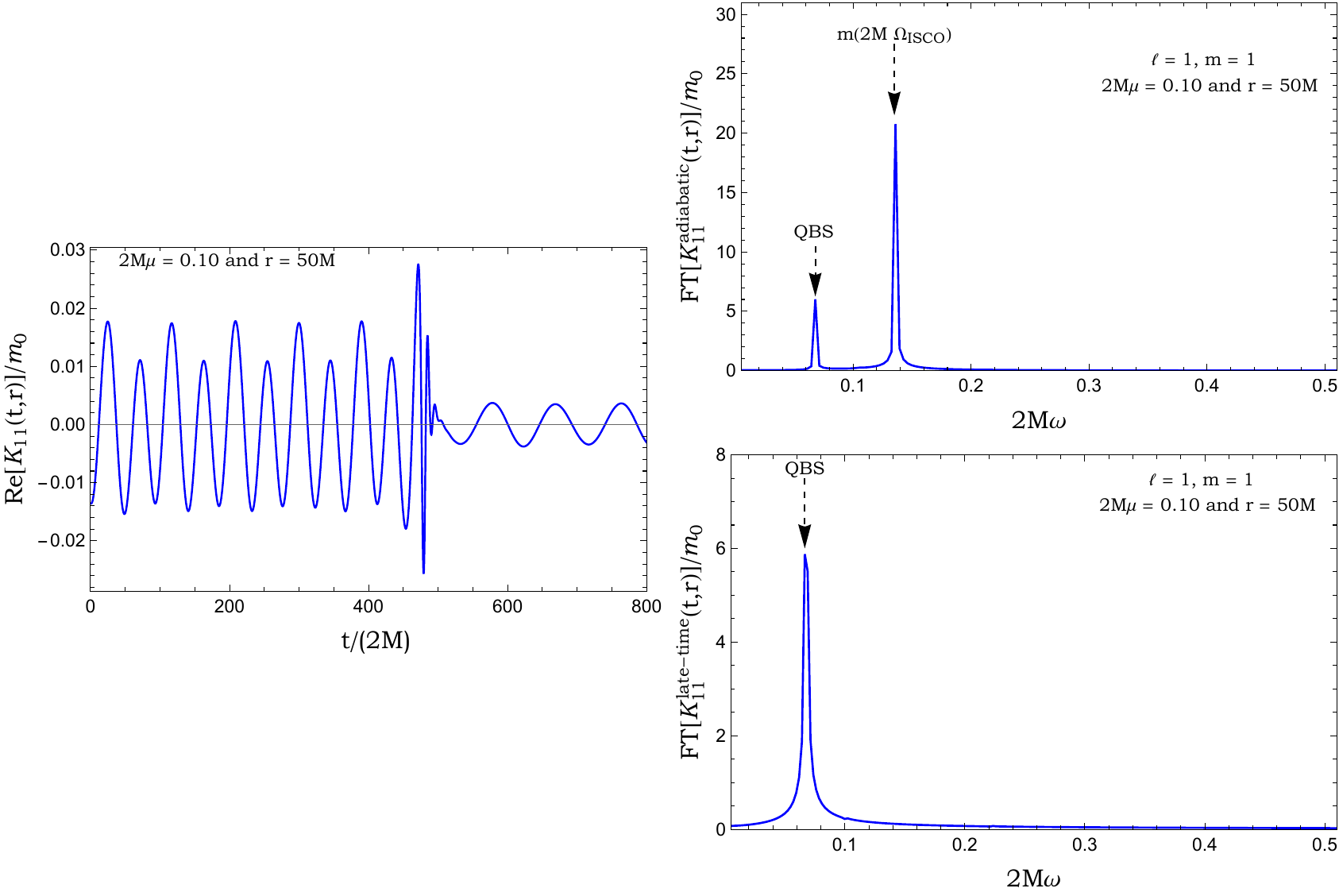}
\caption{\label{Fig:Even_K_Waveform_l_1_m_1} Dipolar waveform of the $K$ component produced by the plunging particle (right panel) and the spectral content of the adiabatic and late-time phases (upper left and lower left panels, respectively). The result is obtained for a massive spin-$2$ field ($2M\mu=0.10$), and the observer is located at $r = 50M$.}
\end{figure*}
\begin{figure*}[htbp]
 \includegraphics[scale=0.50]{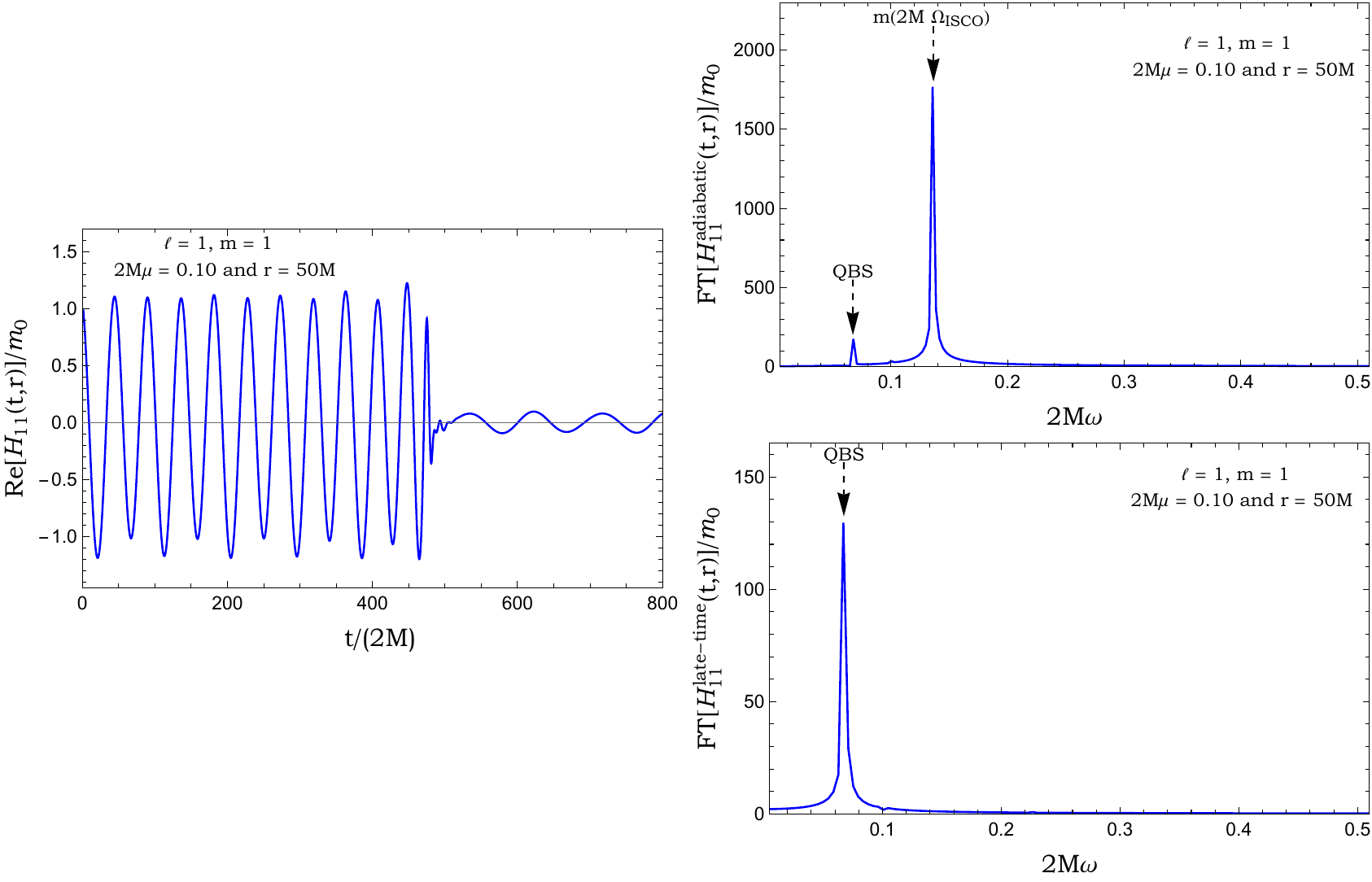}
\caption{\label{Fig:Even_H_Waveform_l_1_m_1} Dipolar waveform of the $H_{r}$ component produced by the plunging particle (right panel) and the spectral content of the adiabatic and late-time phases (upper left and lower left panels, respectively). The result is obtained for a massive spin-$2$ field ($2M\mu=0.10$), and the observer is located at $r = 50M$.}
\end{figure*} 
\begin{figure*}[htbp]
\includegraphics[scale=0.50]{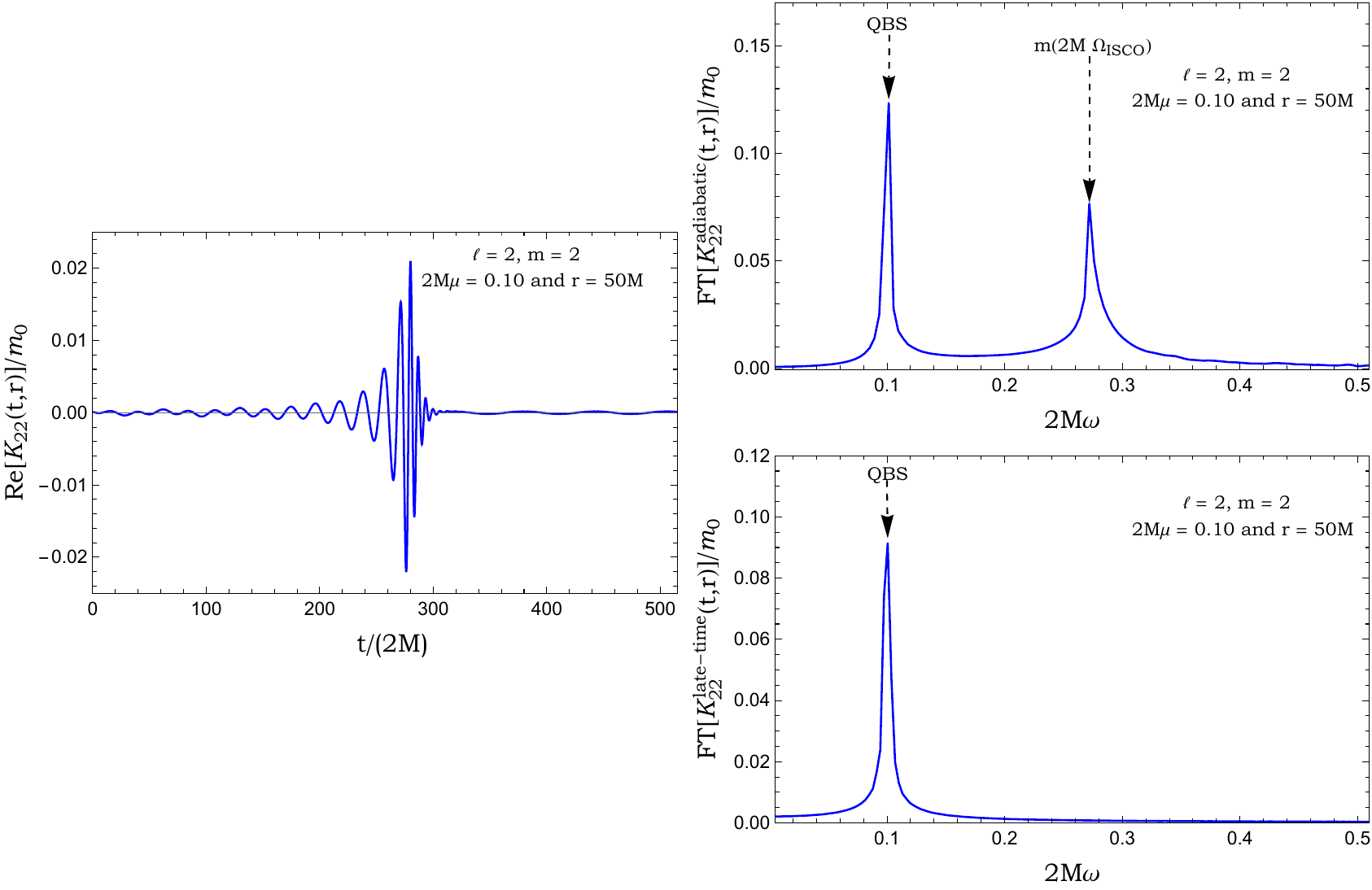}
\caption{\label{Fig:Even_K_Waveform_l_2_m_2} Quadrupolar waveform of the $K$ component produced by the plunging particle (right panel) and the spectral content of the adiabatic and late-time phases (upper left and lower left panels, respectively). The result is obtained for a massive spin-$2$ field ($2M\mu=0.10$), and the observer is located at $r = 50M$.}
\end{figure*}
\begin{figure*}[htbp]
 \includegraphics[scale=0.50]{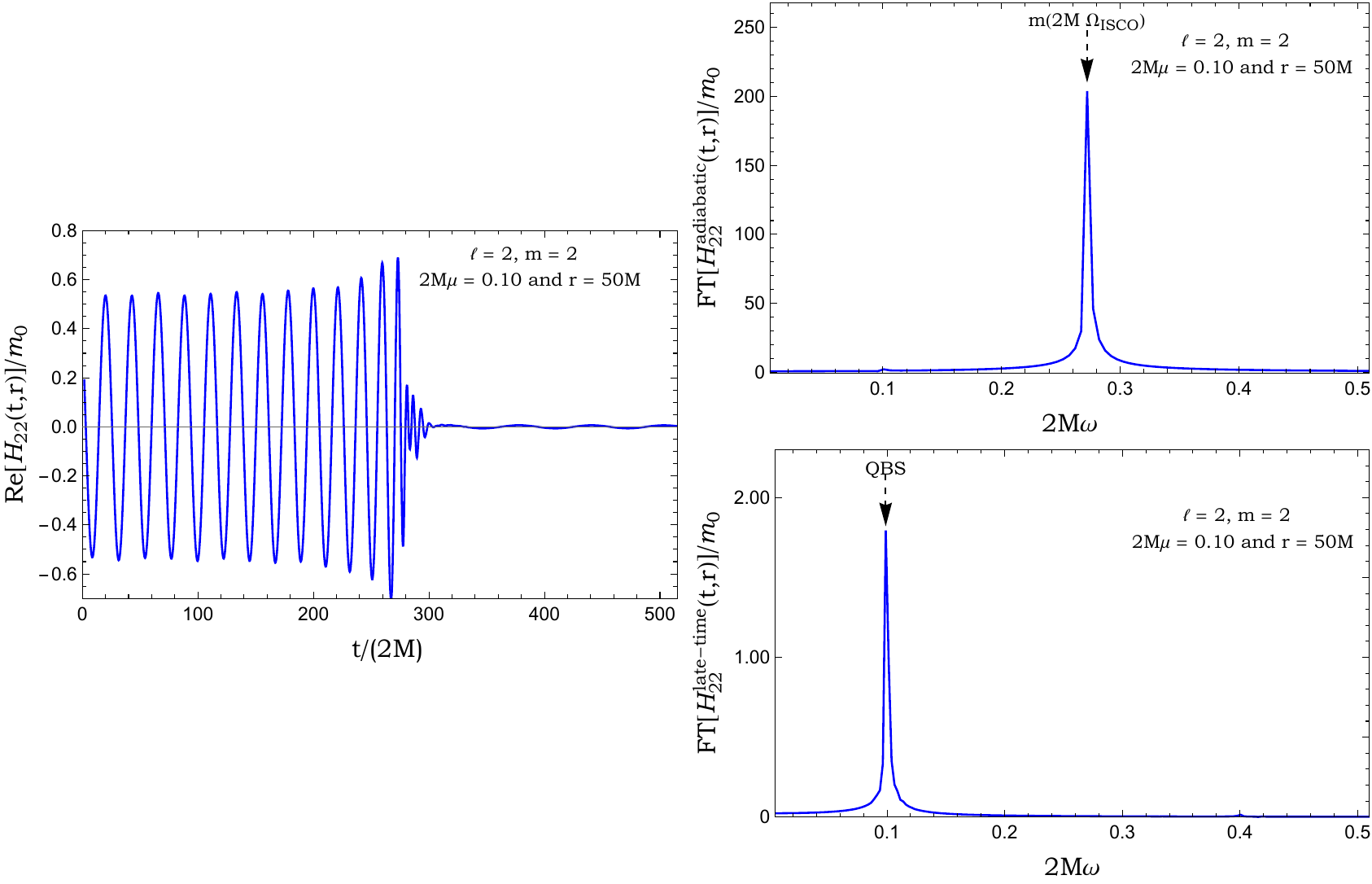}
\caption{\label{Fig:Even_H_Waveform_l_2_m_2}  Quadrupolar waveform of the $H_{r}$ component produced by the plunging particle (right panel) and the spectral content of the adiabatic and late-time phases (upper left and lower left panels, respectively). The result is obtained for a massive spin-$2$ field ($2M\mu=0.10$), and the observer is located at $r = 50M$.}
\end{figure*} 
\begin{figure*}[htbp]
 \includegraphics[scale=0.50]{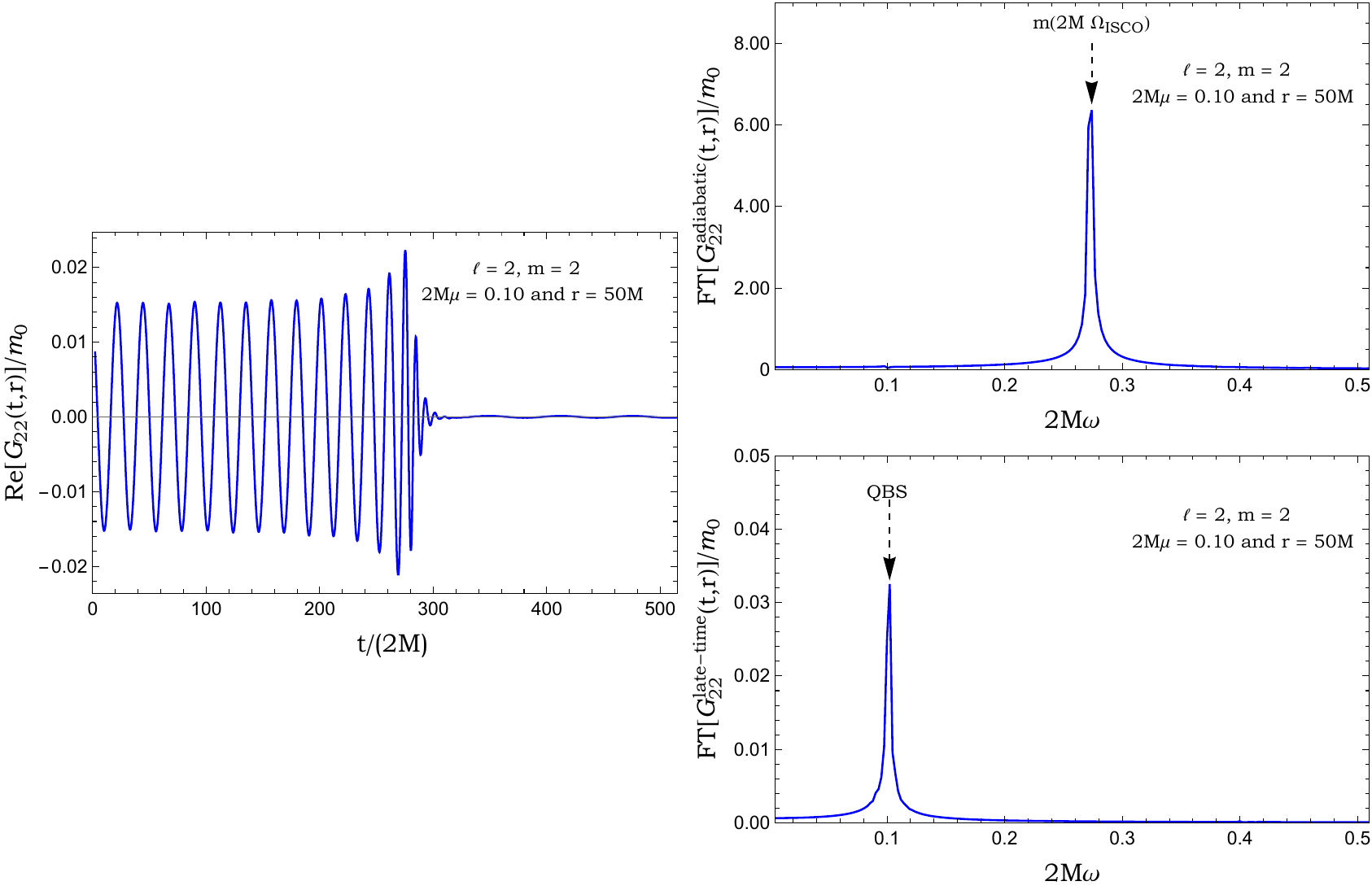}
\caption{\label{Fig:Even_G_Waveform_l_2_m_2}  Quadrupolar waveform of the $G$ component produced by the plunging particle (right panel) and the spectral content of the adiabatic and late-time phases (upper left and lower left panels, respectively). The result is obtained for a massive spin-$2$ field ($2M\mu=0.10$), and the observer is located at $r = 50M$.}
\end{figure*} 


\begin{figure}[htbp]
 \includegraphics[scale=0.50]{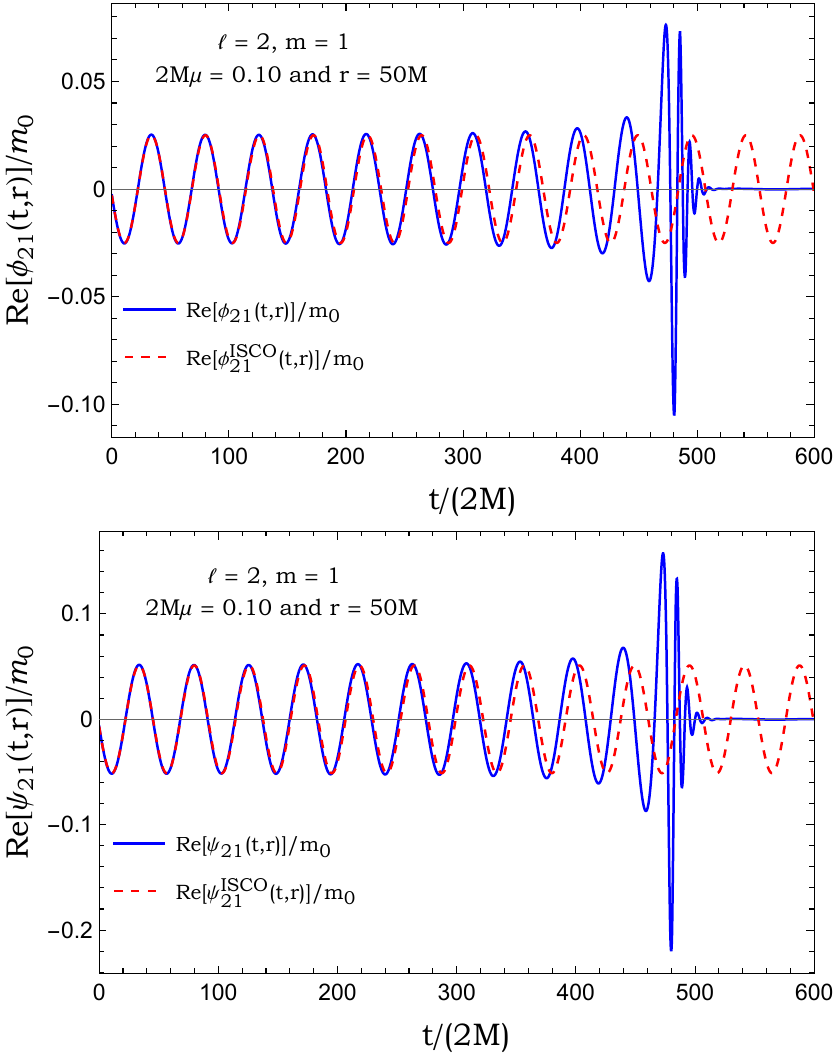}
\caption{\label{Fig:Odd_Circ_Waveform_l_2_m_1} Odd-parity quadrupolar waveforms of the \( \phi_{\ell m} \) (top) and \( \psi_{\ell m} \) (bottom) components, generated by a plunging particle (solid blue line) and a particle orbiting the ISCO (red dashed line). Results are for a massive spin-2 field (\( 2M\mu = 0.10 \)) with the observer at \( r = 50M \).}
\end{figure}
\begin{figure}[htbp]
 \includegraphics[scale=0.50]{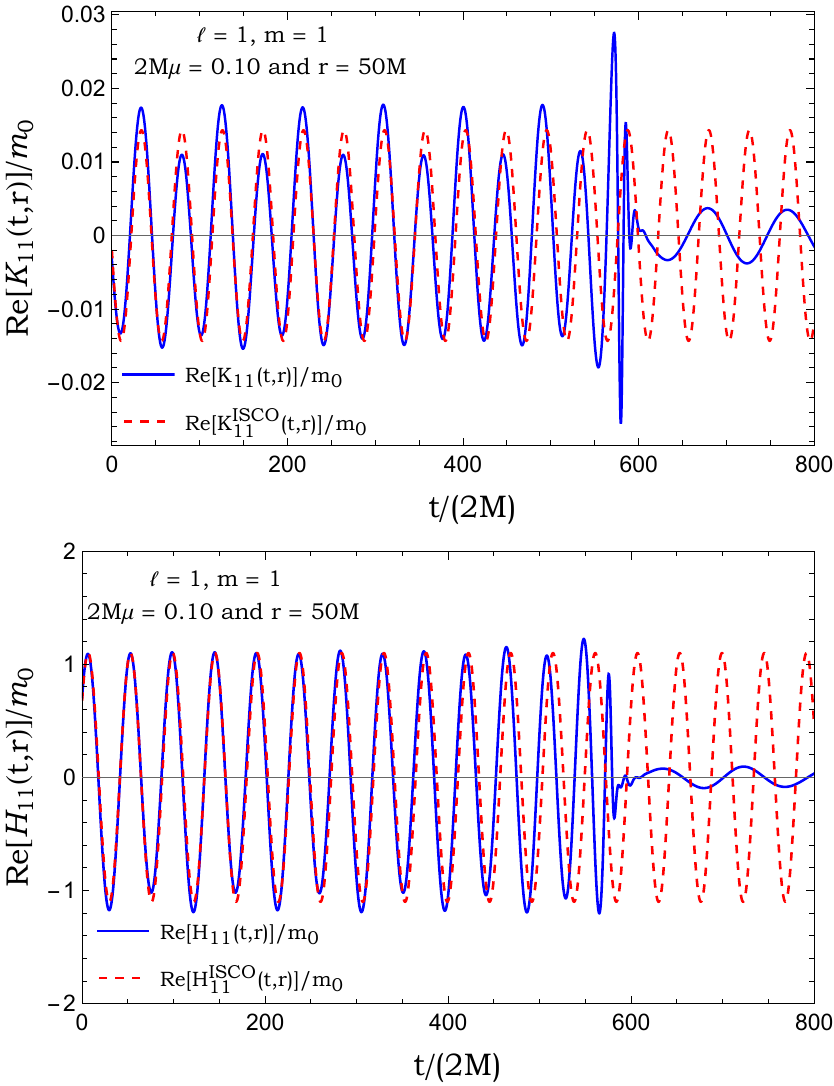}
\caption{\label{Fig:Even_Circ_Waveform_l_1_m_1} Even-parity dipolar waveforms of the \( K \) (top) and \( H_r \) (bottom) components, generated by a plunging particle (solid blue line) and a particle orbiting the ISCO (red dashed line). Results are for a massive spin-2 field (\( 2M\mu = 0.10 \)) with the observer at \( r = 50M \).}
\end{figure}
\begin{figure}[htbp]
 \includegraphics[scale=0.50]{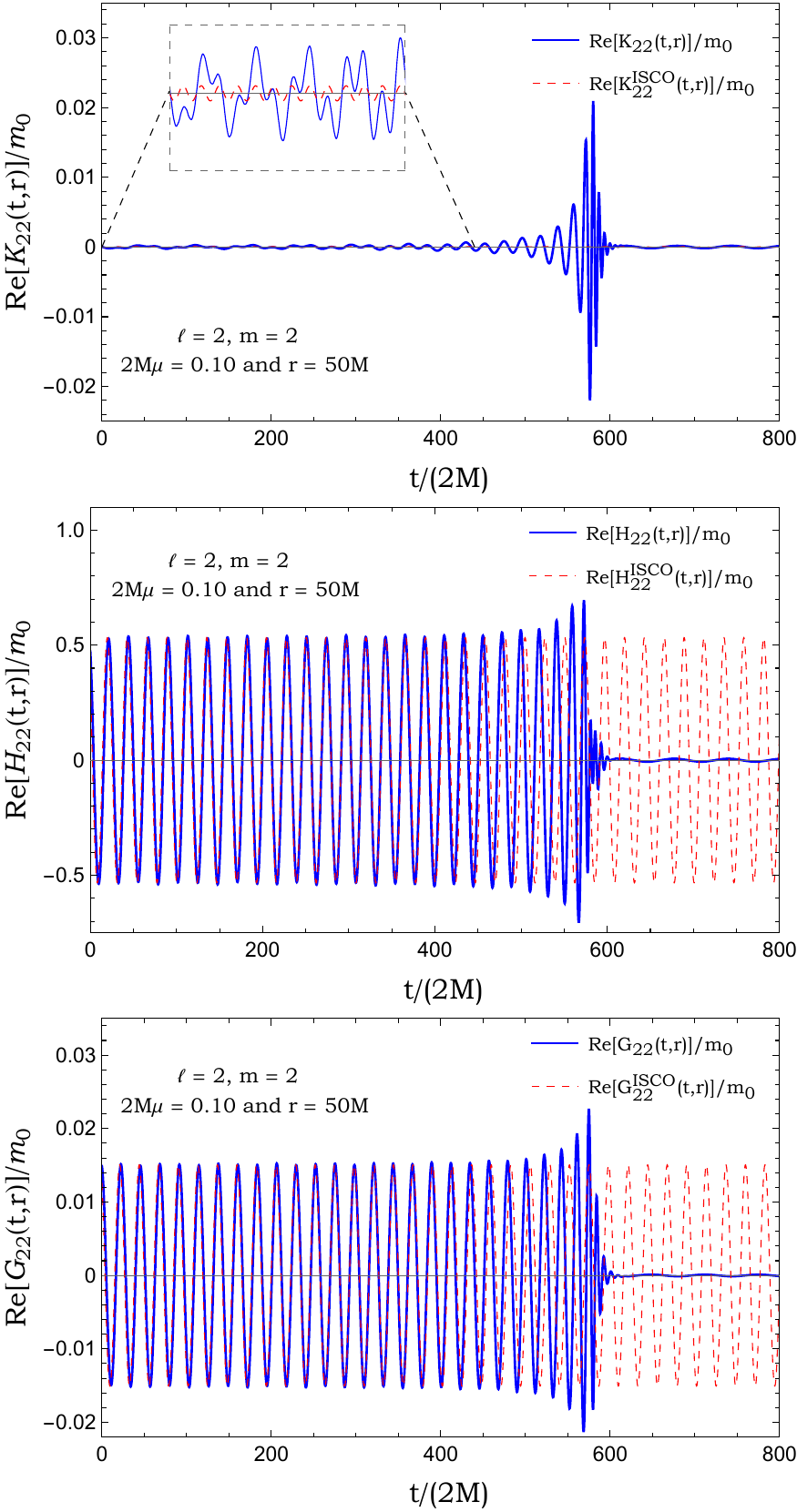}
\caption{\label{Fig:Even_Circ_Waveform_l_2_m_2} Even-parity quadrupolar waveforms of the \( K \) (top), \( H_r \) (middle) and \( G \) (bottom) components produced by a plunging particle (solid blue line) and a particle orbiting ISCO (red dashed line). The results are for a massive spin-2 field (\( 2M\mu = 0.10 \)) with the observer at \( r = 50M \).}
\end{figure}


\begin{figure}[htbp]
 \includegraphics[scale=0.50]{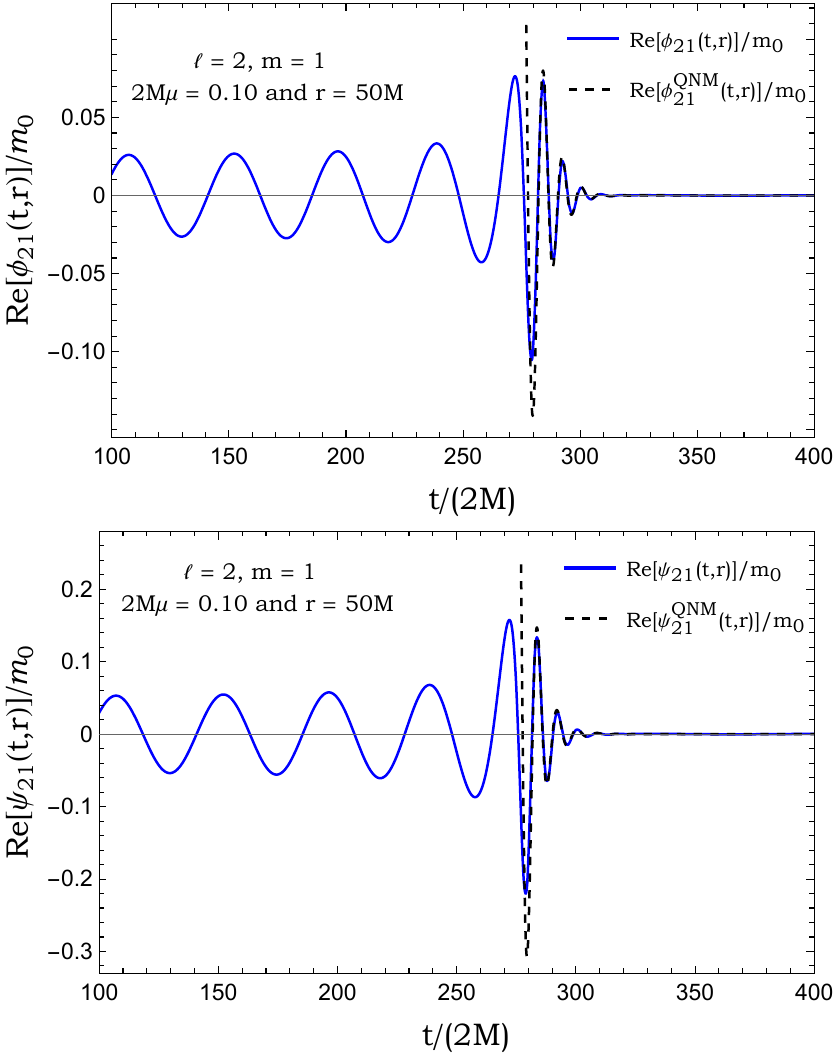}
\caption{\label{Fig:Odd_QNM_Waveform_l_2_m_1} Comparison of the odd-parity quadrupolar waveform generated by the plunging particle (solid blue line) with the odd-parity quadrupolar QNM waveform (black dashed line) for the \( \phi_{\ell m} \) component (top panel) and the \( \psi_{\ell m} \) component (bottom panel). The results are computed for \( 2M\mu = 0.1\), with the observer positioned at \( r = 50M \).}
\end{figure}
\begin{figure}[htbp]
 \includegraphics[scale=0.50]{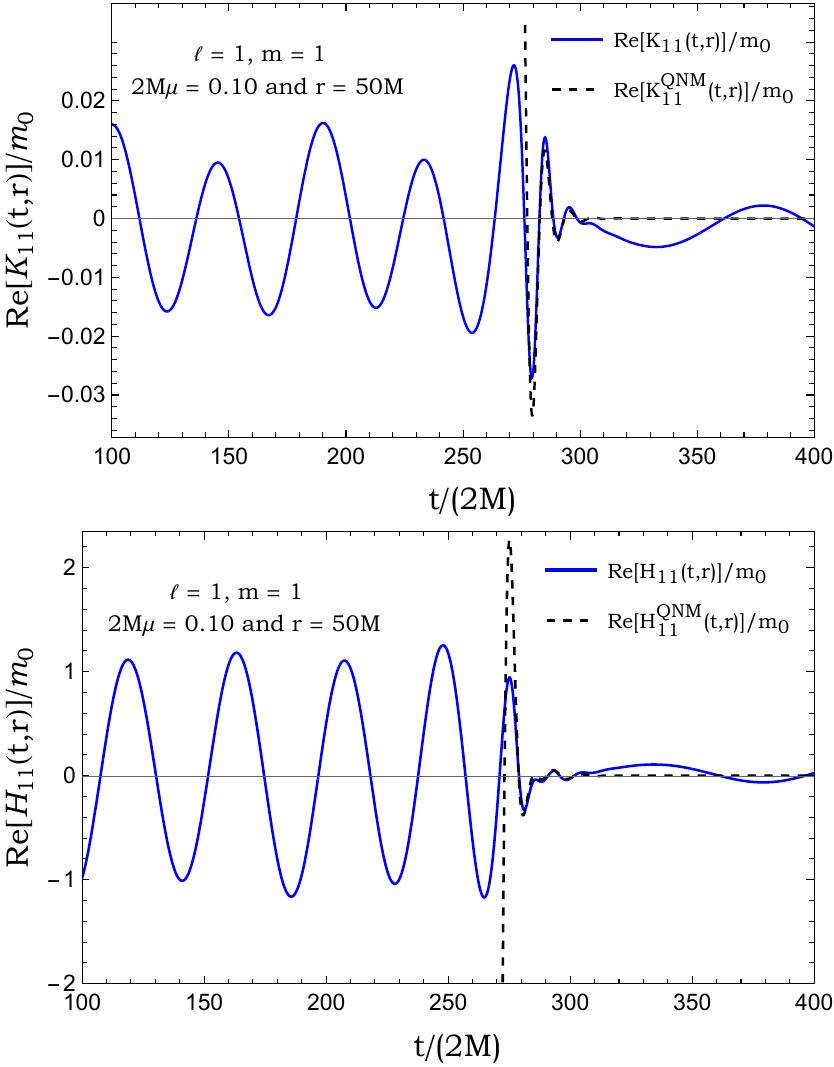}
\caption{\label{Fig:Even_QNM_Waveform_l_1_m_1} Comparison of the regularized even-parity dipolar waveform generated by the plunging particle (solid blue line) with the unregularized even-parity dipolar QNM waveform (black dashed line) for the \(K\) component (top panel) and the \(H_r\) component (bottom panel). The results are computed for \( 2M\mu = 0.1 \), with the observer positioned at \( r = 50M \).}
\end{figure}
\begin{figure}[htbp]
 \includegraphics[scale=0.50]{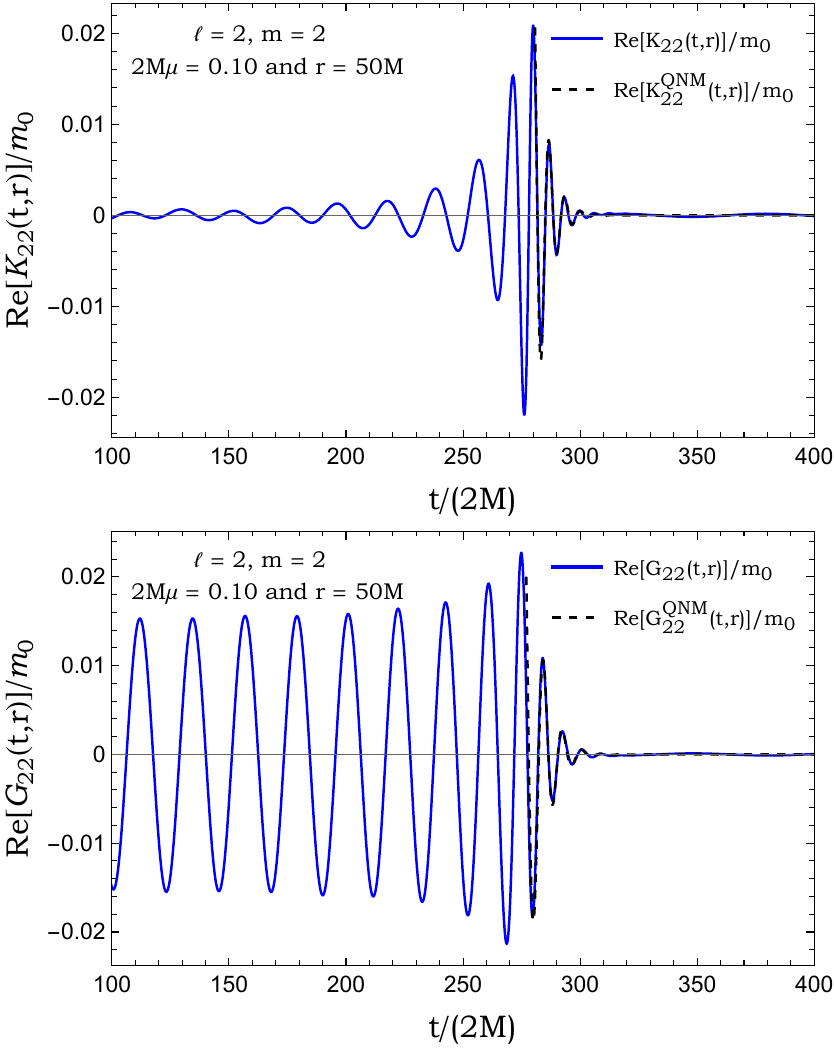}
\caption{\label{Fig:Even_QNM_Waveform_l_2_m_2} Comparison of the regularized even-parity quadrupolar waveform generated by the plunging particle (solid blue line) with the unregularized even-parity quadrupolar QNM waveform (black dashed line) for the \(K\) component (top panel) and the \(G\) component (bottom panel). The results are computed for \( 2M\mu = 0.1 \), with the observer positioned at \( r = 50M \).}
\end{figure}

\begin{figure*}[htbp]
 \includegraphics[scale=0.50]{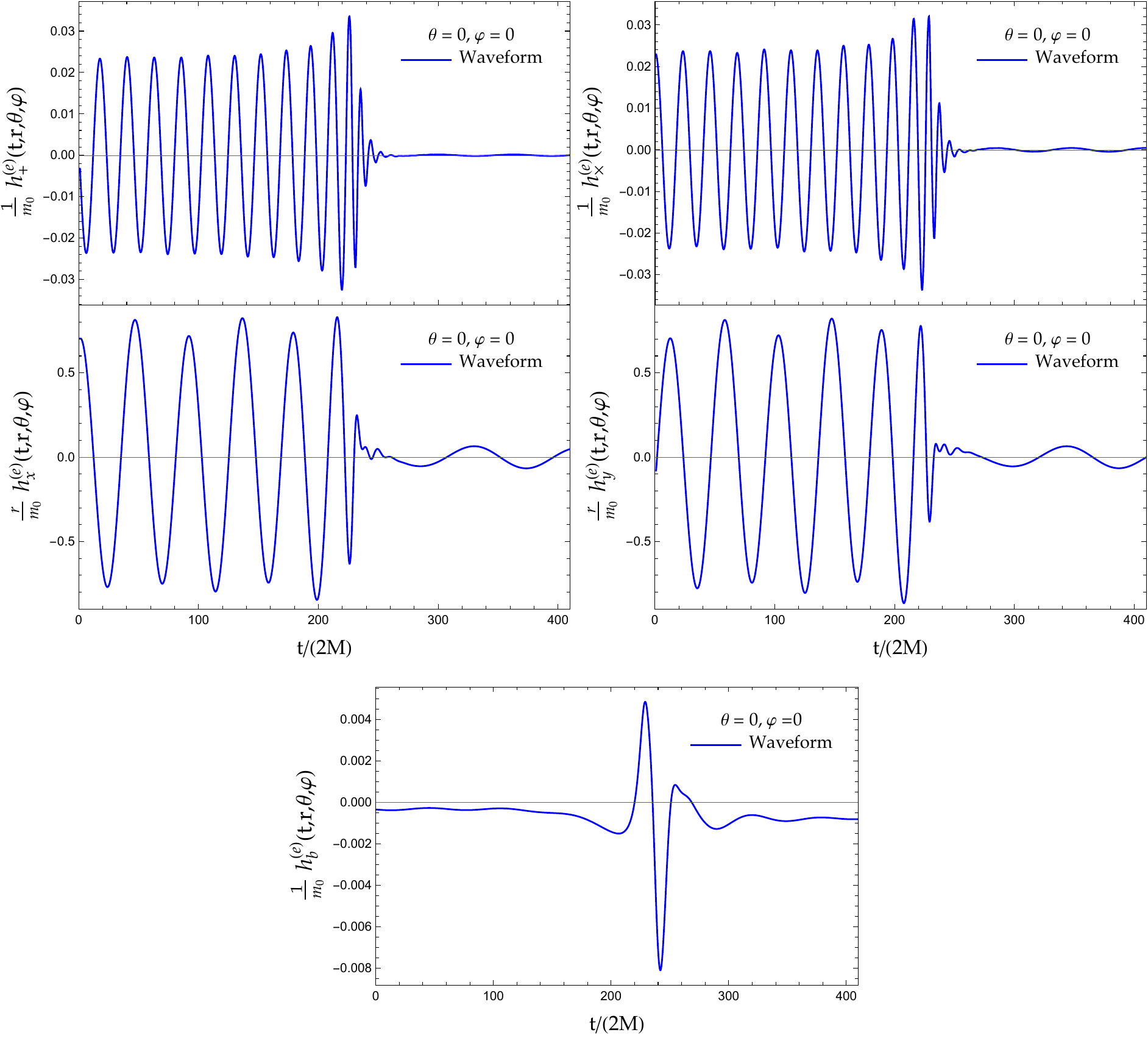}
\caption{\label{Fig:h_P_Even_theta_0-phi_0}  Multipolar gravitational waveforms \(h_\mathfrak{p}^{(\text{e})}\) in the direction (\(\varphi = 0, \theta = 0\)), above the orbital plane of the plunging particle. At \(\theta = 0\), only the \((\ell = 2, m = \pm 2)\) modes contribute to \(h_+^{(\text{e})}\) and \(h_\times^{(\text{e})}\), with \(h_\times^{(\text{e})}\) vanishing at \(\theta = \pm \pi/2\). For \(h_x^{(\text{e})}\), only the \((\ell = 1, m = \pm 1)\) modes contribute at \(\theta = 0\), and the waveform vanishes at \(\theta = \pm \pi/2\). Similarly, for \(h_y^{(\text{e})}\), only the \((\ell = 1, m = \pm 1)\) modes contribute at \(\theta = 0\). Lastly, for \(h_b^{(\text{e})}\), only the \((\ell = 0, m = 0)\) mode contributes to the signal at \(\theta = 0\).}
\end{figure*}
\begin{figure*}[htbp]
 \includegraphics[scale=0.50]{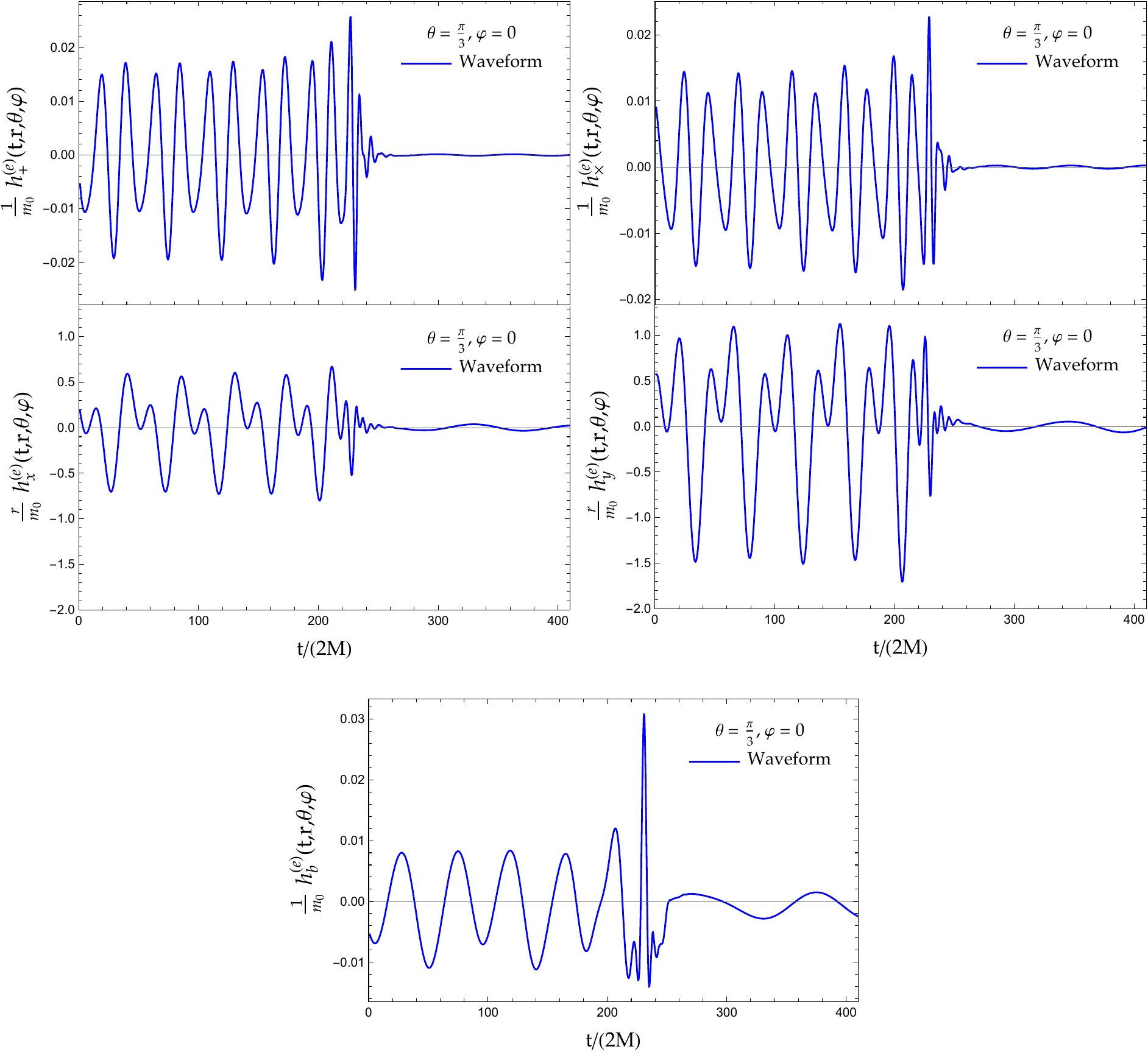}
\caption{\label{Fig:h_P_Even_theta_Pis3-phi_0}  Multipolar gravitational waveforms \(h_\mathfrak{p}^{(\text{e})}\) in the direction \(\varphi = 0\) and \(\theta = \pi/3\), above the orbital plane of the plunging particle.}
\end{figure*}
\begin{figure*}[htbp]
 \includegraphics[scale=0.50]{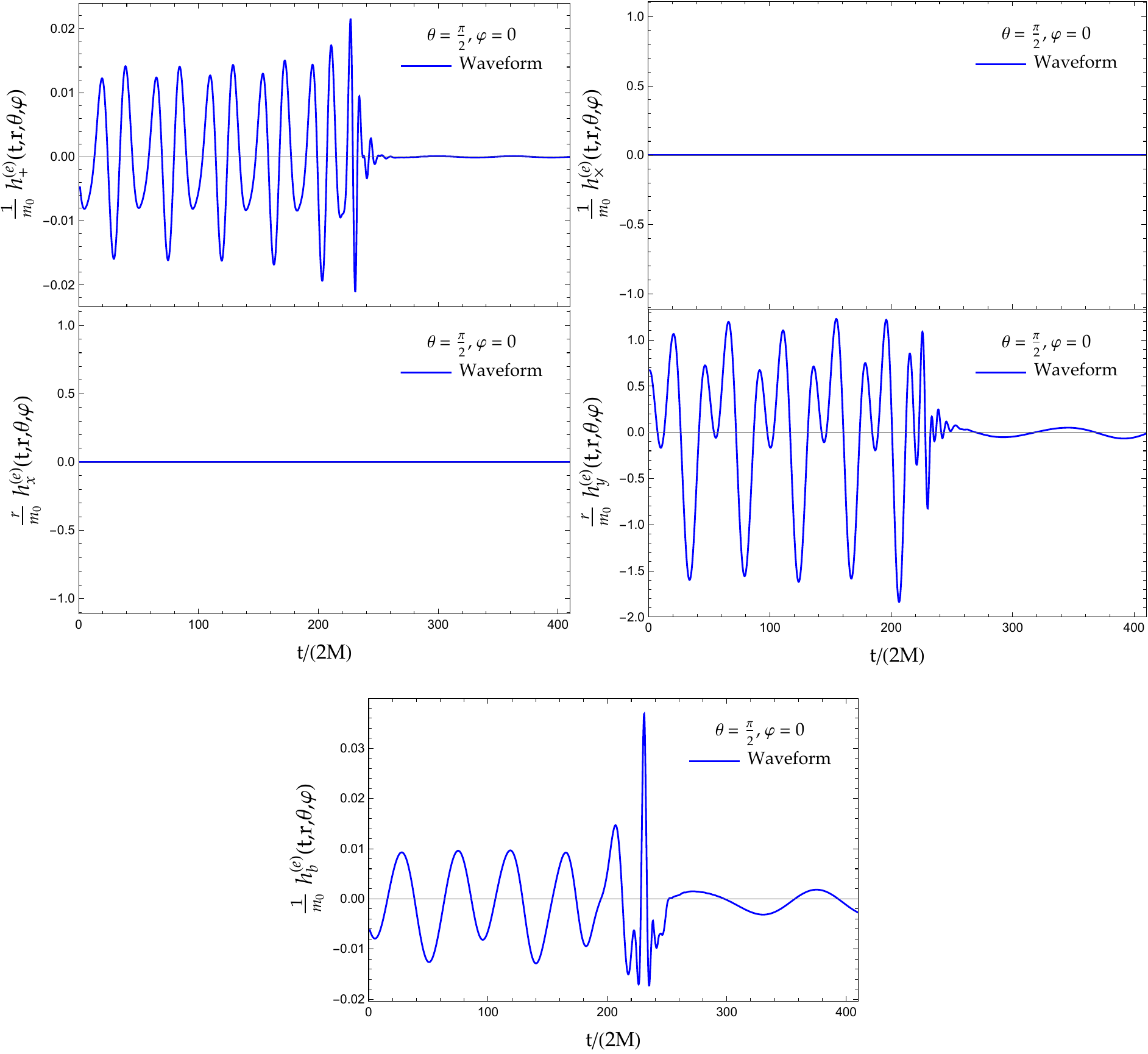}
\caption{\label{Fig:h_P_Even_theta_Pis2-phi_0}   Multipolar gravitational waveforms \(h_\mathfrak{p}^{(\text{e})}\) in the direction $\varphi = 0$ in the orbital plane of the plunging particle ($\theta = \pi/2$).}
\end{figure*}
%
\begin{figure*}[htbp]
 \includegraphics[scale=0.50]{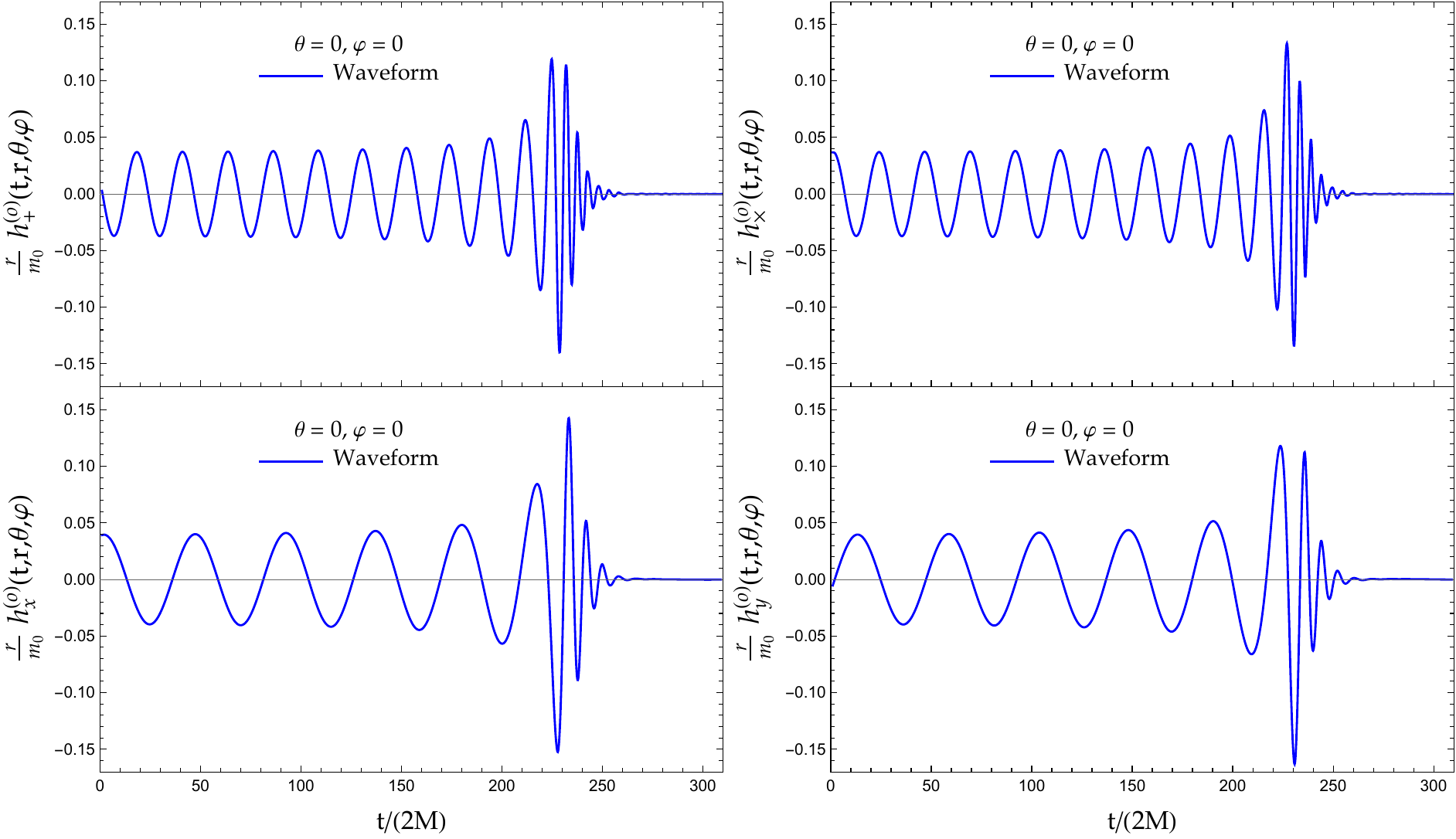}
\caption{\label{Fig:h_P_Odd_theta_0-phi_0} Multipolar gravitational waveforms \(h_\mathfrak{p}^{(\text{o})}\) in the direction (\(\varphi = 0, \theta = 0\)), above the orbital plane of the plunging particle. At \(\theta = 0\), only the \((\ell = 3, m = \pm 2)\) modes contribute to \(h_+^{(\text{o})}\) and \(h_\times^{(\text{o})}\), with \(h_\times^{(\text{o})}\) vanishing at \(\theta = \pm \pi/2\). For \(h_x^{(\text{o})}\), only the \((\ell = 2, m = \pm 1)\) modes contribute at \(\theta = 0\), while the dipole modes (\(\ell = 1\)) do not contribute for any \(\theta\), and the waveform vanishes at \(\theta = \pm \pi/2\). Similarly, for \(h_y^{(\text{o})}\), only the \((\ell = 2, m = \pm 1)\) modes contribute at \(\theta = 0\).}
\end{figure*}
\begin{figure*}[htbp]
 \includegraphics[scale=0.50]{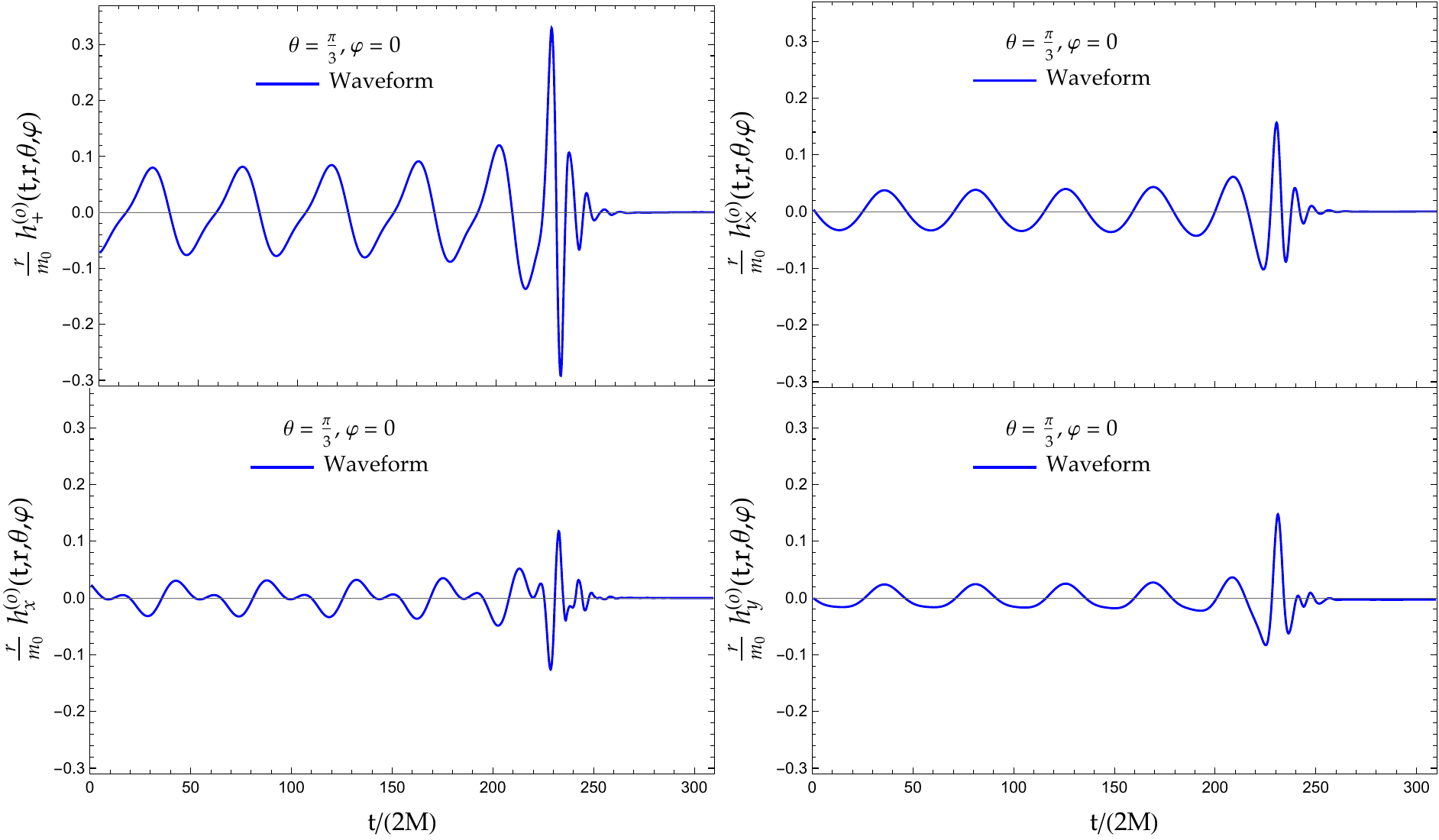}
\caption{\label{Fig:h_P_Odd_theta_Pis3-phi_0} Multipolar gravitational waveforms \(h_\mathfrak{p}^{(\text{o})}\) in the direction \(\varphi = 0\) and \(\theta = \pi/3\), above the orbital plane of the plunging particle.}
\end{figure*}
\begin{figure*}[htbp]
 \includegraphics[scale=0.50]{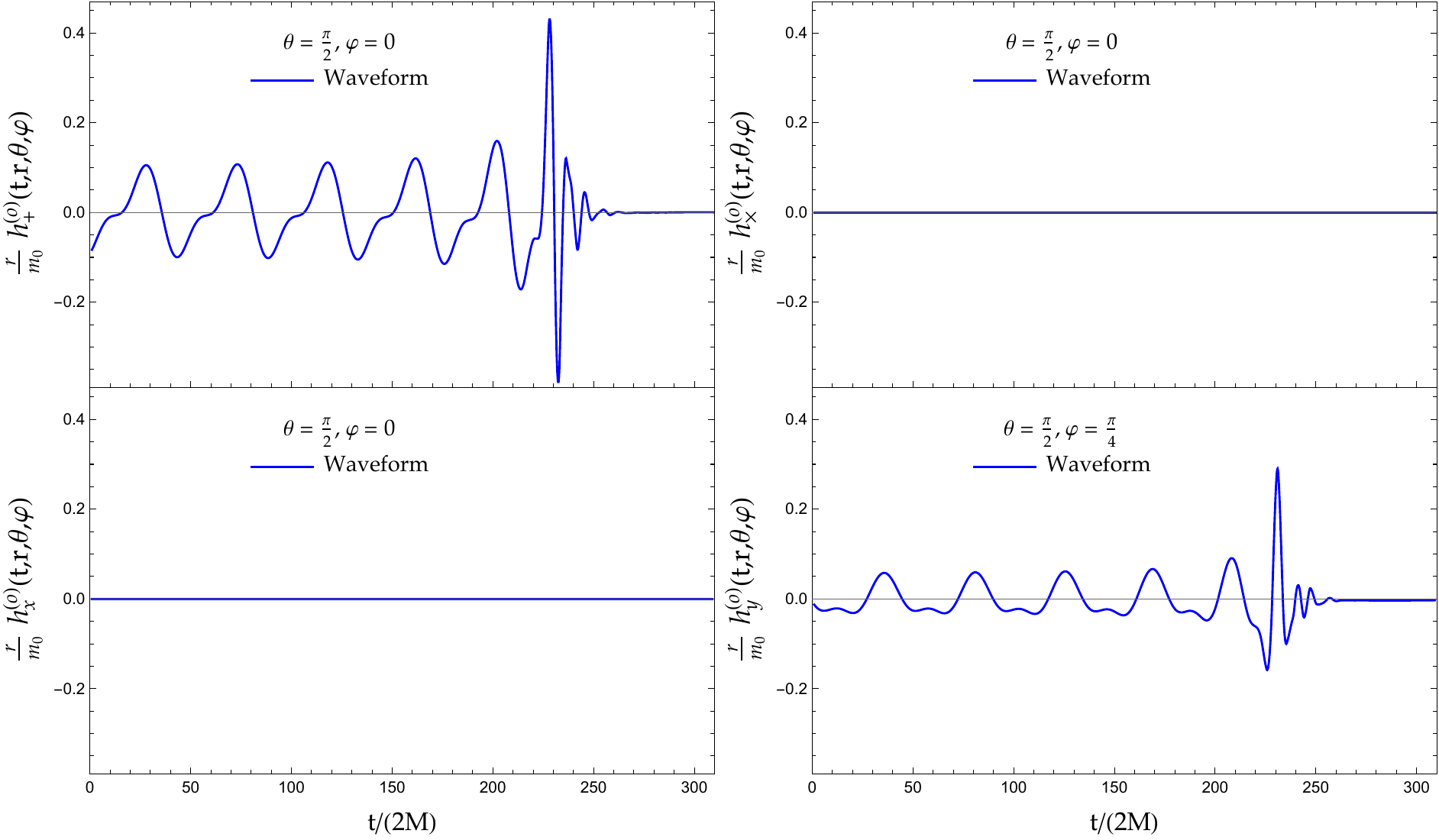}
\caption{\label{Fig:h_P_Odd_theta_Pis2-phi_0} Multipolar gravitational waveforms \(h_\mathfrak{p}^{(\text{o})}\) in the direction $\varphi = 0$ in the orbital plane of the plunging particle ($\theta = \pi/2$).}
\end{figure*}

In Figs.~\ref{Fig:Even_K_Waveform_l_0_m_0}  to \ref{Fig:Even_G_Waveform_l_2_m_2} we show the partial waveforms from the even-parity sector corresponding to the monopole mode $\ell = 0, m = 0$ of the $K$ and $H_{tr}$ components, the dipole mode $\ell = 1, m = 1$ of the $K$ and $H_r$ components, and the quadrupole mode $\ell = 2, m = 2$ of the $K$, $H_r$, and $G$ components for a mass coupling parameter $2M\mu = 0.1$. As in the odd-parity sector, the observer is assumed to be located at $r = 50M$. The waveforms have been constructed assuming that the particle starts at $r = r_{\text{ISCO}} - \epsilon$, with $\epsilon = 10^{-4}$. Furthermore, in Eqs. \eqref{trajectory_plung} and \eqref{trajectory_plung_phi}, we set $\varphi_0 = 0$ and adjusted $t_0/(2M)$ to position the relevant part of the signal in the time window $t/(2M) \in [0, 800]$ for the monopole and dipole waveforms (Figs. \ref{Fig:Even_K_Waveform_l_0_m_0}--\ref{Fig:Even_H_Waveform_l_1_m_1}), and in the window $t/(2M) \in [0, 500]$ for the quadrupole waveforms (Figs. \ref{Fig:Even_K_Waveform_l_2_m_2}--\ref{Fig:Even_G_Waveform_l_2_m_2}).

Figures \ref{Fig:Even_K_Waveform_l_0_m_0} and \ref{Fig:Even_H_Waveform_l_0_m_0} show the partial waveform (left panel) and its spectral content in the adiabatic (top right panel) and late-time phases (bottom right panel) for the monopole mode ($\ell = 0, m = 0$) of the $K$ and $H_{tr}$ components, respectively. During both the adiabatic and late-time phases, a peak is observed at a frequency corresponding to the real part of the complex frequency of the first long-lived QBS mode. In other words, we observe the excitation of the first QBS in both the adiabatic and late-time phases.

The monopole mode deserves special attention, as it is the only mode exhibiting instability associated with massive spin-2 fields on Schwarzschild backgrounds. This instability, characterized by a purely imaginary frequency, grows exponentially over time, with a characteristic timescale given by $\tau = 1 / \text{Im}[\omega]$, depending on the mass coupling $2M\mu$ (see Sec. \ref{sec_4_1_3}). For $2M\mu = 0.1$, the characteristic timescale of the instability is about $\tau / 2M \sim 20$, meaning that after each period of $\tau$, the unstable mode amplitude increases by a factor $e$ (an $e$-folding). Compared to the total observation time of the waveform ($t / 2M \in [-3140, 3140]$), this timescale is very short, theoretically allowing the instability to grow significantly and dominate the signal, leading to exponential amplification. 
However, in our results, Fig.~\ref{Fig:Even_K_Waveform_l_0_m_0} and \ref{Fig:Even_H_Waveform_l_0_m_0}, the waveforms remain remarkably stable, suggesting that the instability does not affect significantly the regime considered here. The absence of any manifestation of the instability can be explained by the weak excitation of the unstable mode by the plunging particle. In other words, unlike other modes (QNMs and QBSs) (see Figs.~\ref{Fig:Even_K_Waveform_l_0_m_0} and \ref{Fig:Even_H_Waveform_l_0_m_0}), the interaction between the plunging particle and the unstable mode is insufficient for the latter to become dominant. Therefore, the initial amplitude of the instability remains negligible, preventing its exponential growth---even after several $e$-foldings---from becoming detectable. Furthermore, we checked the stability by plotting the waveform for an observer located at $r= 500M$, and it also remains stable, showing no evidence of monopolar instability.

In Figs. \ref{Fig:Even_K_Waveform_l_1_m_1}  and \ref{Fig:Even_H_Waveform_l_1_m_1} , we display the partial waveform (left panel) and its spectral content in the adiabatic (top right panel) and late-time phases (bottom right panel) for the dipole mode ($\ell = 1, m = 1$) of the $K$ and $H_r$ components, respectively. During the adiabatic phase, a peak at $\omega = \Omega_{\text{ISCO}}$ is observed for both components, corresponding to the quasicircular motion of the plunging particle near the ISCO, as well as another peak at a frequency corresponding to the real part of the complex frequency of the fundamental QBS mode (i.e. the excitation of the QBS mode in the adiabatic phase). In the waveform (left panel), an interference phenomenon can be seen between the quasibound mode and the quasicircular motion of the particle at the ISCO, leading to the appearance of beats in the adiabatic phase of the waveform.  During the post-QNM and late-time phases, a peak appears at a frequency corresponding to the real part of the complex frequency of the fundamental QBS mode. This indicates the excitation of this mode with an amplified amplitude that decays slowly due to the small imaginary part of the quasibound frequency ($Im[\omega_{QBS}]\sim10^{-5}$). This \emph{resonance phenomenon} can be explained by the fact that the angular velocity of the particle,  $\Omega_{\text{ISCO}} \sim 0.1360$, is ``twice'' the real part of the QBS frequency, $\omega_{QBS} \sim 0.0677$, leading to a \textit{harmonic resonance}  observed in the even-parity dipole mode waveform.

Figures \ref{Fig:Even_K_Waveform_l_2_m_2}--\ref{Fig:Even_G_Waveform_l_2_m_2} show the partial waveform (left panel) and its spectral content in the adiabatic (top right panel) and late-time phases (bottom right panel) for the quadrupole modes ($\ell = 2, m = 2$) of the $K$, $H_r$, and $G$ components. During the adiabatic phase, in addition to the peak at $\omega = 2 \Omega_{\text{ISCO}}$ observed for all three components, another peak appears corresponding to the real part of the complex frequency of the fundamental QBS mode, indicating that the first QBS is excited during the adiabatic phase. Notably, this peak has a particularly high amplitude for the $K$ component, contributing significantly to the waveform in the adiabatic phase, as evidenced by the beats observed in Fig.~\ref{Fig:Even_K_Waveform_l_2_m_2}. In the late phase, a peak appears at a frequency corresponding to the real part of the complex frequency of the long-lived QBS mode, suggesting excitation of the QBS mode.

In Figs. \ref{Fig:Odd_Waveform_l_1_m_0} to \ref{Fig:Even_G_Waveform_l_2_m_2}, which illustrate the partial wavforms, it can be seen that the even-parity dipole waveform of the $H_r$ component has the highest amplitude compared to the other components across all modes (monopole, dipole, quadrupole) and parities. This even-parity dipole waveform reaches an amplitude ratio approximately $2$ times higher than that of the even-parity quadrupole waveform of the $H_r$ component, and up to $55$ times higher than the even-parity quadrupole  waveform of the  $K$ component.

\subsection{Adiabatic phase and circular motion of the particle on the ISCO}
\label{sec_6_2}

In Fig.~\ref{Fig:Odd_Circ_Waveform_l_2_m_1} we compare the odd-parity quadrupolar waveform produced by the plunging particle, obtained in Sec. \ref{sec_6_1_1}, with the quadrupolar waveform produced by a particle orbiting the BH at ISCO, obtained from Eq.~\eqref{Sol_Odd_Coupled_inf_time_ISCO}. During the adiabatic phase, the waveform emitted by the plunging particle is very accurately described by the waveform emitted by the particle at the ISCO. This can be easily understood by noting that the initial position of the plunging particle is very close to the ISCO, causing it to undergo an adiabatic inspiral along a sequence of quasicircular orbits near the ISCO. Such behavior aligns with the analysis presented in Sec. \ref{sec_6_1_1}, where the spectral content of the adiabatic phase waveform was discussed.

In Figs.~\ref{Fig:Even_Circ_Waveform_l_1_m_1} and \ref{Fig:Even_Circ_Waveform_l_2_m_2} we compare the  regularized even-parity dipolar and quadrupolar waveforms produced by the plunging particle, which we obtained in Sec. \ref{sec_6_1_2}, with the corresponding waveforms generated by a particle orbiting the black hole at ISCO, obtained from Eq.~\eqref{Sol_Even_Coupled_inf_time_ISCO}.  As in the  odd-parity case, the waveforms emitted by the plunging particle during the adiabatic phase are accurately described by those emitted by the particle on the ISCO. This similarity can be easily explained by the same  reasons discussed earlier. However, it is important to note that for the \(K\) component, in both the dipole mode \((\ell = 1)\) and the quadrupole mode \((\ell = 2)\), the waveforms produced by the particle on the  ISCO do not perfectly match those produced by the plunging particle. This discrepancy arises because in the adiabatic phase the excitation of QBSs also contributes, as shown in the spectral analysis of the adiabatic phase (see Figs.~\ref{Fig:Even_K_Waveform_l_1_m_1}~and~\ref{Fig:Even_K_Waveform_l_2_m_2}).

\subsection{Ringdown phase and the excitation of QNMs}
\label{sec_6_3}

In Fig.~\ref{Fig:Odd_QNM_Waveform_l_2_m_1}, we compare the odd-parity quadrupolar waveform produced by the plunging particle we have obtained in Sec. \ref{sec_6_1_1}, with the quadrupolar quasinormal waveform \(\sum_{s}\bm{\Phi}^{\text{QNM}}_{s 2 1 0}(t,r)\), given by Eq. ~\eqref{QNM_odd_n_s}. This quasinormal waveform corresponds to the sum of the two fundamental (\(\ell = 2, n = 0\)) QNMs associated with the ``vector'' type (\(s = 1\)) and ``tensor'' type (\(s = 2\)) , and we can see that the quasinormal waveform provides an excellent description of the ringdown phase. 

In Figs. ~\ref{Fig:Even_QNM_Waveform_l_1_m_1} and \ref{Fig:Even_QNM_Waveform_l_2_m_2}, we compare the even-parity dipolar and quadrupolar waveforms produced by the plunging particle obtained in Sec. \ref{sec_6_1_2}, with the even-parity dipolar and quadrupolar quasinormal waveforms \(\sum_{s}\bm{\Psi}^{\text{QNM}}_{s 1 1 0}(t,r)\) and \(\sum_{s}\bm{\Psi}^{\text{QNM}}_{s 2 2 0}(t,r)\) given by Eq. ~\eqref{QNM_even_n_s}. In the dipolar case (\(\ell = 1\)), the quasinormal waveform corresponds to the sum of the two fundamental (\(n = 0\)) QNMs associated with the scalar type  (\(s = 0\)) and vector type  (\(s = 1\)). For the quadrupolar case (\(\ell = 2\)), it corresponds the sum of the three fundamental (\(n = 0\)) QNMs associated with the scalar type  (\(s = 0\)), vector type  (\(s = 1\)), and tensor type  (\(s = 2\)).  The quasinormal waveforms describe the ringdown phase very accurately. It is also important to note, however, that the waveform produced by the plunging particle required regularization , whereas the quasinormal waveform remains unregularized (see Sec. ~\ref{sec_5_2}).

\subsection{Multipolar gravitational waveforms: Even and odd Sectors}
\label{sec_6_4}

In Figs.~\ref{Fig:h_P_Even_theta_0-phi_0}--\ref{Fig:h_P_Odd_theta_Pis2-phi_0} we considered the components \(h_\mathfrak{p}^{(\text{e/o})}\) of the gravitational waves. Without loss of generality, we constructed only the signals corresponding to directions above the orbital plane of the plunging particle. We also assumed that the observer is located in the plane \(\varphi = 0\). For other values of \(\varphi\), the behavior of the signals remains qualitatively similar. Results for arbitrary values of \(\theta\) and \(\varphi\) can be made available to interested readers on request.

In Figs.~\ref{Fig:h_P_Even_theta_0-phi_0}--\ref{Fig:h_P_Even_theta_Pis2-phi_0} we focused on constructing the multipolar waveforms of even parity for the different polarizations. This was achieved by summing the expressions in \eqref{Polarizations_Even} over harmonics beyond the dominant modes, namely \((\ell=2, m=\pm 2)\) for the tensor modes, \((\ell=1, m=\pm 1)\) for the vector modes, and \((\ell=0, m=0)\) for the scalar mode.

Similarly, in Figs.~\ref{Fig:h_P_Odd_theta_0-phi_0}--\ref{Fig:h_P_Odd_theta_Pis2-phi_0}, we focused on constructing the odd-parity multipolar waveforms for the different polarizations. This was achieved by summing the expressions in \eqref{Polarizations_Odd} over harmonics beyond the dominant contributions, specifically \((\ell=2, m=\pm 1)\) for the tensor modes and \((\ell=1, m=0)\) for the vector modes.    

Of course, the truncations chosen for each polarization mode, whether of even or odd parity, provide reliable and robust results.

The distortion of the multipolar waveforms is clearly visible in Figs.~\ref{Fig:h_P_Even_theta_Pis3-phi_0} and \ref{Fig:h_P_Even_theta_Pis2-phi_0} for even parity, and in Figs. ~\ref{Fig:h_P_Odd_theta_Pis3-phi_0} and \ref{Fig:h_P_Odd_theta_Pis2-phi_0} for odd parity. This distortion manifests both during the adiabatic phase, corresponding to the quasicircular motion of the particle near the ISCO (see Fig.~\ref{Fig:Trajectoire}), and during the ringdown phase. It results from the summation over the \((\ell, m)\) modes in the expressions \eqref{Polarizations_Even} and \eqref{Polarizations_Odd} and depends significantly on the direction of the observer.

Finally, we observe that the vectorial polarizations of even parity, \(h_{x/y}^{(\text{e})}\), exhibit the highest amplitudes compared to the other polarizations for both parities. This behavior is mainly explained by the contribution of the dipole mode (\(\ell = 1\)) of the \(H_r\) component, which dominates the modes of the other partial waveforms (see Sec. \ref{sec_6_1}).

\section{Conclusion}
\label{sec_7}

In this paper, we have numerically constructed the spectra of quasinormal and quasibound frequencies for the massive spin-$2$ field on a Schwarzschild BH background using the matrix-valued Hill determinant method and studied their evolution as a function of the coupling mass $2M\mu$. We have presented, for the first time, the quasinormal frequency spectrum for even parity and highlighted, for the monopole mode of this parity, the existence of two new branches: one for the quasinormal frequencies and the other for the quasibound frequencies. In addition, we have confirmed the results obtained in Ref.~\cite{Brito:2013wya} on the instability of the monopole mode using our matrix-valued Hill determinant method.

We have described gravitational radiation emitted by a massive ``point particle'' plunging from slightly below the ISCO into a Schwarzschild BH. To do this, we constructed the associated partial waveforms and analyzed the spectral content of the different phases. As we have shown, the waveforms can be decomposed into three phases: (i) an adiabatic phase corresponding to the quasicircular motion of the particle near the ISCO, (ii) a ringdown phase due to the excitation of QNMs, and (iii) a late-time phase.  In the adiabatic phase, for an observer at a given distance, we highlighted not only the frequency associated with the quasicircular motion of the particle at the ISCO, but also the excitation of QBS modes. In addition, we showed that during this phase, the waveform emitted by the plunging particle is very well described by the waveform emitted by the particle living on the ISCO. For the ringdown phase, we also showed that it is well characterized by the excitation of the first QNMs, specifically the least damped modes. The analysis of the late-time phase also reveals the excitation of QBS modes, whose amplitude, in the case of the even-parity dipole mode $(\ell = 1)$, is amplified by a harmonic resonance phenomenon. This resonance arises because, for this dipole mode, the angular velocity of the particle at the ISCO is twice the real part of the QBS mode frequency, leading to (i) an amplification of the excited QBS mode amplitude, (ii) beats in the adiabatic phase, and (iii) a post-QNM phase characterized by an amplified and a slowly decaying signal. We have also plotted the multipolar gravitational waveforms for arbitrary directions of observation and, in particular, outside the orbital plane of the plunging particle.

Finally, it is important to consider the instabilities associated with massive gravity theories~\cite{Brito:2013wya,Babichev:2013una,Gregory:1993vy}. As mentioned above, Schwarzschild black holes exhibit monopolar instabilities associated with spherically symmetric modes ($\ell = 0$), which can affect Schwarzschild black holes in two different contexts. First, they can affect the background geometry itself, potentially leading to evolution towards different black hole solutions, such as ``hairy'' black holes, on very long timescales~\cite{Gervalle:2020mfr,Brito:2013xaa,East:2023nsk}. However, when the graviton mass is extremely small (\(\mu \sim H_0\), corresponding to the Hubble scale), the characteristic timescale of the instability (\(\tau \sim 1 / H_0\)) becomes comparable to the cosmological timescale. Such a timescale, much longer than the duration of typical astrophysical phenomena, renders this instability harmless in these scenarios.

Even for an ultralight graviton, interactions with supermassive black holes can easily result in mass coupling parameter values $2M\mu$ that fall within the regime studied here. Indeed, for supermassive black holes with masses in the range \(M \sim 10^6 - 2 \times 10^{10} M_\odot\) and a graviton mass of \(\mu \sim 1.35 \times 10^{-55} \, \mathrm{kg}\) (a plausible upper bound from massive gravity theories~\cite{Finn:2001qi}), the coupling parameter \(2M\mu\) can reach values between \(10^{-3}\) and \(22\). These values fall within the regime studied in this work (\(2M\mu = 0.1\)), making the analysis relevant for astrophysical black holes observed in extreme mass ratio inspirals (EMRIs). Such scenarios underline the astrophysical significance of the phenomena analyzed here.

In the second case, monopolar instabilities can also affect spherically symmetric gravitational modes ($\ell = 0$), manifesting in the observed gravitational signal. For mass coupling values $2M\mu = 0.10$, the characteristic timescale is shorter ($\tau / 2M \sim 20$), which could theoretically allow the instability to grow fast enough to influence the observed signal within the time window studied ($t / 2M \sim [-3140, 3140]$). Such exponential growth could amplify the monopolar mode and lead to a divergent signal. However, our results show that this instability does not manifest, due to weak excitation of the unstable mode by the plunging particle or a negligible initial amplitude of the mode. While the monopolar instability might have been excited during the formation process of the black hole, its impact on the specific gravitational waveforms analyzed here remains negligible. Additionally, our analyses show that the waveforms for the monopolar mode remain stable even for observers at larger distances ($r=100M, 500M$).

These observations support the idea that in the small graviton mass regimes considered here, the monopolar instability does not have a significant impact on either the background geometry or the gravitational signals produced by astrophysical phenomena such as a plunging particle. Consequently, the results presented in this work remain robust and valid within the framework studied and can be used as a signature of massive gravity. Furthermore, this work suggests that future gravitational wave observations of EMRIs involving supermassive black holes could provide additional constraints on the graviton mass and further insights into massive gravity theories.

\begin{acknowledgments}
We would like to thank Vitor Cardoso and Richard Brito for kindly providing us with their table of quasibound frequencies for the even-parity dipole mode ($\ell = 1$). This allowed us to compare their data with our results, confirming the consistency of our numerical method and highlighting the agreement between the two approaches. We would also like to thank the referee for their insightful comments and suggestions, which clarified several key points and enriched this paper.
\end{acknowledgments}

\appendix
\section{Perturbation equations for massive spin-2}
\label{appendix_A}

\subsection{Structure of gravitational perturbations}
\label{appendix_A_1}

The field  $h_{\mu\nu}(t,r,\theta,\varphi)$  describing gravitational waves propagating in Schwarzschild spacetime can be expressed in Fourier space as follows
\begin{fleqn}
\begin{multline}\label{décomp_h}
  h_{\mu\nu}(t,r,\theta,\varphi) = \frac{1}{\sqrt{2 \pi}} \int_{-\infty}^{+\infty} d\omega\,e^{-i \omega t}\\
  \times \left[h^{\text{(e)}}_{\mu\nu}(\omega,r,\theta,\varphi)+h^{\text{(o)}}_{\mu\nu}(\omega,r,\theta,\varphi)\right] 
\end{multline}
\end{fleqn}
with
\begin{widetext}
\begin{subequations}
\label{h_even_odd}
\begin{equation}
        \label{h_even}
        h^{\text{(e)}}_{\mu\nu}(\omega,r,\theta, \varphi)   = \sum_{\ell=0}^{+\infty}\sum_{m=-\ell}^{m=+\ell} \begin{pmatrix}
             H^{\ell m}_{t t} Y^{\ell m}     &   H^{\ell m}_{t r} Y^{\ell m}    &     H^{\ell m}_{t} Y^{\ell m}_{\theta}  &     
             H^{\ell m}_{t} Y^{\ell m}_{\varphi}      \\
             sym &   H^{\ell m}_{r r} Y^{\ell m}   &     H^{\ell m}_{r} Y^{\ell m}_{\theta}   &    H^{\ell m}_{r} Y^{\ell m}_{\varphi} \\
             sym & sym &   r^2(K^{\ell m} Y^{\ell m} + G^{\ell m} Y^{\ell m}_{\theta \theta})  &    r^2 G^{\ell m} Y^{\ell m}_{\theta \varphi}   \\
             sym & sym &    sym        &     r^2(K^{\ell m} \sin^2\theta Y^{\ell m} + G^{\ell m} Y^{\ell m}_{\varphi \varphi })
                        \end{pmatrix} 
\end{equation}
and
\begin{equation}
        \label{h_odd}
        h^{\text{(o)}}_{\mu\nu}(\omega,r,\theta, \varphi)  = \sum_{\ell=1}^{+\infty}\sum_{m=-\ell}^{m=+\ell} \begin{pmatrix}
                           0     &   0   &     h^{\ell m}_{t} X^{\ell m}_{\theta}             &     h^{\ell m}_{t} X^{\ell m}_{\varphi}      \\
                           sym &   0   &     h^{\ell m}_{r} X^{\ell m}_{\theta}             &    h^{\ell m}_{r} X^{\ell m}_{\varphi}        \\
                           sym & sym &     h^{\ell m} X^{\ell m}_{\theta\theta}        &    h^{\ell m} X^{\ell m}_{\theta\varphi}   \\
                           sym & sym &                        sym                                     &    h^{\ell m} X^{\ell m}_{\varphi\varphi}
                        \end{pmatrix} 
\end{equation}
\end{subequations}
\end{widetext}
with the following conventions: for  $\ell = 0$, we have $H^{\ell m}_{t} = H^{\ell m}_{r} = 0$; for $\ell = 0,1$, we have $G^{\ell m} = 0$; and for $\ell = 1$, we have $h^{\ell m} = 0$. The indices $(e)$  and $(o)$  denote even (polar) and odd (axial) parity, respectively. When considering~\eqref{h_even_odd}, it is important to keep in mind that the frequency and radial dependencies are contained in the functions $ H^{\ell m}_{t t},  H^{\ell m}_{t r},  H^{\ell m}_{r r},  H^{\ell m}_{t},  H^{\ell m}_{r},  K^{\ell m},  G^{\ell m}_{t t},  h^{\ell m}_{t},  h^{\ell m}_{r}$, and $ h^{\ell m}$, while the angular dependencies are contained in the spherical harmonics $ Y^{\ell m}, Y^{\ell m}_{\theta}, Y^{\ell m}_{\varphi}, Y^{\ell m}_{\theta\theta}, Y^{\ell m}_{\varphi \varphi}, X^{\ell m}_{\theta}, X^{\ell m}_{\varphi}, X^{\ell m}_{\theta \theta} $, and $X^{\ell m}_{\varphi \varphi}$. The scalar, vector, and tensor spherical harmonics that are used here are explicitly described in the appendix of  \cite{Martel:2005ir}.

\subsection{Structure of the stress-energy tensor: Source of gravitational perturbations}
\label{appendix_A_2}

The most general form of the stress-energy tensor generating gravitational perturbations of a Schwarzschild black hole is given in Fourier space by
\begin{fleqn}
\begin{multline} \label{décomp_tau}
  \mathcal{T}_{\mu\nu}(t,r,\theta,\varphi) = \frac{1}{\sqrt{2 \pi}} \int_{-\infty}^{+\infty} d\omega\,e^{-i \omega t} \\
   \times \left[\mathcal{T}^{\text{(e)}}_{\mu\nu}(\omega,r,\theta,\varphi) + \mathcal{T}^{\text{(o)}}_{\mu\nu}(\omega,r,\theta,\varphi)\right]
\end{multline}
\end{fleqn}
with
\begin{widetext}
\begin{subequations}
\begin{equation}
\label{tau_even}
\mathcal{T}^{\text{(e)}}_{\mu\nu} (\omega,r,\theta,\varphi)    = \sum_{\ell=0}^{+\infty}\sum_{m=-\ell}^{m=+\ell} \begin{pmatrix}
     T^{\ell m}_{t t}Y^{\ell m}    &   T^{\ell m}_{t r}Y^{\ell m}    &     T^{\ell m}_{t} Y^{\ell m}_{\theta}  &    T^{\ell m}_{t} Y^{\ell m}_{\varphi}      \\
     sym &   T^{\ell m}_{r r}Y^{\ell m}    &     T^{\ell m}_{r} Y^{\ell m}_{\theta}   &   T^{\ell m}_{r} Y^{\ell m}_{\varphi}        \\
     sym & sym &    T^{\ell m}_{1} Y^{\ell m} + T^{\ell m}_{2}Y^{\ell m}_{\theta \theta}    &  T^{\ell m}_{2} Y^{\ell m}_{\theta \varphi}   \\
     sym & sym &    sym     &    T^{\ell m}_{1} \sin^2\theta Y^{\ell m} + T^{\ell m}_{2}Y^{\ell m}_{\varphi \varphi} 
\end{pmatrix}
\end{equation}
and
\begin{equation}
\label{tau_odd}
\mathcal{T}^{\text{(o)}}_{\mu\nu} (\omega,r,\theta,\varphi)  = \sum_{\ell=0}^{+\infty}\sum_{m=-\ell}^{m=+\ell} \begin{pmatrix}
                           0     &   0   &     L^{\ell m}_{t} X^{\ell m}_{\theta}             &    L^{\ell m}_{t} X^{\ell m}_{\varphi}      \\
                           sym &   0   &     L^{\ell m}_{r} X^{\ell m}_{\theta}             &    L^{\ell m}_{r} X^{\ell m}_{\varphi}        \\
                           sym & sym &    L^{\ell m} X^{\ell m}_{\theta\theta}        &     L^{\ell m} X^{\ell m}_{\theta\varphi}   \\
                           sym & sym &                        sym                                     &   L^{\ell m} X^{\ell m}_{\varphi\varphi}
\end{pmatrix}
\end{equation}
\end{subequations}
\end{widetext}
with the following conventions: for $\ell = 0$, we have $T^{\ell m}_{t} = T^{\ell m}_{r} = 0$, and for  $\ell = 0, 1$,  $T^{\ell m}_{2} = 0$. Additionally, for $\ell = 0$, $L^{\ell m}_{t} = L^{\ell m}_{r} = 0$, and for $\ell = 0, 1$, $L^{\ell m} = 0$. The dependencies in $\omega$ and $r$ are contained in the functions $T^{\ell m}_{t t}, T^{\ell m}_{t r}, T^{\ell m}_{r r}, T^{\ell m}_{t}, T^{\ell m}_{r}, L^{\ell m}_{t}, L^{\ell m}_{r}, T^{\ell m}_{1}, T^{\ell m}_{2}$, and $L^{\ell m}$, which are the known inputs of the problem and depend on the physical process being studied.

\subsection{Odd-parity sector}
\label{appendix_A_3}

The odd-parity sector field equations are derived by substituting the decompositions~\eqref{h_odd} and \eqref{tau_odd} into the linearized field equations~\eqref{eq_Fierz-Pauli}, which gives
\begin{widetext}
\begin{eqnarray}
  & & f(r) \frac{\partial^2 h^{\ell m}_{t} }{\partial r^2} + \left(\frac{\omega^2}{f(r)} -\frac{\Lambda +2}{r^2} +\frac{4M}{r^3} -\mu^2 \right) h^{\ell m}_{t}  -\frac{2 i M \omega}{r^2}h^{\ell m}_{r}   = -16 \pi L^{\ell m}_{t} \label{eq_odd_1} \\
   \nonumber\\[1ex]
  & & f(r) \frac{\partial^2 h^{\ell m}_{r} }{\partial r^2}+ \frac{4M}{r^2}\frac{\partial h^{\ell m}_{r} }{\partial r}+\left(\frac{\omega^2}{f(r)} -\frac{\Lambda +6}{r^2} +\frac{8M}{r^3} -\mu^2 \right) h^{\ell m}_{r} -\frac{2 i M \omega}{r^2 f(r)^2}h^{\ell m}_{t}
  + \frac{\Lambda}{r^3} h^{\ell m}= -16 \pi L^{\ell m}_{r}  \label{eq_odd_2}\\
   \nonumber\\[1ex]
  & & f(r) \frac{\partial^2 h^{\ell m} }{\partial r^2} +\frac{6M-2r}{r^2} \frac{\partial h^{\ell m} }{\partial r} +  \left(\frac{\omega^2}{f(r)} -\frac{\Lambda - 2}{r^2} -\frac{8M}{r^3} -\mu^2 \right) h^{\ell m} +\frac{4 f(r)}{r} h^{\ell m}_{r}  = -16 \pi L^{\ell m}   \label{eq_odd_3}
\end{eqnarray}
\end{widetext}
where $\Lambda =(\ell-1)(\ell+2) = \ell(\ell+1)-2$. Here, Eqs. \eqref{eq_odd_1}, \eqref{eq_odd_2}, and \eqref{eq_odd_3} correspond to the ($t \theta$), ($r \theta$), and ($\theta \theta$) components of the field equations, respectively. By inserting the decomposition~\eqref{h_odd} and \eqref{tau_odd} into the transverse constraint \eqref{eq_Fierz-Pauli_contraintes_1}, we obtain 
\begin{fleqn}
\begin{equation}\label{eq_odd_4_contrainte_1}
  f(r) \frac{\partial h^{\ell m}_ {r} }{\partial r}+\frac{2(r-M)}{r^2} h^{\ell m}_ {r} +\frac{i \omega}{f(r)}  h^{\ell m}_ {t} - \frac{\Lambda}{2 r^2}  h^{\ell m} = 0.
\end{equation}
\end{fleqn}
With respect to trace \eqref{eq_Fierz-Pauli_contraintes_2}, it is in accordance with the following equation [cf., \eqref{h_odd} and \eqref{tau_odd}]
\begin{equation}\label{eq_odd_5_contrainte_2}
  h^{\ell m} = \frac{-16 \pi}{3 \mu^2} L^{\ell m}
\end{equation}

We can simplify the system \eqref{eq_odd_1}--\eqref{eq_odd_3} by reducing it to a pair of coupled differential equations, using the constraint~\eqref{eq_odd_4_contrainte_1} to eliminate $h^{\ell m}_{t}$. This yields the system \eqref{Système_odd_freq_1}--\eqref{Système_odd_freq_2}, where the source terms are expressed in terms of the components of the stress-energy tensor $\mathcal{T}^{\text{(o)}}_{\mu\nu}$
\begin{eqnarray}
  && S^{(\phi)}_{\omega\ell m}(r) =-16\pi f(r)^2 L^{\ell m}_{r}(\omega,r)\label{source_phi} \\
  \nonumber\\[1ex]
  && S^{(\psi)}_{\omega\ell m}(r) =-16\pi \frac{f(r)}{r} L^{\ell m}(\omega,r)\label{source_psi}
\end{eqnarray}

Finally, for a massive point particle plunging from the ISCO into a Schwarzschild BH, the source terms are provided in Eqs. \eqref{soure_phi_tot} and \eqref{soure_psi_tot}.

\subsection{Even-parity sector}
\label{appendix_A_4}

The field equations of the even-parity sector are derived by substituting the decompositions~\eqref{h_even} and \eqref{tau_even} into the linearized field equations~\eqref{eq_Fierz-Pauli}, which gives us

\begin{widetext}
\begin{multline}\label{eq_even_1}
f(r)\frac{\partial^2 H_{tt}^{\ell m}}{\partial r^2} + \frac{2(r-3M)}{r^2}\frac{\partial H_{tt}^{\ell m}}{\partial r}+ \left[\frac{\omega^2}{f(r)} + \frac{2M^2}{(r-2M)r^3}-\frac{\Lambda + 2}{r^2} - \mu^2\right] H_{tt}^{\ell m}  \\
+ f(r) \frac{2M(3M-2r)}{r^4} H_{rr}^{\ell m} - \frac{4 i M \omega}{r^2} H_{tr}^{\ell m} + f(r) \frac{4M}{r^3}K^{\ell m}   = \frac{16 \pi}{3} \left(\frac{2M^2}{\mu^2 r^4} - \frac{\omega^2}{\mu^2}-2 f(r)\right) T^{\ell m}_{tt} \\ 
   + \frac{16 \pi}{3} f(r)^2 \left(\frac{2 M^2}{\mu^2 r^4} + \frac{\omega^2}{\mu^2} - f(r)\right) T^{\ell m}_{rr} + \frac{32 \pi}{3} \frac{f(r)}{\mu^2 r^2} \left[\omega^2 - f(r) \left(\mu^2 +\frac{2M}{r^3}\right)\right] T^{\ell m}_{1} \\
   - \frac{16 \pi M f(r)}{3\mu^2 r^2} \frac{\partial T^{\ell m}_{tt}}{\partial r} +\frac{16 \pi M f(r)^3}{3\mu^2 r^2} \frac{\partial T^{\ell m}_{rr}}{\partial r}  +\frac{32 \pi M f(r)^2}{3\mu^2 r^4} \frac{\partial T^{\ell m}_{1}}{\partial r}
\end{multline}
\begin{multline}\label{eq_even_2}
f(r)\frac{\partial^2 H_{tr}^{\ell m}}{\partial r^2} -\frac{2(M-r)}{r^2} \frac{\partial H_{tr}^{\ell m}}{\partial r} + \left[ \frac{\omega^2}{f(r)}-\frac{4M^2}{(r-2M)r^3}-\frac{\Lambda+4}{r^2} - \mu^2 \right] H_{tr}^{\ell m} \\
 - \frac{2 i M \omega}{r^2}  H_{rr}^{\ell m} - \frac{2 i M \omega}{(r-2M)^2} H_{tt}^{\ell m} 
+ \frac{2(\Lambda + 2) }{r^3} H_{t}^{\ell m}   =  -16 \pi \,  T_{tr}^{\ell m} + \frac{16 \pi i M \omega}{\mu^2 (r - 2M)^2} T_{tt}^{\ell m} + \frac{16 \pi i M \omega}{ 3 \mu^2 r^2} T_{rr}^{\ell m} \\
+ \frac{32 \pi i (3M-2r) \omega}{ 3 \mu^2 (r-2M) r^3} T_{1}^{\ell m} - \frac{16 \pi i \omega}{3 \mu^2 f(r)} \frac{\partial T^{\ell m}_{tt}}{\partial r} + \frac{16 \pi i \omega f(r)}{3 \mu^2 } \frac{\partial T^{\ell m}_{rr}}{\partial r} + \frac{32 \pi i \omega}{3 \mu^2 r^2} \frac{\partial T^{\ell m}_{1}}{\partial r}
\end{multline}
\begin{multline}\label{eq_even_3}
f(r)\frac{\partial^2 H_{t}^{\ell m}}{\partial r^2} + \left[ \frac{\omega^2}{f(r)}+ \frac{4M}{r^3}- \frac{\Lambda + 2}{r^2} - \mu^2 \right]H_{t}^{\ell m}  - \frac{2 i M \omega}{r^2}H_{r}^{\ell m} + \frac{2 f(r)H_{tr}^{\ell m}}{r} = \\ -16 \pi \, T^{\ell m}_{t} -\frac{16\pi i\omega}{3\mu^2 f(r)} T^{\ell m}_{tt}
+\frac{16\pi i\omega f(r)}{3\mu^2 } T^{\ell m}_{rr} +\frac{32\pi i\omega}{3\mu^2 r^2} T^{\ell m}_{1} 
\end{multline}
\begin{multline}\label{eq_even_4}
f(r)\frac{\partial^2 H^{\ell m}_{rr}}{\partial r^2} +\frac{2(M+r)}{r^2} \frac{\partial H_{rr}^{\ell m}}{\partial r} + \left[\frac{\omega^2}{f(r)} + \frac{2M(4r-7M)}{r-2M} -\frac{\Lambda +6}{r^2} - \mu^2\right] H_{rr}^{\ell m}  \\
\hspace{-6ex}- \frac{2M (2 r-3 M)}{r^4f(r)^3} H^{\ell m}_{tt}- \frac{4 i M \omega}{(r-2M)^2}  H_{tr}^{\ell m}  +\frac{4(\Lambda +2)}{r^3}H^{\ell m}_{r} + \frac{4(r-3M) }{r^3f(r)}K^{\ell m}= 
\\ \hspace{2ex}-\frac{16 \pi}{f(r)}\left[\frac{2}{3} - \frac{2M(2r-5M)}{3\mu^2 r^4} -\frac{4M}{3r}\right] T^{\ell}_{rr} 
   - \frac{16 \pi}{f(r)^3}\left[\frac{1}{3} - \frac{2M(2r-M)}{3\mu^2 r^4} -\frac{2M}{3r}\right] T^{\ell}_{tt} \\
   -\frac{16 \pi}{f(r)}\left[\frac{4(3r-7M)}{3\mu^2 r^5} -\frac{2}{3r^2}\right] T^{\ell}_{1} - \frac{80 \pi M}{3\mu^2 r^2} \frac{\partial T^{\ell m}_{rr}}{\partial r}
   - \frac{16 \pi M}{\mu^2 r^2 f(r)^2} \frac{\partial T^{\ell m}_{tt}}{\partial r}\\
    + \frac{32 \pi (4r-9M)}{3\mu^2 r^4 f(r)} \frac{\partial T^{\ell m}_{1}}{\partial r}
   -\frac{16 \pi f(r)}{3\mu^2} \frac{\partial^2 T^{\ell m}_{rr}}{\partial r^2} + \frac{16 \pi }{3\mu^2  f(r)} \frac{\partial^2 T^{\ell m}_{tt}}{\partial r^2}
   -\frac{32 \pi}{3\mu^2 r^2} \frac{\partial^2 T^{\ell m}_{1}}{\partial r^2}
\end{multline}
\begin{multline}\label{eq_even_5}
f(r)\frac{\partial^2 H_{r}^{\ell m}}{\partial r^2} + \frac{4M}{r^2} \frac{\partial H_{r}^{\ell m}}{\partial r} + \left[\frac{\omega^2}{f(r)}+\frac{8M}{r^3}-\frac{\Lambda+6}{r^2} - \mu^2\right]H_{r}^{\ell m} \\
- \frac{2 i M \omega}{(r-2M)^2}H^{\ell m}_{t} +\frac{\Lambda G^{\ell m}}{r} -  \frac{2 K^{\ell m}}{r} + \frac{2f(r) H_{rr}^{\ell m}}{r}= 
-16 \pi \,T^{\ell m}_{r} - \frac{16 \pi}{3 \mu^2 r f(r)^2} T^{\ell m}_{tt} \\
+ \frac{16 \pi(r-4M)}{3 \mu^2 r^2} T^{\ell m}_{tt}
+ \frac{32 \pi}{ \mu^2 r^3} T^{\ell m}_{1} +\frac{16 \pi}{3 \mu^2  f(r)} \frac{\partial T^{\ell m}_{tt}}{\partial r} 
-\frac{16 \pi f(r)}{3 \mu^2} \frac{\partial T^{\ell m}_{rr}}{\partial r} -\frac{32 \pi}{3 \mu^2  r^2} \frac{\partial T^{\ell m}_{1}}{\partial r} 
\end{multline}
\begin{multline}\label{eq_even_6}
f(r)\frac{\partial^2 G^{\ell m}}{\partial r^2} + \frac{2(r-M)}{r^2} \frac{\partial G^{\ell m} }{\partial r} +\left[\frac{\omega^2}{f(r)}-\frac{\Lambda}{r^2}-\mu^2\right]G^{\ell m} + \frac{4 f(r)H_{r}^{\ell m}}{r^2} =\\ -\frac{16 \pi}{r^2} T^{\ell m}_{2} -\frac{32 \pi}{3 \mu^2 r^4} T^{\ell m}_{1} + \frac{16 \pi}{3 \mu^2 r^2 f(r)} T^{\ell m}_{tt} -\frac{16 \pi f(r)}{3 \mu^2 r^2} T^{\ell m}_{rr}
\end{multline}
\begin{multline}\label{eq_even_7}
f(r)\frac{\partial^2 K^{\ell m}}{\partial r^2}   +\frac{2(r-M)}{r^2} \frac{\partial K^{\ell m}}{\partial r} + \left[\frac{\omega^2}{f(r)}+\frac{8M}{r^3} -\frac{\Lambda+4}{r^2} - \mu^2\right]K^{\ell m} \\
\hspace{-25ex}- \frac{2(\Lambda+2)f(r)}{r^3}H_{r}^{\ell m} + \frac{2f(r)(r-3M)}{r^3}H_{rr}^{\ell m} +\frac{2M }{r^3 f(r)} H_{tt}^{\ell m} =\\
- \frac{16 \pi}{3 \mu^2 r^4} \left[\frac{8M}{r} -(\Lambda+6) + \mu^2 r^2\right] T^{\ell m}_{1} 
-\frac{16 \pi}{3 \mu^2 r^2 f(r)} \left[1+\frac{2M}{r} +\frac{\Lambda}{2} +\mu^2 r^2\right] T^{\ell m}_{tt} \\
+ \frac{16 \pi f(r)}{3 \mu^2} \left[\frac{\Lambda +2}{2r^2} -\frac{2M}{r^3} +\mu^2\right] T^{\ell m}_{rr}
+\frac{16 \pi}{3 \mu^2 r} \frac{\partial T^{\ell m}_{tt}}{\partial r} 
-\frac{16 \pi f(r)^2}{3 \mu^2 r} \frac{\partial T^{\ell m}_{rr}}{\partial r} -\frac{32 \pi f(r)}{3 \mu^2 r^3} \frac{\partial T^{\ell m}_{1}}{\partial r}.
\end{multline}
\end{widetext}

Here the components ($tt$), ($tr$), ($t\theta$), ($rr$), ($r\theta$), and ($\theta\varphi$) provide the field equations \eqref{eq_even_1} to \eqref{eq_even_6}, while Eq. \eqref{eq_even_7} is obtained by combining the ($\theta\varphi$) and ($\theta\theta$) components.

By inserting the decompositions \eqref{h_even} and \eqref{tau_even} into the transverse constraint \eqref{eq_Fierz-Pauli_contraintes_1}, we get the following three radial equations:
\begin{widetext}
\begin{multline}\label{Contrainte_1}
   f(r)\frac{\partial H_{tr}^{\ell m}}{\partial r} + \frac{2(r-M)H_{tr}^{\ell m}}{r^2} +\frac{ i\omega}{f(r)} H_{tt}^{\ell m} -\frac{(\Lambda +2)}{r^2} H_{t}^{\ell m}=- \frac{16 \pi i \omega}{3 \mu^2 f(r)} T^{\ell m}_{tt} + \frac{16 \pi i \omega f(r)}{3 \mu^2} T^{\ell m}_{tt} + \frac{32 \pi i \omega}{3 \mu^2 r^2} T^{\ell m}_{1}
\end{multline}
\begin{multline}\label{Contrainte_2}
   f(r)\frac{\partial H_{rr}^{\ell m}}{\partial r} +\frac{2r-M}{r^2}H_{rr}^{\ell m} +\frac{M}{(r-2M)^2}H_{tt}^{\ell m} -\frac{(\Lambda+2)}{r^2}H_{r}^{\ell m}+ \frac{i\omega}{f(r)} H^{\ell m}_{tr}-\frac{2}{r}K^{\ell m} = \\
   -\frac{32 \pi M}{3 \mu^2 r^2 f(r)^2} T^{\ell m}_{tt} -\frac{32 \pi M}{3 \mu^2 r^2 } T^{\ell m}_{rr} +\frac{64 \pi}{3 \mu^2 r^3 } T^{\ell m}_{1}
   +\frac{16 \pi}{3\mu^2f(r)}\frac{\partial T^{\ell m}_{tt}}{\partial r} -\frac{16 \pi f(r)}{3\mu^2}\frac{\partial T^{\ell m}_{rr}}{\partial r}
   -\frac{32 \pi }{3\mu^2 r^2}\frac{\partial T^{\ell m}_{1}}{\partial r}
\end{multline}
\begin{multline}\label{Contrainte_3}
   f(r)\frac{\partial H_{r}^{\ell m}}{\partial r} +\frac{2(r-M)}{r^2}H_{r}^{\ell m} + \frac{i\omega}{f(r)} H_{t}^{\ell m}-\frac{\Lambda }{2}G^{\ell m} + K^{\ell m} =
    \frac{16 \pi}{3 \mu^2 f(r)} T^{\ell m}_{tt} -\frac{16 \pi f(r)}{3 \mu^2 } T^{\ell m}_{rr} - \frac{32 \pi}{3 \mu^2 r^2} T^{\ell m}_{1}
\end{multline}
\end{widetext}
for the $t$, $r$, and $\theta$ components, respectively. Finally, for even-parity, the trace \eqref{eq_Fierz-Pauli_contraintes_2} gives us
\begin{multline}\label{trace_even}
  f(r) H_{rr}^{\ell m}-\frac{H_{tt}^{\ell m}}{f(r)} +2 K^{\ell m} = \\ + \frac{16 \pi}{3 \mu^2 f(r)} T^{\ell m}_{tt} - \frac{16 \pi f(r)}{3 \mu^2} T^{\ell m}_{rr} -\frac{32 \pi}{3 \mu^2 r^2} T^{\ell m}_{1}.
\end{multline}

The system of equations \eqref{eq_even_1}--\eqref{eq_even_7} can be reduced to a system of three equations for $\ell \geq 2$. We have chosen to work with the $K^{\ell m}, H^{\ell m}_{r}$ and $G^{\ell m}$ components, a choice already made in Ref.~\cite{Brito:2013wya} for the same reasons: this choice is particularly useful, since the system directly contains the monopole and dipole cases ($\ell = 0, 1$).

After some tedious algebraic calculations, we find that the even parity is completely described by the equations \eqref{Système_even_freq_1}--\eqref{Système_even_freq_3}, where the functions $ \alpha_i^{(K)}$ in Eq.~\eqref{Système_even_freq_1} are given by

\begin{widetext}
\begin{fleqn}
\begin{multline}\label{alpha_K_1}
\alpha^{(K)}_1(r)= \frac{f(r) \left[ (\Lambda + 2)^2 + (\Lambda^2 - 4) f(r) + \mu^4 (f(r)+1) r^4 \right]
}{
(\Lambda + 2) \left[\Lambda + 2 - 2 f(r)\right] r + 2 \left[\mu^2 (\Lambda + 2)+ \mu^2 f(r) + 2 \omega^2  \right] r^3 + \mu^4 r^5
}\\
\\[-1ex]
+\frac{2f(r)\left[ f(r) \left( \mu^2 (\Lambda + 2)+ 4 \mu^2 f(r) + 6 \omega^2\right)+ \mu^2 (\Lambda + 2) + 2 \omega^2  \right] r^2 
}{
(\Lambda + 2) \left[\Lambda + 2 - 2 f(r)\right] r + 2 \left[\mu^2 (\Lambda + 2)+ \mu^2 f(r) + 2 \omega^2  \right] r^3 + \mu^4 r^5
}
\end{multline}
\end{fleqn}
\begin{fleqn}
\begin{multline}\label{alpha_K_2}
\alpha^{(K)}_2(r) = \frac{
- (\Lambda + 2) f(r) \left[ 2 (8 + \Lambda - 6 f(r)) f(r) + \Lambda^2 - 4 \right]
}{
(\Lambda + 2) \left[\Lambda + 2 - 2 f(r)\right] r^2 + 2 \left[\mu^2 (\Lambda + 2)+ \mu^2 f(r) + 2 \omega^2  \right] r^4 + \mu^4 r^6
}\\
\\[-1ex]
\hspace{15ex}+ \frac{
\left[ (\Lambda + 2)^2 \omega^2 - f(r) \left( 2 \left( 5 \mu^2 (\Lambda + 2) + 6 \omega^2 \right) f(r)+ \mu^2 (\Lambda + 2) (3 \Lambda-2) + 2 (5 \Lambda+4) \omega^2  \right) \right] r^2
}{
(\Lambda + 2) \left[\Lambda + 2 - 2 f(r)\right] r^2 + 2 \left[\mu^2 (\Lambda + 2)+ \mu^2 f(r) + 2 \omega^2  \right] r^4 + \mu^4 r^6
} \\
\\[-1ex]
\hspace{25ex}+ \frac{
\left[ 2 \mu^2 (\Lambda + 2) \omega^2  - \mu^2 f(r) \left( 8 \mu^2 f(r) +\mu^2 ( 3 \Lambda + 2) + 6 \omega^2 \right) + 4 \omega^4 \right] r^4
}{
(\Lambda + 2) \left[\Lambda + 2 - 2 f(r)\right] r^2 + 2 \left[\mu^2 (\Lambda + 2)+ \mu^2 f(r) + 2 \omega^2  \right] r^4 + \mu^4 r^6
} \\
\\[-2ex]
+ \frac{
\mu^4 \left[\omega^2 - \mu^2 f(r)\right] r^6
}{
(\Lambda + 2) \left[\Lambda + 2 - 2 f(r)\right] r^2 + 2 \left[\mu^2 (\Lambda + 2)+ \mu^2 f(r) + 2 \omega^2  \right] r^4 + \mu^4 r^6
}
\end{multline}
\end{fleqn}
\begin{fleqn}\label{alpha_K_3}
\begin{equation}\label{alpha_K_3}
\alpha^{(K)}_3(r) =\frac{
4 (\Lambda + 2)  f(r)^3  \left[\Lambda  + 3 - 3 f(r) + \mu^2 r^2 \right]
}{
(\Lambda + 2) \left[\Lambda + 2 - 2 f(r)\right] r^2 + 2 \left[\mu^2 (\Lambda + 2)+ \mu^2 f(r) + 2 \omega^2  \right] r^4 + \mu^4 r^6
}
\end{equation}
\end{fleqn}
\begin{fleqn}
\begin{equation}\label{alpha_K_4}
\alpha^{(K)}_4(r) = -\frac{
2 (\Lambda + 2) f(r)^2 \left[ 6 f(r)^2 + f(r) \left(-6 - 3 \Lambda + \mu^2 r^2\right) + (\Lambda + \mu^2 r^2) \left(\Lambda + 3 + \mu^2 r^2\right)
\right]
}{
(\Lambda + 2) \left[\Lambda + 2 - 2 f(r)\right] r^3 + 2 \left[\mu^2 (\Lambda + 2)+ \mu^2 f(r) + 2 \omega^2  \right] r^5 + \mu^4 r^7
}
\end{equation}
\end{fleqn}

\begin{fleqn}
\begin{equation}\label{alpha_K_5}
\alpha^{(K)}_5(r) = \frac{
2 \Lambda (\Lambda + 2) f(r)^3
}{
(\Lambda + 2) \left[\Lambda + 2 - 2 f(r)\right] r + 2 \left[\mu^2 (\Lambda + 2)+ \mu^2 f(r) + 2 \omega^2  \right] r^3 + \mu^4 r^5
}
\end{equation}
\end{fleqn}
\begin{fleqn}\label{alpha_K_6}
\begin{equation}\label{alpha_K_6}
\alpha^{(K)}_6(r) = -\frac{
\Lambda (\Lambda + 2) f(r)^2 \left[\Lambda + 4 - 6 f(r) + \mu^2 r^2\right]
}{
(\Lambda + 2) \left[\Lambda + 2 - 2 f(r)\right] r^2 + 2 \left[\mu^2 (\Lambda + 2)+ \mu^2 f(r) + 2 \omega^2  \right] r^4 + \mu^4 r^6
}
\end{equation}
\end{fleqn}
those of  $\alpha_i^{(H)}$ in Eq.~\eqref{Système_even_freq_2} are
\begin{fleqn}
\begin{multline}\label{alpha_H_1}
\alpha^{(H)}_1(r) = \frac{
f(r) \left[ (\Lambda + 2) \left(6 + 3 \Lambda + \Lambda f(r) - 6 f(r)^2\right)  \right]
}{
(\Lambda + 2) \left[\Lambda + 2 - 2 f(r)\right] r + 2 \left[\mu^2 (\Lambda + 2)+ \mu^2 f(r) + 2 \omega^2  \right] r^3 + \mu^4 r^5
}\\
\\[-1ex]
\hspace{25ex}
+\frac{
2f(r) \left[ 3 \mu^2 (\Lambda + 2) + 6 \omega^2 - f(r) \left(\mu^2 (\Lambda - 1)+ 3 \mu^2 f(r) + 6 \omega^2  \right)  \right] r^2
}{
(\Lambda + 2) \left[\Lambda + 2 - 2 f(r)\right] r + 2 \left[\mu^2 (\Lambda + 2)+ \mu^2 f(r) + 2 \omega^2  \right] r^3 + \mu^4 r^5
}\\
\\[-2.5ex]
+\frac{
- 3 \mu^4 f(r) \left(f(r)-1 \right) r^4 
}{
(\Lambda + 2) \left[\Lambda + 2 - 2 f(r)\right] r + 2 \left[\mu^2 (\Lambda + 2)+ \mu^2 f(r) + 2 \omega^2  \right] r^3 + \mu^4 r^5
}
\end{multline}
\end{fleqn}
\begin{fleqn}
\begin{multline}\label{alpha_H_2}
\alpha^{(H)}_2(r) = \frac{
f(r) \left[(\Lambda + 2)(2 - 2 \Lambda - \Lambda^2) + \left(2 (\Lambda + 2) \omega^2 - (\Lambda (3 \Lambda + 10) + 6) \mu^2 \right) r^2 - \mu^2 \left(3 \mu^2 (\Lambda + 2) + 2 \omega^2\right) r^4 - \mu^6 r^6\right]
}{
(\Lambda + 2) \left[\Lambda + 2 - 2 f(r)\right] r^2 + 2 \left[\mu^2 (\Lambda + 2)+ \mu^2 f(r) + 2 \omega^2  \right] r^4 + \mu^4 r^6
}\\
\\[-1ex]
\hspace{25ex}
+\frac{
 (1 + \omega^2 r^2) \left[(\Lambda + 2)^2 + 2 \left(\mu^2 (\Lambda + 2) + 2 \omega^2\right) r^2 + \mu^4 r^4\right]
}{
(\Lambda + 2) \left[\Lambda + 2 - 2 f(r)\right] r^2 + 2 \left[\mu^2 (\Lambda + 2)+ \mu^2 f(r) + 2 \omega^2  \right] r^4 + \mu^4 r^6
} \\
\\[-1ex]
\hspace{35ex}
+\frac{
-f(r)^2 \left[(\Lambda + 2)^2 + 4 \left(2 \mu^2 (\Lambda + 2) + 5 \omega^2\right) r^2 + 7 \mu^4 r^4\right] 
}{
(\Lambda + 2) \left[\Lambda + 2 - 2 f(r)\right] r^2 + 2 \left[\mu^2 (\Lambda + 2)+ \mu^2 f(r) + 2 \omega^2  \right] r^4 + \mu^4 r^6
}\\
\\[-1ex]
+\frac{
-2 f(r)^3 \left(\Lambda + 2 + 5 \mu^2 r^2\right)
}{
(\Lambda + 2) \left[\Lambda + 2 - 2 f(r)\right] r^2 + 2 \left[\mu^2 (\Lambda + 2)+ \mu^2 f(r) + 2 \omega^2  \right] r^4 + \mu^4 r^6
}
\end{multline}
\end{fleqn}
\begin{fleqn}
\begin{equation}\label{alpha_H_3}
\alpha^{(H)}_3(r) = \frac{-2 f(r) \left[\Lambda + 2 + \left(\mu^2 - 4 \omega^2\right) r^2 - f(r) \left(\Lambda + 2 + 3 \mu^2 r^2\right)\right]}{
(\Lambda + 2) \left[\Lambda + 2 - 2 f(r)\right] + 2 \left[\mu^2 (\Lambda + 2) + \mu^2 f(r) + 2 \omega^2\right] r^2 + \mu^4 r^4}
\end{equation}
\end{fleqn}
\begin{fleqn}
\begin{equation}\label{alpha_H_4}
\alpha^{(H)}_4(r) = \frac{-\left(\Lambda + 2 + \mu^2 r^2\right) \left[\left(\Lambda + 2 - 6 f(r)\right) \left(f(r)-1\right) + \left(3 \mu^2 f(r)-\mu^2 + 4 \omega^2 \right) r^2\right]}
{(\Lambda + 2) \left[\Lambda + 2 - 2 f(r)\right] r + 2 \left[\mu^2 (\Lambda + 2)  + \mu^2 f(r) + 2 \omega^2\right] r^3 + \mu^4 r^5}
\end{equation}
\end{fleqn}
\begin{fleqn}
\begin{equation}\label{alpha_H_5}
\alpha^{(H)}_5(r) = \frac{2 \Lambda (\Lambda + 2) f(r)^2}
{(\Lambda + 2) \left[\Lambda + 2 - 2 f(r)\right] + 2 \left[\mu^2 (\Lambda + 2) + \mu^2 f(r) + 2 \omega^2 \right] r^2 + \mu^4 r^4}
\end{equation}
\end{fleqn}
\begin{fleqn}
\begin{multline}\label{alpha_H_6}
\alpha^{(H)}_6(r) = \frac{
2 \Lambda \left[f(r) \left(  3 \mu^2 f(r) + \mu^2 (2 \Lambda + 3) + 6 \omega^2\right) - \mu^2 (\Lambda + 2) - 2 \omega^2 \right] r^2 
}{
2 (\Lambda + 2) \left[\Lambda + 2 - 2 f(r)\right] r + 4 \left[\mu^2 (\Lambda + 2) + \mu^2 f(r) + 2 \omega^2 \right] r^3 + 2 \mu^4 r^5
}\\
+\frac{
\Lambda \left[(\Lambda + 2) \left( f(r) \left(6 f(r) + \Lambda\right) - \Lambda - 2 \right)  + \mu^4 \left(3 f(r) - 1\right) r^4\right]
}{
2 (\Lambda + 2) \left[\Lambda + 2 - 2 f(r)\right] r + 4 \left[\mu^2 (\Lambda + 2) + \mu^2 f(r) + 2 \omega^2 \right] r^3 + 2 \mu^4 r^5
}
\end{multline}
\end{fleqn}
\end{widetext}
and the functions $\alpha_i^{(G)}$ in Eq.~\eqref{Système_even_freq_3} are
\begin{equation}\label{alpha_G_1}
\alpha^{(G)}_1(r)  = \frac{2 f(r) (r-M)}{r^2},
\end{equation}
\begin{equation}\label{alpha_G_2}
\alpha^{(G)}_2(r)  = \omega^2 - f(r)\left(\frac{\Lambda}{r^2} + \mu^2\right),
\end{equation}
\begin{equation}\label{alpha_G_3}
\alpha^{(G)}_3(r)  = 0,
\end{equation}
\begin{equation}\label{alpha_G_4}
\alpha^{(G)}_4(r)  = 0,
\end{equation}
\vspace{4pt}
\begin{equation}\label{alpha_G_5}
\alpha^{(G)}_5(r)  = 0,
\end{equation}
\begin{equation}\label{alpha_G_6}
\alpha^{(G)}_6(r) = \frac{4 f(r)^2}{r^3}.
\end{equation}

Furthermore, the source terms are expressed in terms of the stress-energy tensor components $\mathcal{T}^{\text{(e)}}_{\mu\nu}$
\begin{widetext}
\begin{fleqn}
\begin{multline}\label{S_K_even_tot}
S^{(K)}_{\omega\ell m}(r) = \xi^{(K)}_1(r) \frac{\partial T^{\ell m}_{tt}(r,\omega)}{\partial r} +  \xi^{(K)}_2(r)   T^{\ell m}_{tt}(r,\omega) +  \xi^{(K)}_3(r) \frac{\partial T^{\ell m}_{rr}(r,\omega)}{\partial r} \\ +  \xi^{(K)}_4(r)   T^{\ell m}_{rr}(r,\omega) + \xi^{(K)}_5(r) \frac{\partial T^{\ell m}_{1}(r,\omega)}{\partial r} +  \xi^{(K)}_6(r)   T^{\ell m}_{1}(r,\omega) +  \xi^{(K)}_7(r)   T^{\ell m}_{r}(r,\omega),
\end{multline}
\end{fleqn}
\begin{fleqn}
\begin{multline}\label{S_H_even_tot}
S^{(H_r)}_{\omega\ell m}(r) = \xi^{(H_r)}_1(r) \frac{\partial T^{\ell m}_{tt}(r,\omega)}{\partial r} +  \xi^{(H_r)}_2(r)   T^{\ell m}_{tt}(r,\omega) +  \xi^{(H_r)}_3(r) \frac{\partial T^{\ell m}_{rr}(r,\omega)}{\partial r} \\ +  \xi^{(H_r)}_4(r)   T^{\ell m}_{rr}(r,\omega) + \xi^{(H_r)}_5(r) \frac{\partial T^{\ell m}_{1}(r,\omega)}{\partial r} +  \xi^{(H_r)}_6(r)   T^{\ell m}_{1}(r,\omega) +  \xi^{(H_r)}_7(r)   T^{\ell m}_{r}(r,\omega),
\end{multline}
\end{fleqn}
\begin{fleqn}
\begin{multline}\label{S_G_even_tot}
S^{(G)}_{\omega\ell m}(r) = \xi^{(G)}_1(r) \frac{\partial T^{\ell m}_{tt}(r,\omega)}{\partial r} +  \xi^{(G)}_2(r)   T^{\ell m}_{tt}(r,\omega) +  \xi^{(G)}_3(r) \frac{\partial T^{\ell m}_{rr}(r,\omega)}{\partial r} \\ +  \xi^{(G)}_4(r)   T^{\ell m}_{rr}(r,\omega) + \xi^{(G)}_5(r) \frac{\partial T^{\ell m}_{1}(r,\omega)}{\partial r} +  \xi^{(G)}_6(r)   T^{\ell m}_{1}(r,\omega) +  \xi^{(G)}_7(r)   T^{\ell m}_{2}(r,\omega).
\end{multline}
\end{fleqn}

The  functions $\xi^{(K)}_i$ in Eq.~\eqref{S_K_even_tot} are given by
\begin{fleqn}
\begin{equation}\label{X_1_K}
  \xi^{(K)}_1(r) = \frac{16 \pi f(r)}{3\mu^2 r},
\end{equation}
\end{fleqn}
\begin{fleqn}
\begin{multline}\label{X_2_K}
\xi^{(K)}_2(r) = \frac{
-32 \pi f(r)^2 \left(2 \mu^2 r^2 + 3 \Lambda + 6 \right)
}{
3 \mu^2\left[ (\Lambda + 2) \left[\Lambda + 2 - 2 f(r)\right] r^2 + 2 \left[\mu^2 (\Lambda + 2)+ \mu^2 f(r) + 2 \omega^2  \right] r^4 + \mu^4 r^6\right]
}\\
\\[-1ex]
\hspace{15ex}
+ \frac{
16 \pi f(r) \left[(\Lambda + 2) (3 \Lambda + 8)+ \mu^2 (6 + \Lambda) r^2 - 4 \mu^4 r^4\right]
}{
3 \mu^2\left[ (\Lambda + 2) \left[\Lambda + 2 - 2 f(r)\right] r^2 + 2 \left[\mu^2 (\Lambda + 2)+ \mu^2 f(r) + 2 \omega^2  \right] r^4 + \mu^4 r^6\right]
}\\
\\[-1ex]
+ \frac{
 - 8 \pi  \left(2 \mu^2 r^2 + \Lambda + 2 \right) \left[(\Lambda + 2)^2 + 2 \left(\mu^2 (\Lambda + 2) + 2 \omega^2\right) r^2 + \mu^4 r^4\right]
}{
3 \mu^2\left[ (\Lambda + 2) \left[\Lambda + 2 - 2 f(r)\right] r^2 + 2 \left[\mu^2 (\Lambda + 2)+ \mu^2 f(r) + 2 \omega^2  \right] r^4 + \mu^4 r^6\right]
},
\end{multline}
\end{fleqn}
\begin{fleqn}
\begin{equation}\label{X_3_K}
\xi^{(K)}_3(r) = -\frac{16 \pi f(r)^3}{3 \mu^2 r} + \frac{64 \pi f(r)^4 r}{(\Lambda + 2) \left[\Lambda + 2 - 2 f(r)\right] + 2 \left[\mu^2 (\Lambda + 2) + 2 \omega^2 + \mu^2 f(r)\right] r^2 + \mu^4 r^4},
\end{equation}
\end{fleqn}

\begin{fleqn}
\begin{multline}\label{X_4_K}
\xi^{(K)}_4(r) = \frac{
32 \pi f(r)^4 \left[\Lambda + 2 + 4 \mu^2 r^2\right]
}{
3 \mu^2 \left[(\Lambda + 2) \left[\Lambda + 2 - 2 f(r)\right] r^2 + 2 \left[\mu^2 (\Lambda + 2)+ \mu^2 f(r) + 2 \omega^2  \right] r^4 + \mu^4 r^6\right]}
\\
\\[-1ex]
\hspace{15ex}
+\frac{
16 \pi f(r)^3  \left[\left(10 \mu^2 - 3 \Lambda \mu^2 + 8 \omega^2\right) r^2-\Lambda (\Lambda + 2) \right]
}{
3 \mu^2 \left[(\Lambda + 2) \left[\Lambda + 2 - 2 f(r)\right] r^2 + 2 \left[\mu^2 (\Lambda + 2)+ \mu^2 f(r) + 2 \omega^2  \right] r^4 + \mu^4 r^6
\right]}\\
\\[-1ex]
+\frac{
8 \pi f(r)^2 \left(2 \mu^2 r^2+ \Lambda - 2  \right) \left[(\Lambda + 2)^2 + 2 \left(\mu^2 (\Lambda + 2) + 2 \omega^2\right) r^2 + \mu^4 r^4\right]
}{
3 \mu^2 \left[(\Lambda + 2) \left[\Lambda + 2 - 2 f(r)\right] r^2 + 2 \left[\mu^2 (\Lambda + 2)+ \mu^2 f(r) + 2 \omega^2  \right] r^4 + \mu^4 r^6
\right]},
\end{multline}
\end{fleqn}
\begin{fleqn}
\begin{equation}\label{X_5_K}
\xi^{(K)}_5(r)= - \frac{32 \pi f(r)^2}{3 \mu^2 r^3},
\end{equation}
\end{fleqn}
\begin{fleqn}
\begin{multline}\label{X_6_K}
\xi^{(K)}_6(r) = \frac{
16 \pi f(r) \left[ 8 f(r) r^2 \left(2 \mu^2 (\Lambda + 2) + 3 \omega^2 + \mu^4 r^2\right) - 16 \mu^2 f(r)^2 r^2 \right]
}{
3 \mu^2 \left[(\Lambda + 2) \left[\Lambda + 2 - 2 f(r)\right]  r^4 + 2 \left[\mu^2 (\Lambda + 2) + 2 \omega^2 + \mu^2 f(r)\right]  r^6 + \mu^4 r^8\right]
}\\
\\[-1ex]
+\frac{
16 \pi f(r) (\Lambda - \mu^2 r^2) \left[(\Lambda + 2)^2 + 2 \left(\mu^2 (\Lambda + 2) + 2 \omega^2\right) r^2 + \mu^4 r^4\right]
}{
3 \mu^2 \left[(\Lambda + 2) \left[\Lambda + 2 - 2 f(r)\right]  r^4 + 2 \left[\mu^2 (\Lambda + 2) + 2 \omega^2 + \mu^2 f(r)\right]  r^6 + \mu^4 r^8\right]
},
\end{multline}
\end{fleqn}
\begin{fleqn}
\begin{equation}\label{X_7_K}
\xi^{(K)}_7(r) =- \frac{128 \pi (\Lambda + 2) f(r)^3}{(\Lambda + 2) \left[\Lambda + 2 - 2 f(r)\right] r + 2 \left[\mu^2 (\Lambda + 2) + 2 \omega^2 + \mu^2 f(r)\right] r^3 + \mu^4 r^5}.
\end{equation}
\end{fleqn}
Those of   $\xi^{(H)}_i$ in Eq.~\eqref{S_H_even_tot} are 
\begin{fleqn}
\begin{equation}\label{X_1_H}
\xi^{(H)}_1(r) =\frac{16 \pi }{3 \mu^2},
\end{equation}
\end{fleqn}
\begin{fleqn}
\begin{equation}\label{X_2_H}
\xi^{(H)}_2(r) = \frac{16 \pi \left[(\Lambda + 2) \left(\Lambda + 4 - 4 f(r)\right) - 2 \left(3 \mu^2 f(r) + \mu^2 \Lambda + 2 \omega^2 \right) r^2 - 3 \mu^4 r^4\right]}
{3 \mu^2 \left[(\Lambda + 2) \left[\Lambda + 2 - 2 f(r)\right] r + 2 \left[\mu^2 (\Lambda + 2) + 2 \omega^2 + \mu^2 f(r)\right] r^3 + \mu^4 r^5\right]},
\end{equation}
\end{fleqn}
\begin{fleqn}
\begin{equation}\label{X_3_H}
\xi^{(H)}_3(r)  = \frac{16 \pi f(r)^2}{3} \left[\frac{12 f(r) r^2}{(\Lambda + 2) \left[\Lambda + 2 - 2 f(r)\right] + 2 \left[\mu^2 (\Lambda + 2) + 2 \omega^2 + \mu^2 f(r)\right] r^2 + \mu^4 r^4} - \frac{1}{\mu^2} \right],
\end{equation}
\end{fleqn}
\begin{fleqn}
\begin{multline}\label{X_4_H}
\xi^{(H)}_4(r) = \frac{
16 \pi f(r) \left[(\Lambda + 2) \left[(\Lambda + 4) f(r) -2 (\Lambda + 2) \right] - \mu^4 (f(r)+2) r^4\right]
}{
3 \mu^2 \left[(\Lambda + 2) \left[\Lambda + 2 - 2 f(r)\right] r + 2 \left[\mu^2 (\Lambda + 2) + 2 \omega^2 + \mu^2 f(r)\right] r^3 + \mu^4 r^5\right]
}\\
\\[-1ex]
+\frac{
32 \pi f(r)  \left[ f(r) \left( 5 \mu^2 f(r) + 8 \mu^2 + 6 \omega^2\right) -2 \mu^2 (\Lambda + 2) - 4 \omega^2 \right] r^2 
}{
3 \mu^2 \left[(\Lambda + 2) \left[\Lambda + 2 - 2 f(r)\right] r + 2 \left[\mu^2 (\Lambda + 2) + 2 \omega^2 + \mu^2 f(r)\right] r^3 + \mu^4 r^5\right]
},
\end{multline}
\end{fleqn}
\begin{fleqn}
\begin{equation}\label{X_5_H}
\xi^{(H)}_5(r) = - \frac{32 \pi f(r)}{3 \mu^2 r^2},
\end{equation}
\end{fleqn}
\begin{fleqn}
\begin{multline}\label{X_6_H}
\xi^{(H)}_6(r) = \frac{
32 \pi \left[(\Lambda + 2) \left[2 f(r) (\Lambda + 2 - f(r))  - \Lambda  -2 \right]  + \mu^4 ( 6 f(r) - 1) r^4\right]
}{
3 \mu^2  \left[(\Lambda + 2) \left[\Lambda + 2 - 2 f(r)\right] r^3 + 2 \left[\mu^2 (\Lambda + 2) + 2 \omega^2 + \mu^2 f(r)\right] r^5 + \mu^4 r^7\right]
}\\
\\[-1ex]
+\frac{
64 \pi  \left[-\mu^2 (\Lambda + 2) - 2 \omega^2 + f(r) \left(4 \mu^2 (\Lambda + 2) + 8 \omega^2 - 3 \mu^2 f(r)\right)\right] r^2
}{
3 \mu^2  \left[(\Lambda + 2) \left[\Lambda + 2 - 2 f(r)\right] r^3 + 2 \left[\mu^2 (\Lambda + 2) + 2 \omega^2 + \mu^2 f(r)\right] r^5 + \mu^4 r^7\right]
},
\end{multline}
\end{fleqn}
\begin{fleqn}
\begin{equation}\label{X_7_H}
\xi^{(H)}_7(r) = - 16 \pi f(r) \left[1 + \frac{8 (\Lambda + 2) f(r)}{(\Lambda + 2) \left[\Lambda + 2 - 2 f(r)\right] + 2 \left[\mu^2 (\Lambda + 2) + 2 \omega^2 + \mu^2 f(r)\right] r^2 + \mu^4 r^4}\right],
\end{equation}
\end{fleqn}
\end{widetext}
and the functions  $\xi^{(G)}_i$ in Eq.~\eqref{S_G_even_tot} are 
\begin{equation}\label{X_1_G}
\xi^{(G)}_1(r) = 0,
\end{equation}
%
\begin{equation}\label{X_2_G}
\xi^{(G)}_2(r) = \frac{16 \pi}{3 \mu^2 r^2},
\end{equation}
%
\begin{equation}\label{X_3_G}
\xi^{(G)}_3(r) = 0,
\end{equation}
%
\begin{equation}\label{X_4_G}
\xi^{(G)}_4(r) = - \frac{16 \pi f(r)^2}{3 \mu^2 r^2},
\end{equation}
%
\begin{equation}\label{X_5_G}
\xi^{(G)}_5(r) = 0,
\end{equation}
%
\begin{equation}\label{X_6_G}
\xi^{(G)}_6(r) =  - \frac{32 \pi f(r)}{3 \mu^2 r^4},
\end{equation}
%
\begin{equation}\label{X_7_G}
\xi^{(G)}_7(r) = - \frac{16 \pi f(r)}{ r^2}.
\end{equation}

Thus, for $\ell \geq 2 $, the source terms are given by \eqref{source_K_main}, \eqref{source_H_main}, and \eqref{source_G_main} for a massive point particle plunging from ISCO into a Schwarzschild BH, with the functions $\mathcal{B}^{(K)}_{\omega\ell m}$, $\mathcal{C}^{(K)}_{\omega\ell m}$, $\mathcal{D}^{(K)}_{\omega\ell m}$, $\mathcal{E}^{(K)}_{\omega\ell m}$, $\mathcal{F}^{(K)}_{\omega\ell m}$, $\mathcal{I}^{(K)}_{\omega\ell m}$, $\mathcal{J}^{(K)}_{\omega\ell m}$ in the source term \eqref{source_K_main} given by 
\begin{widetext}
\begin{equation}\label{B_K_S}
\hspace{-125pt} \mathcal{B}^{(K)}_{\omega\ell m}(r) = 9 \mu^2 r^7  \Bigl[ \Lambda^2 + 4 + 4 \mu^2 r^2  + \mu^4 r^4  + 2 \Lambda (\mu^2 r^2 + 2) + 4 \omega^2 r^2  - 2 f(r) (\Lambda + 2 - \mu^2 r^2 )\Bigr],
\end{equation}
\begin{fleqn}
\begin{multline}\label{C_K_S}
   \mathcal{C}^{(K)}_{\omega\ell m}(r) = -4 i \, r^2 \Bigl[ 4\, \omega \, r^2 (\Lambda + 2 + 5 \, \mu^2 \, r^2)  
+ 3 \sqrt{6} \, m \, M \Big(\Lambda^2 + 4 + \mu^4 \, r^4 \\ + 2 \, \Lambda \, (\mu^2 \, r^2 + 2) + 4 \, r^2 (\mu^2 + \omega^2) \Big) 
- 6 \sqrt{6} \, m \, M \, f(r) \, (\Lambda + 2 - 7 \, \mu^2 \, r^2) \Bigr],
\end{multline}
\end{fleqn}
\begin{fleqn}
\begin{multline}\label{D_K_S}
   \mathcal{D}^{(K)}_{\omega\ell m}(r) = -8 i \Biggl[
   -12 \sqrt{6} \, m \, M \, r^2 \Bigl( 4 + \Lambda^2 + \mu^4 \, r^4 + 2 \Lambda (2 + \mu^2 \, r^2) + 4 \, r^2 (\mu^2 + \omega^2) \Bigr) \\
   + r^4 \, \omega \Bigl( 4 + 9 \, \Lambda^2 + 9 \, \mu^4 \, r^4 + 2 \, \Lambda (10 + 9 \, \mu^2 \, r^2) + 4 \, r^2 (13 \, \mu^2 + 9 \, \omega^2) \Bigr) \\
   + 6 M  f(r) \biggl(\sqrt{6} \, m \Bigl( 4 \, r^2 (2 + \Lambda - \mu^2 \, r^2) 
      + 27 \, M^2 \bigl(2 \Lambda (2 + \mu^2 \, r^2) + 4 \, r^2 (\mu^2 + \omega^2) + \Lambda^2 + 4 + \mu^4 \, r^4  \bigr) \Bigr) \\+ 36 \, M \, r^2 (2 + \Lambda - \mu^2 \, r^2) \, \omega \biggr)
   - 324 \sqrt{6} \, m \, M^3 (2 + \Lambda - \mu^2 \, r^2) f(r)^2 
\Biggr],
\end{multline}
\end{fleqn}
\begin{fleqn}
\begin{equation}\label{E_K_S}
   \mathcal{E}^{(K)}_{\omega\ell m}(r) = 72 \sqrt{2} M \Big(2 r^2 - 27 M^2 f(r)\Big) \Biggl[2 \Lambda (\mu^2 r^2 + 2) + 4 r^2 \omega^2 - 2 f(r) (\Lambda + 2 - \mu^2 r^2) + \Lambda^2 + 4  + 4 \mu^2  r^2 + \mu^4 r^4   \Biggr],
\end{equation}
\end{fleqn}
\begin{fleqn}
\begin{multline}\label{F_K_S}
   \mathcal{F}^{(K)}_{\omega\ell m}(r) = - \sqrt{2}\Biggl[864 M^3 \Big(2 \Lambda (r^2 \mu^2 + 2) + 4 r^2 (\mu^2 + \omega^2) +  \Lambda^2 + 4 + \mu^4  r^4\Big) + 72 M r^2 \Bigl(2 \Lambda (r^2 \mu^2 + 2) + 4 r^2 (\mu^2 + \omega^2) + 
\Lambda^2 + 4 + \mu^4 r^4\Bigr)\\ 
+ 108 M^2 r \bigl(3 \mu^2 r^2 - 2 \bigr) \Bigl(2 \Lambda (\mu^2 r^2 + 2 ) + 4 r^2 (\mu^2 + \omega^2) + \Lambda^2 + 4 +  \mu^4 r^4 \Bigr) + r^3 \biggl(9 \Lambda^3 + \Lambda^2 \bigl(36 \mu^2 r^2  + 34\bigr) \\
+ 2 \Bigl(-36 + 9 r^6 \mu^6 + r^2 (80 \mu^2 + 92 \omega^2) + r^4 (91 \mu^4 + 36 \mu^2 \omega^2)\Bigr) + \Lambda \bigl(4 r^2 (46 \mu^2 + 9 \omega^2) + 45 \mu^4 r^4 - 4  \bigr) \biggr) \\
- 16 f(r) \biggl(27 M^2 r \Bigl(\Lambda (4 + 6 r^2 \mu^2) + 2 r^2 (6 \mu^2 + 5 \omega^2) + \Lambda^2 + 4  + 3 \mu^4 r^4 \Bigr) -2 r^3 (2 + \Lambda + 4 r^2 \mu^2)\biggr) \\ + 864 M^2 r f(r)^2 (\Lambda + 2 + \mu^2 r^2 ) \Biggr],
\end{multline}
\end{fleqn}
\begin{fleqn}
\begin{equation}\label{I_K_S}
   \mathcal{I}^{(K)}_{\omega\ell m}(r) = 18 \sqrt{2} M r^2  \Biggl[2 \Lambda (r^2 \mu^2 + 2) + 4 \omega^2 r^2  - 2 f(r) \bigl(\Lambda + 2 + 5 \mu^2 r^2  \bigr)  +  \Lambda^2 + 4 + 4 \mu^2 r^2  + \mu^4 r^4 \Biggr],
\end{equation}
\end{fleqn}
\begin{multline}\label{J_K_S}
\hspace{-12pt}   \mathcal{J}^{(K)}_{\omega\ell m}(r) = 2 \sqrt{2} r^2 \Biggl[r \biggl( \Lambda (\mu^2 r^2 + 6 ) + 2 r^2 (9 \mu^2 + 8 \omega^2) + \Lambda^2 + 8 + 2 \mu^4 r^4 \biggr)  \\ 
   - 8 M \bigl(\Lambda + 2 + 5 r^2 \mu^2 \bigl) - 2 r f(r) \bigl(\Lambda +2 + 6 \mu^2 r^2  \bigr)\Biggr],
\end{multline}
and those of the source  \eqref{source_H_main} are given by
\begin{equation}\label{B_H_S}
\hspace{-393pt} \mathcal{B}^{(H)}_{\omega\ell m}(r) = \frac{(\Lambda + 2)}{r} \mathcal{B}^{(K)}_{\omega\ell m}(r),
\end{equation}
\begin{fleqn}
\begin{equation}\label{C_H_S}
   \mathcal{C}^{(H)}_{\omega\ell m}(r) =  - \frac{(\Lambda + 2)}{2} \mathcal{C}^{(K)}_{\omega\ell m}(r) + 18 i \sqrt{6} m  M r^4 \mu^2 \Bigl[\Lambda^2 + 4 + 4 \mu^2  r^2 + \mu^4 r^4 + 2 \Lambda (\mu^2 \, r^2 + 2) + 4 r^2 \omega^2 - 2 f(r) (\Lambda + 2 - \mu^2 \, r^2) \Bigr],
\end{equation}
\end{fleqn}
\begin{fleqn}
\begin{equation}\label{D_H_S}
   \mathcal{D}^{(H)}_{\omega\ell m}(r) =  - \frac{(\Lambda + 2)}{2} \mathcal{D}^{(K)}_{\omega\ell m}(r),
\end{equation}
\end{fleqn}
\begin{fleqn}
\begin{equation}\label{E_H_S}
   \mathcal{E}^{(H)}_{\omega\ell m}(r) =  - \frac{(\Lambda + 2)}{2} \mathcal{E}^{(K)}_{\omega\ell m}(r),
\end{equation}
\end{fleqn}
\begin{fleqn}
\begin{multline}\label{F_H_S}
   \mathcal{F}^{(H)}_{\omega\ell m}(r) =  -4 \sqrt{2} r (\Lambda + 2) \Biggl[-27 M^2 \Bigl(\Lambda^2 + 4 +\mu^4 r^4 + 2 \Lambda (\mu^2 r^2  + 2)\\ 
+ 4 r^2 (\mu^2 + \omega^2)\Bigr) - 2 r^2 \Bigl(\Lambda^2 + \Lambda (6 \mu^2 r^2 + 2) + r^2 (8 \mu^2 + 12 \omega^2 + 5 \mu^4 r^2)\Bigr) \\ + 4 f(r)\Biggl( 27 M^2 \biggl(\Lambda^2 + 4 + 2 \mu^4 r^4  + \Lambda (3  \mu^2 r^2 + 4) 
+ 6 r^2 (\mu^2 + \omega^2)\biggr)-5 r^4 \mu^2 \Biggr)  - 54 M^2 f(r)^2 (3 \Lambda  + 6 + \mu^2 r^2 ) \Biggr],
\end{multline}
\end{fleqn}
\begin{fleqn}
\begin{equation}\label{I_H_S}
   \mathcal{I}^{(H)}_{\omega\ell m}(r) =  - \frac{(\Lambda + 2)}{2} \mathcal{I}^{(K)}_{\omega\ell m}(r),
\end{equation}
\end{fleqn}
\begin{equation}\label{J_H_S}
\hspace{-65pt}   \mathcal{J}^{(H)}_{\omega\ell m}(r) =  - \frac{(\Lambda + 2)}{2} \mathcal{J}^{(K)}_{\omega\ell m}(r) + \sqrt{2} \, r^3 \, (\Lambda + 2) \Bigg[ r^2 \Bigl((\Lambda + 2) \, \mu^2 + 3 \, r^2 \, \mu^4 + 4 \, \omega^2\Bigr) + 2 \, f(r)  (\Lambda + 2 - r^2 \, \mu^2) \Bigg],
\end{equation}
\end{widetext}

\subsubsection{Even-parity monopole mode}
\label{appendix_A_4_1}

For the monopole $\ell=0$, the vector and tensor spherical harmonics vanish, so that the components $H^{\ell m}_{t}$, $H^{\ell m}_{r}$, and $G^{\ell m}$ of the perturbation tensor \eqref{h_even} are undefined. There remain the components $K^{\ell m}$ and $H^{\ell m}_{tr}$, which satisfy a pair of coupled of differential equations \eqref{l_0_even_freq_1} and \eqref{l_0_even_freq_2}, where the coefficients $ \alpha^{(K)}_i$ are given by
\begin{equation}\label{alpha_K_1_l_0}
\alpha^{(K)}_1(r)  = \frac{2 f(r) (r-M)}{r^2},
\end{equation}
\begin{equation}\label{alpha_K_2_l_0}
\alpha^{(K)}_2(r)  = \omega^2 - f(r)\left(\frac{6}{r^2 }- \frac{20M}{r^3} + \mu^2\right),
\end{equation}
\begin{equation}\label{alpha_K_3_l_0}
\alpha^{(K)}_3(r)  = \frac{2i f(r)^3}{\omega r^2},
\end{equation}
\begin{equation}\label{alpha_K_4_l_0}
\alpha^{(K)}_4(r)  = \frac{4 i f(r)^2 (r-M)}{\omega r^4},
\end{equation}
and the coefficients $\alpha^{(H)}_i$ given by
\begin{equation}\label{alpha_Htr_1_l_0}
\alpha^{(H)}_1(r)  = \frac{2 f(r) (r+M)}{r^2},
\end{equation}
\begin{equation}\label{alpha_Htr_2_l_0}
\alpha^{(H)}_2(r)  = \omega^2 - f(r)\left(\frac{2(r-3M)}{(r-2M)r^2} -\frac{6M}{r^3} + \mu^2\right),
\end{equation}
\begin{equation}\label{alpha_Htr_3_l_0}
\alpha^{(H)}_3(r)  = 0,
\end{equation}
\begin{equation}\label{alpha_Htr_4_l_0}
\alpha^{(H)}_4(r)  = \frac{4 i M \omega}{ r^2}.
\end{equation}

The source terms in~\eqref{l_0_even_freq_1} and \eqref{l_0_even_freq_2} are expressed in terms of the components of the stress-energy tensor $\mathcal{T}^{\text{(e)}}_{\mu\nu}$
\begin{widetext}
\begin{fleqn}
\begin{multline}\label{S_K_monopole}
S^{(K)}_{\omega\ell m}(r) = \xi^{(K)}_1(r) \frac{\partial T^{\ell m}_{tt}(r,\omega)}{\partial r} +  \xi^{(K)}_2(r)   T^{\ell m}_{tt}(r,\omega) +  \xi^{(K)}_3(r) \frac{\partial T^{\ell m}_{rr}(r,\omega)}{\partial r} \\ +  \xi^{(K)}_4(r)   T^{\ell m}_{rr}(r,\omega) + \xi^{(K)}_5(r) \frac{\partial T^{\ell m}_{1}(r,\omega)}{\partial r} +  \xi^{(K)}_6(r)   T^{\ell m}_{1}(r,\omega),
\end{multline}
\end{fleqn}
\begin{fleqn}
\begin{multline}\label{S_H_monopole}
S^{(H_{tr})}_{\omega\ell m}(r) = \xi^{(H_{tr})}_1(r) \frac{\partial T^{\ell m}_{tt}(r,\omega)}{\partial r} +  \xi^{(H_{tr})}_2(r)   T^{\ell m}_{tt}(r,\omega) +  \xi^{(H_{tr})}_3(r) \frac{\partial T^{\ell m}_{rr}(r,\omega)}{\partial r} \\ +  \xi^{(H_{tr})}_4(r)   T^{\ell m}_{rr}(r,\omega) + \xi^{(H_{tr})}_5(r) \frac{\partial T^{\ell m}_{1}(r,\omega)}{\partial r} +  \xi^{(H_{tr})}_6(r)   T^{\ell m}_{1}(r,\omega) +  \xi^{(H_{tr})}_7(r)   T^{\ell m}_{tr}(r,\omega),
\end{multline}
\end{fleqn}
\end{widetext}
where 
\begin{equation}\label{X_1_K_0}
\xi^{(K)}_1(r) = \frac{16 \pi f(r)}{3\mu^2 r},
\end{equation}
\begin{equation}\label{X_2_K_0}
\xi^{(K)}_2(r) = -\frac{16 \pi}{3},
\end{equation}
\begin{equation}\label{X_3_K_0}
\xi^{(K)}_3(r) = -\frac{16 \pi f(r)^3}{3 \mu^2 r },
\end{equation}
\begin{equation}\label{X_4_K_0}
\xi^{(K)}_4(r) = 16 \pi f(r)^2 \left( \frac{1}{3} - \frac{4 M}{3 \mu^2 r^3} \right),
\end{equation}
\begin{equation}\label{X_5_K_0}
\xi^{(K)}_5(r) = -\frac{32 \pi f(r)^2}{3 r^3 \mu^2},
\end{equation}
\begin{equation}\label{X_6_K_0}
\xi^{(K)}_6(r) = -16 \pi f(r) \left( \frac{12 M - 4 r + r^3 \mu^2}{3 r^5 \mu^2} \right),
\end{equation}
and 
\begin{equation}\label{X_1_Htr_0}
\xi^{(H_{tr})}_1(r) = -\frac{16 i \pi \omega}{3 \mu^2},
\end{equation}
\begin{equation}\label{X_2_Htr_0}
\xi^{(H_{tr})}_2(r) = \frac{16 i M \pi \omega}{3 \mu^2 f(r) r^2},
\end{equation}
\begin{equation}\label{X_3_Htr_0}
\xi^{(H_{tr})}_3(r) = \frac{16 i \pi \omega f(r)^2}{3 \mu^2},
\end{equation}
\begin{equation}\label{X_4_Htr_0}
\xi^{(H_{tr})}_4(r) = \frac{16 i M \pi \omega f(r)}{r^2 \mu^2},
\end{equation}
\begin{equation}\label{X_5_Htr_0}
\xi^{(H_{tr})}_5(r) = \frac{32 i \pi \omega f(r)}{3 r^2 \mu^2},
\end{equation}
\begin{equation}\label{X_6_Htr_0}
\xi^{(H_{tr})}_6(r) =  \frac{32 i \pi (5 M - 2 r) \omega}{3 r^4 \mu^2},
\end{equation}
\begin{equation}\label{X_7_Htr_0}
\xi^{(H_{tr})}_7(r) = -16 \pi f(r).
\end{equation}

\subsubsection{Even-parity dipole mode}
\label{appendix_A_4_2}

In the case of the even-particle dipole mode  ($\ell = 1$), the system of Eqs. \eqref{Système_even_freq_1}--\eqref{Système_even_freq_3} reduces to two equations, \eqref{Système_even_freq_1_bis} and \eqref{Système_even_freq_2_bis}.

\section{Quasinormal modes and quasibound states: A numerical resolution}
\label{appendix_B}

\subsection{Matrix-valued Hill determinant method}

We present the numerical method used to compute the resonance spectra for both odd and even parities. Various numerical techniques for calculating the quasinormal and quasibound frequency spectra of black holes have been discussed in the literature (see Ref.\cite{Berti:2004md} for an overview and Ref.\cite{Kokkotas:1999bd} for further details). Among these, the \textit{continued-fraction method} introduced by Leaver~\cite{Leaver:1985ax,leaver1986solutions,Leaver:1990zz} remains the most widely used for determining black hole resonance spectra, including its matrix-valued version, which is applied to coupled systems \cite{Rosa:2011my,Pani:2012bp,Brito:2013wya}. However, we use the \textit{Hill determinant method} \cite{mp}, along with its matrix-valued extension for coupled systems. The Hill determinant method has already proven to be effective in calculating the Regge pole spectrum of black holes and compact objects  (see also Refs.~\cite{Folacci:2018sef,Folacci:2019cmc,Folacci:2019vtt,Folacci:2021uld,OuldElHadj:2019kji,Torres:2022fyf,OuldElHadj:2023anu}).

To implement this approach, an appropriate ansatz is required for each parity. For the odd-parity case, we seek the resonant mode solutions satisfying the correct boundary conditions (see Sec.~\ref{sec_5_1}) in the form
\begin{multline}\label{ansatz_odd}
\mathcal{Q}_j^{(o)} (r)= e^{-2iM\bigl[p(\omega)+\omega\bigr]} \left(\frac{r}{2M}\right)^{i\left[2M p(\omega)+\frac{M\mu^2}{p(\omega)}\right]}\\
\times  f(r)^{-2iM\omega} e^{i p(\omega) r_*} \sum_{n=0}^{+\infty} a_n^{(j)}f(r)^n
\end{multline}
where $p(\omega)$  is defined by the expression given below Eq. \eqref{Asymp_Behavior}. It is important to note that the quasinormal frequency spectrum lies on the lower part of the first Riemann sheet associated with the function $p(\omega)$, while the quasibound frequency spectrum is located on the second Riemann sheet associated with $p(\omega)$ . In the latter case, it is necessary to use $-p(\omega)$ instead of $p(\omega)$ (for more details, see Ref.~\cite{Decanini:2015yba}).

For the even-parity case, the situation is slightly different, since the different equations of the system exhibit different behaviors at the boundary conditions (see Sec.~\ref{sec_5_2}). Therefore, it is necessary to introduce ansatz that correctly satisfies these conditions. Hence we have
\begin{multline}\label{ansatz_even_1}
\mathcal{Q}_j^{(e,1)} (r)= e^{-2iM\bigl[p(\omega)+\omega\bigr]} \left(\frac{r}{2M}\right)^{i\left[2M p(\omega)+\frac{M\mu^2}{p(\omega)}\right]-1}\\
\times  f(r)^{-2iM\omega} e^{i p(\omega) r_*} \sum_{n=0}^{+\infty} a_n^{(j)}f(r)^n
\end{multline}
and
\begin{multline}\label{ansatz_even_2}
\mathcal{Q}_j^{(e,2)} (r)= e^{-2iM\bigl[p(\omega)+\omega\bigr]} \left(\frac{r}{2M}\right)^{i\left[2M p(\omega)+\frac{M\mu^2}{p(\omega)}\right]-1}\\
\times  f(r)^{-2iM\omega-1} e^{i p(\omega) r_*} \sum_{n=0}^{+\infty} a_n^{(j)}f(r)^n
\end{multline}

\subsection{Odd-parity modes}

\subsubsection{Odd-parity dipole mode $(\ell = 1)$}

The dipole mode $ \ell = 1$ is governed by homogeneous Eq.~\eqref{dipole_odd_parity}. Inserting ansatz \eqref{ansatz_odd} in this equation then leads to the three-term recurrence relation
\begin{align}\label{three-term_rec_dipole_odd}
\alpha_0 a_1 +\beta_0 a_0  =& 0, \nonumber\\
\alpha_n a_{n+1} + \beta_n a_{n} +\gamma_n a_{n-1}=& 0, \qquad \forall n\geq 1
\end{align}
where
\begin{subequations}
\begin{align}
  \alpha_n &= (n + 1)(n + 1 - 4 i M \omega), \\
  \beta_n  &= \frac{M\bigl(p(\omega) + \omega\bigr)}{p(\omega)} \Bigl[4M\bigl(p(\omega) + \omega\bigr)^2 \nonumber \\
              & + i (2n + 1) \bigl(3 p(\omega) + \omega\bigr)\Bigr]- 2 n (n + 1) + 2\\
  \gamma_n &= \Biggl[n - \frac{i M \bigl(p(\omega) + \omega\bigr)^2}{p(\omega)}\Biggr]^2 - 9.
\end{align}
\end{subequations}

We solve the recurrence relation \eqref{three-term_rec_dipole_odd} using the Hill determinant method, and nontrivial solutions arise when the Hill determinant is zero,
\begin{equation}
\label{Determinant_Hill_3_termes}
\small
D   =  \begin{vmatrix}
\beta_0 &  \alpha_0 &  0 & 0 & 0 &  \ldots  & \ldots  &  \ldots \\
\gamma_1 & \beta_1 & \alpha_1 & 0  &  0  &  \ldots  &  \ldots  &  \ldots  \\
0      &  \gamma_2 &  \beta_2  &  \alpha_2 &  0  &  \ldots  &  \ldots  &  \ldots  \\
\vdots & \ddots & \ddots  &  \ddots  & \ddots  &  \ddots  &  \ldots  &  \ldots \\
\vdots & \vdots & \ddots      & \gamma_{n-1} & \beta_{n-1} & \alpha_{n-1} & \ddots & \ldots \\
\vdots & \vdots & \vdots & 0 & \gamma_{n} & \beta_{n} & \alpha_{n} & \ddots   \\
\vdots & \vdots & \vdots & \vdots & \ddots & \ddots & \ddots & \ddots
\end{vmatrix} = 0.
\end{equation}
Considering $D_n$ as the determinant of the $n \times n$ submatrix of $D$,
\begin{equation}
\label{det_recurrence_3_termes}
D_n=\beta_n D_{n-1} - \gamma_{n}\alpha_{n-1}D_{n-2},
\end{equation}
with the initial conditions
\begin{equation}
\label{Determinant_initial_conds}
\begin{split}
D_0 &=\beta_0, \\
D_1 &=\beta_1\beta_0-\gamma_1\alpha_0,
\end{split}
\end{equation}
or, equivalently
\begin{align}
\label{recurrence_Hill_RN_4_termes}
D_n    & =\left(\prod_{k=1}^{n+1} k^2\right) P_{n+1} \nonumber\\
       &  = 1\times 2^2 \ldots (n-2)^2(n-1)^2n^2(n+1)^2 P_{n+1},
\end{align}
where
\begin{eqnarray}
\label{recurrence_Hill_RN_4_termes_bis}
P_n=
&&\frac{\beta_{n-1}}{n^2} P_{n-1}-\frac{\gamma_{n-1}}{n^2}\,\frac{\alpha_{n-2}}{(n-1)^2}P_{n-2}
\end{eqnarray}
with the following initial conditions
 \begin{equation}
 \begin{split}
    P_0&= 1,\\
    P_1&=\beta_0
 \end{split}
 \end{equation}

\subsubsection{Odd-parity  modes $(\ell \geq 2)$}

The odd-parity modes for $\ell \geq $ are governed by a pair of coupled homogeneous differential equations, Eqs.~\eqref{Système_odd_freq_1} and \eqref{Système_odd_freq_2}. Substituting the ansatz \eqref{ansatz_odd} into this system leads to a matrix-valued three-term recurrence relation,
\begin{align}\label{matrix_three-term_rec_odd}
\boldsymbol{\alpha_0 U_1 + \beta_0 U_0 } = & 0, \nonumber\\
\bm{\alpha_n} \mathbf{U_{n+1} }+ \bm{\beta_n} \mathbf{U_{n}} +\bm{\gamma_n} \mathbf{U_{n-1}}=& 0, \qquad \forall n\geq 1
\end{align}
where $ \mathbf{U_n}= \begin{pmatrix} a_n^{(\phi)} & a_n^{(\psi)}  \end{pmatrix}^\top$ and the matrix coefficients are given by
\begin{eqnarray}
\boldsymbol{\alpha_n} & =& \begin{pmatrix} 
\alpha_n^{(\phi)} & 0 \\ 
0 & \alpha_n^{(\psi)} 
\end{pmatrix}, \quad
\boldsymbol{\beta_n} = \begin{pmatrix} 
\beta_n^{(\phi)} & \beta_n^{(\psi\phi)} \\ 
\beta_n^{(\phi\psi)} & \beta_n^{(\psi)} 
\end{pmatrix} \nonumber \\
\boldsymbol{\gamma_n} &=& \begin{pmatrix} 
\gamma_n^{(\phi)} & \gamma_n^{(\psi\phi)} \nonumber\\ 
0 & \gamma_n^{(\psi)} 
\end{pmatrix}
\end{eqnarray}
where
\begin{subequations}
\begin{align}
  \alpha_n^{(\phi)} &= (n + 1)(n + 1 - 4 i M \omega), \\
  \beta_n^{(\phi)} &= \frac{M\bigl(p(\omega) + \omega\bigr)}{p(\omega)} \Bigl[4M\bigl(p(\omega) + \omega\bigr)^2 \nonumber \\
                   &\hspace{20pt}+ i (2n + 1) \bigl(3 p(\omega) + \omega\bigr)\Bigr] \nonumber \\
                   &\hspace{60pt}- 2 n (n + 1) -(\Lambda-2),\\
  \gamma_n^{(\phi)} &= \Biggl[n - \frac{i M \bigl(p(\omega) + \omega\bigr)^2}{p(\omega)}\Biggr]^2 - 9,\\
  \beta_n^{(\psi\phi)}& =-\frac{\Lambda}{2},\\
  \delta_n^{(\psi\phi)}& =\frac{3 \Lambda}{2},\\
\end{align}
\end{subequations}
and
\begin{subequations}
\begin{align}
  \alpha_n^{(\psi)} &=  \alpha_n^{(\phi)} , \\
  \beta_n^{(\psi)} &=  \beta_n^{(\phi)} - 3,\\
  \gamma_n^{(\psi)} &= \gamma_n^{(\phi)} + 9,\\
  \beta_n^{(\phi\psi)}& = 4.
\end{align}
\end{subequations}

The matrix-valued three-term recurrence relation \eqref{matrix_three-term_rec_odd} can be solved using the matrix-valued Hill determinant. The nontrivial solutions arise when
\begin{equation}
\label{Determinant_Hill_3_termes}
\small
D   =  \begin{vmatrix}
\bm{\beta_0} &  \bm{\alpha_0} &  0 & 0 & 0 &  \ldots  & \ldots  &  \ldots \\
\bm{\gamma_1} & \bm{\beta_1} & \bm{\alpha_1} & 0  &  0  &  \ldots  &  \ldots  &  \ldots  \\
0      & \bm{ \gamma_2} &  \bm{\beta_2}  &  \bm{\alpha_2} &  0  &  \ldots  &  \ldots  &  \ldots  \\
\vdots & \ddots & \ddots  &  \ddots  & \ddots  &  \ddots  &  \ldots  &  \ldots \\
\vdots & \vdots & \ddots      & \bm{\gamma_{n-1}} & \bm{\beta_{n-1}} & \bm{\alpha_{n-1}} & \ddots & \ldots \\
\vdots & \vdots & \vdots & 0 & \bm{\gamma_{n}} & \bm{\beta_{n}} & \bm{\alpha_{n}} & \ddots   \\
\vdots & \vdots & \vdots & \vdots & \ddots & \ddots & \ddots & \ddots
\end{vmatrix} = 0.
\end{equation}

The Hill determinant $D$ of a block tridiagonal matrix of dimension  $n \times n$, where each element $\bm{\beta_n}$, $\bm{\gamma_n}$, and $\bm{\alpha_n}$ is a square matrix of dimension $2 \times 2$, can be computed using LU factorization and the Schur complement~\cite{SALKUYEH2006442,MOLINARI20082221}. This leads to a recursive relation:
\begin{equation}\label{Sous_det}
  \mathbf{D_k} = \boldsymbol{\beta_k} - \boldsymbol{\gamma_k} \mathbf{D_{k-1}^{-1}}  \boldsymbol{\alpha_{k-1}}
\end{equation}
starting with 
\begin{equation}\label{conditon_Sous_det}
  \mathbf{D_0}  = \boldsymbol{\beta_0}
\end{equation}
This recursive approach is continued until all the blocks have been processed, and the Hill determinant is obtained as the product of the determinants at each step
\begin{equation}\label{Det_Hill_Matrix}
  D = \prod_{k=1}^{n} \det(\mathbf{D_k})  
\end{equation}
Alternatively, the normalized form can be used
\begin{equation}\label{Sous_det_normalization}
  \mathbf{P_k} = \frac{\bm{\beta_{k-1}}}{k^2} - \frac{\bm{\alpha_{k-1}}}{k^2} \mathbf{P_{k-1}^{-1}}  \frac{\bm{\alpha_{k-2}}}{(k-1)^2}
\end{equation}
with the initial condition 
\begin{equation}\label{conditon_Sous_det_normalization}
  \mathbf{P_1}  = \bm{\beta_0}
\end{equation}
The normalized Hill determinant is then given by
\begin{equation}\label{Det_Hill_Matrix_norm}
  \widetilde{D} = \prod_{k=1}^{n} \det(\mathbf{P_k})  
\end{equation}

\subsection{Even-parity modes}

\subsubsection{Even-parity  monopole mode $(\ell  = 0)$}

The monopole mode ($\ell = 0$) is governed by a pair of coupled homogeneous differential equations, \eqref{l_0_even_freq_1}  and \eqref{l_0_even_freq_2} . By inserting the ansatz \eqref{ansatz_even_1} into Eq. \eqref{l_0_even_freq_1}  and \eqref{ansatz_even_2} into Eq. \eqref{l_0_even_freq_2} , this leads to a matrix-valued six-term recurrence relation
\begin{multline}\label{matrix_six-term_rec_even_monopole}
\bm{\alpha_n} \mathbf{U_{n+1} } + \bm{\beta_n} \mathbf{U_{n}} +\bm{\gamma_n} \mathbf{U_{n-1}} 
  + \bm{\delta_n} \mathbf{U_{n-2}} \\ +\bm{\epsilon_n} \mathbf{U_{n-3}} + \bm{\eta_n} \mathbf{U_{n-4}}= 0, \quad \forall n\geq 4
\end{multline}
where $ \mathbf{U_n}= \begin{pmatrix} a_n^{(K)} & a_n^{(H)}  \end{pmatrix}^\top$ and the matrix coefficients are given by
\begin{eqnarray}\label{coeffs_monopole_matrix}
\boldsymbol{\alpha_n} & = & \begin{pmatrix} 
\alpha_n^{(K)} & 0 \\ 
0 & \alpha_n^{(H)} 
\end{pmatrix}, \quad
\hspace{0.55cm} \boldsymbol{\beta_n} = \begin{pmatrix} 
\beta_n^{(K)} & \beta_n^{(HK)} \\ 
\beta_n^{(KH)} & \beta_n^{(H)} 
\end{pmatrix}, \nonumber \\
\boldsymbol{\gamma_n} & = & \begin{pmatrix} 
\gamma_n^{(K)} & \gamma_n^{(HK)} \\ 
\gamma_n^{(KH)} & \gamma_n^{(H)} 
\end{pmatrix}, \quad
\hspace{0.2cm} \boldsymbol{\delta_n} = \begin{pmatrix} 
\delta_n^{(K)} & \delta_n^{(HK)} \\ 
\delta_n^{(KH)} & \delta_n^{(H)} 
\end{pmatrix}, \nonumber \\
\boldsymbol{\epsilon_n} & = & \begin{pmatrix} 
\epsilon_n^{(K)} & \epsilon_n^{(HK)} \\ 
0 & \epsilon_n^{(H)} 
\end{pmatrix}, \quad
\hspace{0.55cm} \boldsymbol{\eta_n} = \begin{pmatrix} 
0 & \eta_n^{(HK)} \\ 
0 & 0 
\end{pmatrix}.
\end{eqnarray}
To simplify the notation, $H_{tr}$ is denoted by $H$ in \eqref{coeffs_monopole_matrix}. The expressions for the coefficients are long. They can be provided upon request.

The matrix-valued six-term recurrence relation \eqref{matrix_six-term_rec_even_monopole} can be solved using matrix-valued Hill determinant. The nontrivial solutions arise when
\begin{equation}\label{six_term_monopole}
\small
\hspace{-10pt}D  = \begin{vmatrix}
\bm{\beta_0} & \bm{\alpha_0} & 0 & 0 & 0 & 0 & 0 & \cdots \\
\bm{\gamma_1} & \bm{\beta_1} & \bm{\alpha_1} & 0 & 0 & 0 & 0 & \cdots \\
\bm{\delta_2} & \bm{\gamma_2} & \bm{\beta_2} & \bm{\alpha_2} & 0 & 0 & 0 & \cdots\\
\bm{\epsilon_3} & \bm{\delta_3} & \bm{\gamma_3} & \bm{\beta_3} & \bm{\alpha_3} & 0 & 0 & \cdots \\
\bm{\eta_4} & \bm{\epsilon_4} & \bm{\delta_4} & \bm{\gamma_4} & \bm{\beta_4} & \bm{\alpha_4} & 0 & \cdots \\
0 & \bm{\eta_5} & \bm{\epsilon_5} & \bm{\delta_5} & \bm{\gamma_5} & \bm{\beta_5} & \bm{\alpha_5} & \cdots \\
\vdots & \ddots & \ddots & \ddots & \ddots & \ddots & \ddots &\ddots \\
\vdots & \vdots & \vdots &  \bm{\eta_{n-1}} & \bm{\epsilon_{n-1}} & \bm{\delta_{n-1}} & \bm{\gamma_{n-1}} & \ddots \\
\vdots & \vdots & \vdots & 0 &  \bm{\eta_n} & \bm{\epsilon_n} & \bm{\delta_n} & \bm{\gamma_n}\\
\vdots & \vdots & \vdots & \vdots &  \ddots & \ddots & \ddots & \ddots
\end{vmatrix} = 0.
\end{equation}

To compute the Hill determinant of a block band matrix of dimension $n\times n$ with upper and lower bandwidths of $1$ and $4$, respectively, where each element $\bm{\beta_n}$, $\bm{\gamma_n}$, $\bm{\delta_n}$, $\bm{\epsilon_n}$, $\bm{\eta_n}$, and $\bm{\alpha_n}$ is a square matrix of dimension $m \times m$ , we generalized the procedure used for a block tridiagonal matrix. Specifically, we applied LU factorization and the Schur complement, resulting in a recursive relation
\begin{equation}\label{Sous_det_dipole}
\hspace{-6pt}\mathbf{D_k} = \bm{\beta_k} + \sum_{j=1}^{4} (-1)^{j} \left[ \bm{\gamma^{(j)}_k} \prod_{i=0}^{j-1} \left( \mathbf{D_{k-(j-i)}^{-1}} \bm{\alpha_{k-(j-i)}} \right) \right]
\end{equation}
where $\bm{\gamma^{(1)}_k} = \bm{\gamma_k}$, $\bm{\gamma^{(2)}_k} = \bm{\delta_k}$, $\bm{\gamma^{(3)}_k} = \bm{\epsilon_k}$, and $\bm{\gamma^{(4)}_k} = \bm{\eta_k}$. Explicitly expanded, and keeping in mind that we are dealing with matrix products, this expression becomes
\begin{fleqn}
\begin{eqnarray}
   \mathbf{D_k}&& = \bm{\beta_k} - \bm{\gamma_k} \mathbf{D_{k-1}^{-1}} \bm{\alpha_{k-1}} \nonumber\\
    & & +\bm{\delta_k} \mathbf{D_{k-2}^{-1}}  \bm{\alpha_{k-2}} \mathbf{D_{k-1}^{-1}}  \bm{\alpha_{k-1}} \nonumber\\
    & & -\bm{\epsilon_k}  \mathbf{D_{k-3}^{-1}}  \bm{\alpha_{k-3}} \mathbf{D_{k-2}^{-1}}  \bm{\alpha_{k-2}} \mathbf{D_{k-1}^{-1}}  \bm{\alpha_{k-1}}\nonumber\\
    & & +\bm{\eta_k} \mathbf{D_{k-4}^{-1}}  \bm{\alpha_{k-4}} \mathbf{D_{k-3}^{-1}}  \bm{\alpha_{k-3}} \mathbf{D_{k-2}^{-1}}  \bm{\alpha_{k-2}} \mathbf{D_{k-1}^{-1}}  \bm{\alpha_{k-1}} \nonumber\\
\end{eqnarray}
\end{fleqn}
with the initial conditions 
\begin{align}
\mathbf{D_0} &= \bm{\beta_0}, \\
\mathbf{D_1} &= \bm{\beta_1} - \bm{\gamma_1} \mathbf{D_0}^{-1} \bm{\alpha_0}, \\
\mathbf{D_2} &= \bm{\beta_2} - \bm{\gamma_2} \mathbf{D_1}^{-1} \bm{\alpha_1} + \bm{\delta_2} \mathbf{D_0}^{-1} \bm{\alpha_0} \mathbf{D_1}^{-1} \bm{\alpha_1}, \\
\mathbf{D_3} &= \bm{\beta_3} - \bm{\gamma_3} \mathbf{D_2}^{-1} \bm{\alpha_2} + \bm{\delta_3} \mathbf{D_1}^{-1} \bm{\alpha_1} \mathbf{D_2}^{-1} \bm{\alpha_2} \nonumber\\
             & - \bm{\epsilon_3} \mathbf{D_0}^{-1} \bm{\alpha_0} \mathbf{D_1}^{-1} \bm{\alpha_1} \mathbf{D_2}^{-1} \bm{\alpha_2}.
\end{align}
Hence, the Hill determinant given by Eq.~\eqref{six_term_monopole} is expressed as
\begin{equation}\label{Det_Hill_Matrix_monopole} 
 D = \prod_{k=1}^{n} \det(\mathbf{D_k})
\end{equation}
Similarly, the normalized form is expressed as
\begin{align}\label{Sous_det_dipole_norm}
&\hspace{-10pt}\mathbf{P_k} = \frac{\bm{\beta_{k-1}}}{k^2} \nonumber\\
&\hspace{-9pt} + \sum_{j=1}^{4} (-1)^{j} \left[ \frac{\bm{\gamma^{(j)}_{k-1}}}{k^2} \prod_{i=0}^{j-1} \left( \mathbf{P_{k-(j-i)}^{-1}} \frac{\bm{\alpha_{k-(j-i)-1}}}{(k-(j-i))^2} \right) \right]
\end{align}
and the corresponding normalized Hill determinant is
\begin{equation}\label{Det_Hill_Matrix_norm_monopole}
  \widetilde{D} = \prod_{k=1}^{n} \det(\mathbf{P_k})  
\end{equation}

\subsubsection{Even-parity  dipole mode $(\ell  = 1)$}

The even-parity dipole mode $\ell  = 1 $  is described by a pair of coupled homogeneous differential equations, Eqs.\eqref{Système_even_freq_1_bis} and \eqref{Système_even_freq_2_bis}. Inserting the ansatz \eqref{ansatz_even_1} into Eq.\eqref{Système_even_freq_1_bis} and \eqref{ansatz_even_2}  into Eq.~\eqref{Système_even_freq_2_bis} yields a matrix-valued twelve-term recurrence relation,
\begin{align}\label{matrix_twelve-term_rec_even_dipole}
\bm{\alpha_n} \mathbf{U_{n+1} } &+ \bm{\beta_n} \mathbf{U_{n}} +\bm{\gamma_n} \mathbf{U_{n-1}} + \bm{\delta_n} \mathbf{U_{n-2}}\nonumber \\
&+\bm{\epsilon_n} \mathbf{U_{n-3}} + \bm{\eta_n} \mathbf{U_{n-4}} + \bm{\theta_n} \mathbf{U_{n-5}}\nonumber\\ 
&+\bm{\kappa_n} \mathbf{U_{n-6}}+ \bm{\lambda_n} \mathbf{U_{n-7}} + \bm{\mu_n} \mathbf{U_{n-8}}\nonumber\\
&+\bm{\nu_n} \mathbf{U_{n-9}} + \bm{\rho_n} \mathbf{U_{n-10}}= 0, \quad \forall n\geq 10 
\end{align}
where $ \mathbf{U_n}= \begin{pmatrix} a_n^{(K)} & a_n^{(H_r)} \end{pmatrix}^\top$.  The $2\times 2$ matrix coefficients and their coefficient expressions are long. They can be provided upon request.

The matrix-valued twelve-term recurrence relation \eqref{matrix_twelve-term_rec_even_dipole} can be solved using the matrix-valued Hill determinant, with nontrivial solutions arising when this determinant equals zero. This Hill determinant corresponds to a block band matrix of dimension $n \times n$ with upper and lower bandwidths of $1$ and $10$, respectively. It can be computed, \textit{mutatis mutandis}, using \eqref{Sous_det_dipole} and \eqref{Det_Hill_Matrix_monopole}, by summing over $j$ from $1$ to $10$. Alternatively, the normalized form given in \eqref{Sous_det_dipole_norm} and \eqref{Det_Hill_Matrix_norm_monopole} can also be applied.

\subsubsection{Even-parity  modes $(\ell  \geq 2)$}

The even-parity modes for $\ell \geq 2$ are governed by three coupled homogeneous differential equations, eqs. \eqref{Système_even_freq_1}, \eqref{Système_even_freq_2}, and \eqref{Système_even_freq_3}. By inserting the ansatz \eqref{ansatz_even_1} into Eqs.\eqref{Système_even_freq_1} and \eqref{Système_even_freq_3}, as well as ansatz \eqref{ansatz_even_2} into Eq.~\eqref{Système_even_freq_2}, also leads to a matrix-valued twelve-term recurrence relation with a vectorial coefficient  
\begin{equation}
\mathbf{U_n} = \begin{pmatrix} a_n^{(K)} \\ a_n^{(H_r)} \\ a_n^{(G)}  \end{pmatrix}
\end{equation}

The $3\times 3$ matrix coefficients have long expressions that can be provided upon request. \textit{Mutatis mutandis}, the resolution procedure is exactly the same as that used for the dipole mode, and the zeros of the matrix-valued Hill determinant correspond to the spectra of the quasinormal and quasibound frequencies.

\section{Regularization of even-parity partial wave amplitudes}
\label{appendix_C}

In this appendix, we explain the regularization of the partial amplitudes for even-parity modes. Indeed, the exact waveforms given in \eqref{Sol_Even_l_0_Coupled_inf_frequency}, \eqref{Sol_Even_l_1_Coupled_inf_frequency}, and \eqref{Sol_Even_l_2_Coupled_inf_frequency}, expressed as integrals over the radial Schwarzschild coordinate, exhibit strong divergences near the ISCO. This divergence arises from the behavior of the sources \eqref{source_K_monopole_even} and \eqref{source_H_monopole_even} for  $\ell =0$, and \eqref{source_K_main} and \eqref{source_H_main} for $\ell \geq 1$ as $r \to 6M$.

As we have already encountered this type of divergence in the context of general relativity—specifically in the study of gravitational waves generated by a massive particle plunging into a Schwarzschild black hole \cite{Folacci:2018cic}, and also in the case of electromagnetic waves generated by a charged particle plunging into the same black hole \cite{Folacci:2018vtf}—we will simply outline the key points of how to apply the regularization algorithm using successive integrations by parts. The reader is encouraged to refer to the article where the electromagnetic case is treated in detail. As with electromagnetic perturbations, we reduced the degree of divergence of the integrals through successive integrations by parts and then numerically regularized them using Levin’s algorithm.

We will first regularize the monopole ($\ell = 0$) and dipole ($\ell = 1$) modes, separating them from the higher modes $\ell \geq 2$. The monopole and dipole are each governed by a pair of coupled differential equations, while the modes for $\ell \geq 2$ are governed by three coupled differential equations. For the monopole mode ($\ell = 0$) described by \eqref{Sol_Even_l_0_Coupled_inf_frequency},  the amplitudes are given by $\Psi^{(1)} = K$ and $\Psi^{(2)} = H_{tr}$, while for the dipole mode ($\ell = 1$) described by \eqref{Sol_Even_l_1_Coupled_inf_frequency}, we have $\Psi^{(1)} = K$ and $\Psi^{(2)} = H_r$. In both cases, the partial amplitudes can be rewritten as  
\begin{equation}\label{Psi_i_unified_l_0_1}
  \hspace{-10pt}\bm{\Psi}_{\omega\ell m }(r) \equiv \begin{pmatrix} \Psi^{(1)} \\ \Psi^{(2)} \end{pmatrix} = \int_{-\infty}^{+\infty} dr'_* \bm{\mathcal{G}}(r_*,r'_*) \bm{\mathcal{S}}_{\omega\ell m}(r'_*)
\end{equation}
where $\bm{\mathcal{S}}_{\omega\ell m}$ is the source vector, and  \( \bm{\mathcal{G}}(r_*,r'_*) \) is the \( 2 \times 2 \) Green’s matrix
\begin{equation}
  \bm{\mathcal{G}}(r_*,r'_*)  = \begin{pmatrix}
    \mathcal{G}_{11}(r_*,r'_*) & \mathcal{G}_{12}(r_*,r'_*) \\
    \mathcal{G}_{21}(r_*,r'_*) & \mathcal{G}_{22}(r_*,r'_*) 
  \end{pmatrix}
\end{equation}
The components $\Psi^{(1)}$ and $\Psi^{(2)}$ can therefore be written explicitly as
\begin{equation}\label{Psi_i_unified}
  \Psi_{\omega\ell m}^{(i)}(r) = \int_{-\infty}^{+\infty} dr'_* \sum_{j=1}^{2} \mathcal{G}_{ij}(r_*,r'_*) S_{\omega\ell m}^{(j)}(r'_*)
\end{equation}
with \( i, j = 1, 2 \).

Before proceeding with the regularization, we will first rewrite the source terms  $ S_{\omega\ell m}^{(j)}$  in \eqref{Psi_i_unified} into a more tractable form, which allow us to regularize the partial amplitudes. We have
\begin{equation}\label{Source_i_unified}
  S_{\omega\ell m}^{(j)}(r) = \gamma \, \kappa^{(j)}(r) \widetilde{A}^{(j)}(r)  e^{\mathrm{i} \Phi(r')}
\end{equation}
where the functions $\kappa^{(j)}$, for $ \ell = 0$, are given by
\begin{align}
  \kappa^{(1)}(r) &= -\frac{\sqrt{2}}{3 \mu^2} \frac{f(r)}{r}, \\
  \kappa^{(2)}(r) &= \frac{\mathrm{i} \sqrt{2} \omega}{3 \mu^2},
\end{align}
and for  $ \ell = 1$, we have
\begin{align}
  \kappa^{(1)}(r) &= -\frac{\sqrt{2}}{3 \mu^2} \frac{f(r)}{r}, \\
  \kappa^{(2)}(r) &= -\frac{\sqrt{2}}{3 \mu^2}.
\end{align}
The phase is given by
\begin{equation}\label{phase} 
  \Phi(r) = \omega t_p(r) - m \varphi_p(r)
\end{equation}
and the factor $\gamma$ is given by 
\begin{equation}
 \gamma = 8 m_0 \sqrt{2\pi} A(\ell,m).
\end{equation}

Taking into account \eqref{Source_i_unified}, we can rewrite \eqref{Psi_i_unified} in Schwarzschild coordinates as
\begin{equation}\label{Psi_i_unified_bis}
  \Psi_{\omega\ell m}^{(i)}(r) = \gamma \int_{2M}^{6M} dr' \sum_{j=1}^{2} \widetilde{\mathcal{G}}_{ij}(r,r') \widetilde{A}^{(j)}(r')  e^{\mathrm{i} \Phi(r')}
\end{equation}
where \( \widetilde{\mathcal{G}}_{ij}(r, r') \) is defined by
\begin{align}
    \widetilde{\mathcal{G}}_{ij}(r,r') = \kappa^{(j)}(r') \frac{\mathcal{G}_{ij}(r,r')}{f(r')}.
\end{align}

To apply the regularization process, we will decompose the amplitudes \( \widetilde{A}^{(j)}(r) \) into a divergent part and a regular part (or more precisely, a part regularized by the phase oscillations). We thus write
\begin{equation}\label{Amp_K_H_l_1}
  \widetilde{A}^{(j)}(r) = \widetilde{A}_{\mathrm{div}}^{(j)}(r) + \widetilde{A}_{\mathrm{reg}}^{(j)}(r)
\end{equation}
with 
\begin{equation}\label{decomp}
  \widetilde{A}_{\mathrm{div}}^{(j)} = \frac{c_1}{(6M - r)^3} + \frac{c_2}{(6M - r)^{5/2}} + \frac{c_3}{(6M - r)^2}
\end{equation}
where the coefficients  \( c_1, c_2, c_3 \) are 
\begin{align}
  c_1 &= i\left(\sqrt{6} m - 36M\omega\right) \label{decomp_coeff_1}, \\
  c_2 &= \frac{\sqrt{3}}{2\sqrt{M}} \label{decomp_coeff_2},\\
  c_3 &= \frac{i m}{2\sqrt{6}M} + 6 i \omega \label{decomp_coeff_3}.
\end{align}

To perform the regularization, we will follow the steps described in \cite{Folacci:2018vtf}. Specifically, to reduce the order of the divergence, we insert \eqref{Amp_K_H_l_1} into \eqref{Psi_i_unified_bis} using Eq. (A. 20) from \cite{Folacci:2018vtf}, and we obtain:
\begin{widetext}
\begin{align}\label{Psi_i_regularized}
  \Psi_{\omega\ell m}^{(i)}(r) = \gamma \Biggl\{&-\frac{1}{\sqrt{3M}}\sum_{j=1}^{2} \frac{\widetilde{\mathcal{G}}_{ij}(r, 2M)}{(4M)^{3/2}}  e^{\mathrm{i} \Phi(2M)}\nonumber\\
   &+\int_{2M}^{6M} dr' \sum_{j=1}^{2} \widetilde{\mathcal{G}}_{ij}(r,r') \widetilde{A}_{\mathrm{reg}}^{(j)}(r') e^{\mathrm{i} \Phi(r')} \nonumber\\
   &- \frac{1}{\sqrt{3M}} \int_{2M}^{6M} dr' \sum_{j=1}^{2} \left[\frac{d}{dr'}\widetilde{\mathcal{G}}_{ij}(r,r') + i \widetilde{\mathcal{G}}_{ij}(r,r') \Theta_{\text{reg}}(r') \right] \frac{ e^{\mathrm{i} \Phi(r')}}{(6M - r')^{3/2}}\Biggr\}
\end{align}
\end{widetext}
with the function $\Theta_{\text{reg}}(r)$ constructed from the phase \eqref{phase} (see the appendix of Ref.~\cite{Folacci:2018vtf} for more details). We recall
\begin{equation}\label{Theta_reg}
  \Theta_{\text{reg}}(r) = \frac{d}{dr}\left[\Phi(r)- \frac{c}{\sqrt{6M-r}}\right]-\frac{d}{\sqrt{6M-r}}
\end{equation} 
where
\begin{equation}
  c =6\sqrt{2M} (m - 6\sqrt{6} M\omega)
\end{equation}
and
\begin{equation}
  d =\frac{m + 12\sqrt{6}M\omega}{2\sqrt{2M}}
\end{equation}

For $\ell \geq 2$ modes , the partial amplitudes are described by \eqref{Sol_Even_l_2_Coupled_inf_frequency} and can be rewritten in the form
\begin{equation}\label{Psi_i_unified_l_2}
\hspace{-9pt}  \bm{\Psi}_{\omega\ell m }(r) \equiv \begin{pmatrix} \Psi^{(1)} \\ \Psi^{(2)} \\  \Psi^{(3)} \end{pmatrix} = \int_{-\infty}^{+\infty} dr'_* \bm{\mathcal{G}}(r_*,r'_*) \bm{\mathcal{S}}_{\omega\ell m}(r'_*)
\end{equation}
where the partial amplitudes are given by $\Psi^{(1)} = K$, $\Psi^{(2)} = H_r$, and $\Psi^{(3)} = G$.  The \( \bm{\mathcal{G}}(r_*,r'_*) \) is the \( 3 \times 3 \) Green's matrix
\begin{equation}
\hspace{-9pt} \bm{\mathcal{G}}(r_*,r'_*) = 
\begin{pmatrix}
\mathcal{G}_{11}(r_*,r'_*) & \mathcal{G}_{12}(r_*,r'_*) & \mathcal{G}_{13}(r_*,r'_*) \\
\mathcal{G}_{21}(r_*,r'_*) & \mathcal{G}_{22}(r_*,r'_*) & \mathcal{G}_{23}(r_*,r'_*) \\
\mathcal{G}_{31}(r_*,r'_*) & \mathcal{G}_{32}(r_*,r'_*) & \mathcal{G}_{33}(r_*,r'_*)
\end{pmatrix}
\end{equation}

The partial amplitudes $\Psi^{(i)}$, can be explicitly written 
\begin{equation}\label{Psi_i_unified_2}
  \Psi_{\omega\ell m}^{(i)}(r) = \int_{-\infty}^{+\infty} dr'_* \sum_{j=1}^{3} \mathcal{G}_{ij}(r_*,r'_*) S_{\omega\ell m}^{(j)}(r'_*)
\end{equation}
where the source terms can be written in the form \eqref{Source_i_unified}, with the functions $\kappa^{(j)}$ given by
\begin{align}
  \kappa^{(1)}(r) &= -\frac{\sqrt{2}}{3 \mu^2} \frac{f(r)}{r},\\
  \kappa^{(2)}(r) &= -\frac{\sqrt{2}}{3 \mu^2},\\
  \kappa^{(3)}(r) &= f(r).
\end{align}

The regularization process follows the same steps as previously outlined. We express the partial amplitudes \eqref{Psi_i_unified_2} in the Schwarzschild coordinate, yielding \eqref{Psi_i_unified_bis} with the sum running from $1$ to $3$. The decomposition of \(\widetilde{A}^{(j)}\) into divergent and regular parts is applied only to the components \(\Psi^{(1)}\) and \(\Psi^{(2)}\) (i.e., \(K\) and \(H_r\)), as their source terms are divergent [see \eqref{source_K_main} and \eqref{source_H_main}]. However, the source term for \(\Psi^{(3)}\) (i.e., \(G\)) given by \eqref{source_G_main} does not require this treatment. The terms \(\widetilde{A}^{(1)}\) and \(\widetilde{A}^{(2)}\) are then expressed in the form \eqref{decomp}, with the coefficients given by \eqref{decomp_coeff_1}--\eqref{decomp_coeff_3}. Successive integration by parts is applied, leading to
\begin{widetext}
\begin{align}\label{Psi_i_regularized_l_2}
  \Psi_{\omega\ell m}^{(i)}(r) = \gamma \Biggl\{&-\frac{1}{\sqrt{3M}}\sum_{j=1}^{2} \frac{\widetilde{\mathcal{G}}_{ij}(r, 2M)}{(4M)^{3/2}}  e^{\mathrm{i} \Phi(2M)}\nonumber\\
   &+\int_{2M}^{6M} dr' \sum_{j=1}^{2} \widetilde{\mathcal{G}}_{ij}(r,r') \widetilde{A}_{\mathrm{reg}}^{(j)}(r') e^{\mathrm{i} \Phi(r')}
    +\int_{2M}^{6M} dr' \widetilde{\mathcal{G}}_{i3}(r,r') \widetilde{A}^{(3)}(r') e^{\mathrm{i} \Phi(r')} \nonumber\\
   &- \frac{1}{\sqrt{3M}} \int_{2M}^{6M} dr' \sum_{j=1}^{2} \left[\frac{d}{dr'}\widetilde{\mathcal{G}}_{ij}(r,r') + i \widetilde{\mathcal{G}}_{ij}(r,r') \Theta_{\text{reg}}(r') \right] \frac{ e^{\mathrm{i} \Phi(r')}}{(6M - r')^{3/2}}\Biggr\}
\end{align}
\end{widetext}

\section{Sources due to a point particle on a circular orbit and associated waveforms}
\label{appendix_D}

In this appendix, we provide an exact expression for the waveform emitted by a particle located on the ISCO, giving insight into the adiabatic phase of the waveform generated by a point particle on a plunging trajectory (see Sec. \ref{sec_5}).

In this context, we assume that the particle is on a stable circular orbit with a constant radius \( r_0 \), such that \( r_p(\tau) = r_0 = \text{Const} \) according to the geodesic equations \eqref{geodesic_eq}. Once integrated, these equations give expressions for the angular momentum and energy of the particle
\begin{equation}\label{L_ISCO}
  \widetilde{L} = \Bigg(\frac{M r_0}{1-\frac{3M}{r_0}}\Bigg)^{1/2}
\end{equation}
and
\begin{equation}\label{E_ISCO}
  \widetilde{E} = \frac{\left(1-\frac{2M}{r_0}\right)}{\left(1-\frac{3M}{r_0}\right)^{1/2}}
\end{equation}
By describing the particle's motion in terms of the proper time $\tau$, we find that the angular coordinate $\varphi_p$ is given by 
\begin{equation}\label{phi_ISCO_tau}
  \varphi_p(\tau) = \left(\frac{M}{r_0^3\left(1-\frac{3M}{r_0}\right)}\right)^{1/2} \tau.
\end{equation}
Alternatively, if we use the Schwarzschild time $t$, we have
\begin{equation}\label{phi_ISCO_t}
  \varphi_p(t) = \Omega t
\end{equation}
where
\begin{equation}\label{Angular_velocity}
  \Omega  = \sqrt{\frac{M}{r_0^3}}
\end{equation}
denotes the angular velocity of the particle (i.e., its orbital frequency).

\subsection{Odd-parity sector}
\label{appendix_D_1}

The odd-parity source terms \( S^{(\phi)}_{\omega \ell m}(r) \) and \( S^{(\psi)}_{\omega \ell m}(r) \) can be constructed from the components of the stress-energy tensor (\eqref{source_phi} and \eqref{source_psi}). By using the stress-energy tensor \eqref{SET_moving} and the orthonormalization properties of the spherical harmonics (scalar, vector, and tensor), we obtain 
\begin{equation}\label{soure_phi_tot_ISCO}
S_{\omega \ell m}^{(\phi)}(r)  =  0 
\end{equation}
and
\begin{multline}\label{soure_psi_tot_ISCO}
 S_{\omega \ell m}^{(\psi)}(r) = -i\, m\, 32\, \sqrt{2}\, \pi ^{3/2} \frac{B(\ell,m)}{\Lambda(\Lambda + 2)} \frac{M m_0/r_0}{\left(1-\frac{3M}{r_0}\right)}  \\ 
 \times  \frac{f(r)}{r} \delta(r-r_0) \delta(\omega-m\Omega)
\end{multline}

The $(\ell,m)$  odd-parity waveform generated by the particle orbiting the BH on a circular orbit with radius $r_0$ can now be derived by substituting the source terms \eqref{soure_phi_tot_ISCO} and \eqref{soure_psi_tot_ISCO} in Eq. \eqref{Sol_Odd_Coupled_inf_time}. After integration, we have
\begin{multline}\label{Sol_Odd_Coupled_inf_time_ISCO}
  \bm{\Phi}_{\ell m}(t,r) =\frac{e^{-i m \Omega t}}{\sqrt{2\pi}}  \\ \times \left\{\frac{1}{f(r_0)} \bm{U} \bm{W^{(\textbf{up})}}(r) \bm{W}^{-1}(r_0) \bm{L}  \bm{\mathcal{S}}_{\omega\ell m}(r_0)\right\}
\end{multline}  
where vector amplitude  \( \bm{\Phi}_{\ell m}(t, r) = \begin{pmatrix} \phi_{\ell m} & \psi_{\ell m} \end{pmatrix}^\top \), and the source vector \( \bm{\mathcal{S}}_{\omega \ell m} \) has components \( S_{\omega \ell m}^{(\phi)} \) and \( S_{\omega \ell m}^{(\psi)} \), given by \eqref{soure_phi_tot_ISCO} and \eqref{soure_psi_tot_ISCO}, respectively.

\subsection{Even-parity sector}
\label{appendix_D_1}

The even-parity source terms (\( S^{(K)}_{\omega \ell m}(r) \), \( S^{(H)}_{\omega \ell m}(r) \), and \( S^{(G)}_{\omega \ell m}(r) \) ) can be constructed from the components of the stress-energy tensor             
\eqref{S_K_even_tot}, \eqref{S_H_even_tot}, and \eqref{S_G_even_tot}. By using the stress-energy tensor \eqref{SET_moving} and the orthonormalization properties of the spherical harmonics, we obtain

\begin{widetext}
\begin{equation}\label{Source_X_Circ}
S^{(i)}_{\omega \ell m}(r) = \frac{16 \sqrt{2} \, m_0 \, \pi^{3/2} \, A(\ell, m)}{3 \sqrt{1 - \frac{3M}{r_0}} \, r_0^4 \, \mu^2} \Biggl( \mathcal{C}^{(i)}(r)  \delta(r - r_0) + \mathcal{D}^{(i)}(r)\delta'(r - r_0) \Biggr)\, \delta(\omega - m \Omega)
\end{equation}
where $i$ denotes $K$, $H$, and $G$. The coefficients $\mathcal{C}^{(K)}$ and $\mathcal{D}^{(K)}$ of the source term \( S^{(K)}_{\omega \ell m}\) are given by
\begin{align}
\mathcal{C}^{(K)}(r) &= \frac{1}{2 r^4 \biggl[\Bigl(\mu^2 r^2   + \Lambda + 2\Bigr)^2 + 4 r^2 \omega^2 - 2 \Bigl(\Lambda + 2 - \mu^2 r^2 \Bigr) f(r)\biggr]} \notag \\
&\quad \times \Biggl\{-r^2 \Bigl(r_0 - 2 M \Bigr)^2 \biggl[(2 \mu^2 r^2 + \Lambda + 2) \Bigl(4 \omega^2 r^2  + \left(\mu^2  r^2 + \Lambda + 2 \right)^2 \Bigr) \notag \\
&\qquad - 2 \Bigl( (3 \Lambda + 8) (\Lambda + 2) + \mu^2 r^2 ( \Lambda + 6)  - 4 \mu^4 r^4  \Bigr) f(r) + 4 \left(2 \mu^2 r^2 + 3 \Lambda + 6\right) f(r)^2\biggr] \notag \\
&\qquad + M r_0^3 f(r) \biggl[(\Lambda - \mu^2 r^2 ) \Bigl(  \left(\Lambda + 2  + \mu^2r^2 \right)^2 + 4 r^2 \omega^2\Bigr) \notag \\
&\qquad\quad + 8 r^2 f(r) \Bigl(2 \mu^2 (\Lambda + 2) + \mu^4 r^2  + 3 \omega^2 - 2 \mu^2 f(r)\Bigr)\biggr]\Biggr\}, \label{C_Source_K_Circ} \\
\mathcal{D}^{(K)}(r) &= \frac{r_0^2 f(r)}{r^3} \left(r^2 f(r_0)^2 - M r_0 f(r)\right), \label{D_Source_K_Circ}
\end{align}
those for the source term \( S^{(H)}_{\omega \ell m}\) are
\begin{align}
\mathcal{C}^{(H)}(r) &= \frac{r_0^2}{r^3 \Bigl[\Bigl(\mu^2 r^2 + \Lambda + 2\Bigr)^2 + 4 r^2 \omega^2 - 2 \Bigl(\Lambda + 2 - r^2 \mu^2\Bigl) f(r)\Bigr]} \notag \\
&\quad \times \Biggl\{ \frac{r^2}{r_0^2} \Bigl(r_0 - 2 M\Bigl)^2 \biggl[(\Lambda + 2) (\Lambda + 4 - 4 f(r)) - 3 \mu^4 r^4 - 2 r^2 (\Lambda \mu^2 + 2 \omega^2 + 3 \mu^2 f(r))\biggr] \notag \\
&\qquad - M r_0 \biggl[\Bigl(\mu^2 r^2 + \Lambda + 2\Bigl)^2 + 4 \omega^2 r^2 - 2 f(r) \Bigl(\Lambda^2 + 4 + 3 \mu^4 r^4 + 4 \Lambda (\mu^2 r^2 + 1) + 8 r^2 (\mu^2 + \omega^2) \notag \\
&\qquad\qquad - (\Lambda + 2 + 3 \mu^2 r^2) f(r)\Bigl)\biggr]\Biggr\}, \label{C_Source_H_Circ} \\
\mathcal{D}^{(H)}(r) &= \frac{r}{f(r)} \mathcal{D}^{(K)}(r), \label{D_Source_H_Circ}
\end{align}
and the coefficients for the source \( S^{(G)}_{\omega \ell m} \) are
\begin{align}
\mathcal{C}^{(G)}(r) &=  \frac{r_0^2 f(r_0)^2}{r^2} - \frac{M r_0^3 \Bigl[\Lambda (\Lambda + 2) + 3 \mu^2 r^2 \left(2 (1 - m^2) + \Lambda\right)\Bigr] f(r)}{r^4 \Lambda (\Lambda + 2)}, \label{C_Source_G_Circ} \\
\mathcal{D}^{(G)}(r) &=  0. \label{D_Source_G_Circ}
\end{align}
\end{widetext}

The even-parity \((\ell \geq 2, m)\) waveform generated by a particle orbiting the BH in a circular orbit of radius \(r_0\) can now be derived by substituting the source terms from Eq. \eqref{Source_X_Circ} into Eq. \eqref{Sol_Even_l_2_Coupled_inf_time}. After integration by parts, we obtain
\begin{multline}\label{Sol_Even_Coupled_inf_time_ISCO}
\hspace{-12pt}\bm{\Psi}_{\ell m}(t,r) = \frac{e^{-i m \Omega t} }{\sqrt{2\pi}}  \\
\hspace{-55pt}\times \Biggl\{\frac{1}{f(r_0)} \bm{U} \bm{W^{(\textbf{up})}}(r) \bm{W}^{-1}(r_0) \bm{L}  \bm{\mathcal{S}}^{(1)}_{\omega\ell m}(r_0)\\
\hspace{-15pt}-\frac{d}{dr'} \left[\frac{1}{f(r')} \bm{U} \bm{W^{(\textbf{up})}}(r) \bm{W}^{-1}(r') \bm{L}  \bm{\mathcal{S}}^{(2)}_{\omega\ell m}(r')\right]_{r' = r_0}  
  \Biggr\}
\end{multline}  
where the vector amplitude \( \bm{\Psi}_{\ell m}(t, r) = \begin{pmatrix} K^{\ell m} & H_r^{\ell m} & G^{\ell m} \end{pmatrix}^\top \), and the source vectors are defined as \( \bm{\mathcal{S}}^{(1)}_{\omega \ell m} = \begin{pmatrix} S^{(K,1)}_{\omega \ell m} & S^{(H,1)}_{\omega \ell m} & S^{(G,1)}_{\omega \ell m} \end{pmatrix}^\top \) and \( \bm{\mathcal{S}}^{(2)}_{\omega \ell m} = \begin{pmatrix} S^{(K,2)}_{\omega \ell m} & S^{(H,2)}_{\omega \ell m} & S^{(G,2)}_{\omega \ell m} \end{pmatrix}^\top \), with
\begin{fleqn}
\begin{equation}\label{S_1_i}
  S^{(i,1)}_{\omega \ell m}(r) = \frac{16 \sqrt{2} \, m_0 \, \pi^{3/2} \, A(\ell, m)}{3 \sqrt{1 - \frac{3M}{r_0}} \, r_0^4 \, \mu^2} \, \mathcal{C}^{(i)}(r) \, \delta(\omega - m \Omega)
\end{equation}
\end{fleqn}
and
\begin{fleqn}
\begin{multline}\label{S_2_i}
  S^{(i,2)}_{\omega \ell m}(r) = \frac{16 \sqrt{2} \, m_0 \, \pi^{3/2} \, A(\ell, m)}{3 \sqrt{1 - \frac{3M}{r_0}} \, r_0^4 \, \mu^2} \, \mathcal{D}^{(i)}(r) \, \delta(\omega - m \Omega).
\end{multline}
\end{fleqn}

It should be noted that for the dipole mode (\(\ell = 1\)), the source terms are also given by \eqref{Source_X_Circ}  with $ \ell = 1$  (i.e., $\Lambda = 0$) and  $m = 1$. The waveforms are obtained using \eqref{Sol_Even_Coupled_inf_time_ISCO}, with the amplitude vector \( \bm{\Psi}_{\ell m}(t, r) = \begin{pmatrix} K^{\ell m} & H_r^{\ell m} \end{pmatrix}^\top \).

Now, for a point particle located on the ISCO, we set \( r_0 = r_{\text{ISCO}} = 6M \) in the waveform expressions \eqref{Sol_Odd_Coupled_inf_time_ISCO} and \eqref{Sol_Even_Coupled_inf_time_ISCO} for odd- and even-parity, respectively. Also note that Eq. \eqref{Angular_velocity} is simplified to
\begin{equation}\label{Angular_Velocity_ISCO}
  \Omega_\text{ISCO} = \sqrt{\frac{M}{r_\text{ISCO}^3}} = \frac{1}{6\sqrt{6} M}.
\end{equation}

\bibliography{Waveforms_Massive_Gravity_EMRI}

\end{document}